\documentclass[conf]{new-aiaa}
\usepackage[utf8]{inputenc}
\usepackage{graphicx}
\usepackage{tabularx}
\usepackage{caption}
\usepackage{subcaption}
\usepackage{amsmath}
\usepackage[version=4]{mhchem}
\usepackage{siunitx}
\usepackage{longtable,tabularx}
\usepackage{color}  
\newcommand\nnfootnote[1]{%
  \begin{NoHyper}
  \renewcommand\thefootnote{}\footnote{#1}%
  \addtocounter{footnote}{-1}%
  \end{NoHyper}
}

\title{Numerical Investigation of Diffusion Flame in Transonic Flow with Large Pressure Gradient}

\author{Yalu Zhu\footnote{Assistant Specialist, Department of Mechanical and Aerospace Engineering, yalu.zhu@uci.edu (Corresponding Author), Member AIAA.}, 
Feng Liu\footnote{Professor, Department of Mechanical and Aerospace Engineering, Fellow AIAA.} 
and William A. Sirignano\footnote{Distinguished Professor, Department of Mechanical and Aerospace Engineering, Honorary Fellow AIAA.}} 
\affil{University of California, Irvine, Irvine, CA, 92697-3975}

\begin{document}

\nnfootnote{Presented as Paper 2024-xxxx at the AIAA SciTech 2024 Forum, Orlando, FL, January 8–12, 2024.}

\maketitle


\begin{abstract}
A finite-volume method for the steady, compressible, reacting, turbulent Navier-Stokes equations is developed by using a steady-state preserving splitting scheme for the stiff source terms in chemical reaction. Laminar and turbulent reacting flows in a mixing layer with large streamwise pressure gradient are studied and compared to boundary-layer solutions. It reveals that chemical reaction strongly enhances turbulent transport due to intensive production of turbulence by the increased velocity gradients and thus produces large turbulent viscosity in the reaction region. Influence of vitiated air on the combustion process and aerodynamic performance is also investigated for the cases of turbulent mixing layer and highly-loaded transonic turbine cascade. Both cases indicate viability for the turbine-burner concept.
\end{abstract}

\section{Nomenclature}
{\renewcommand\arraystretch{1.0}
\noindent\begin{longtable*}{@{}l @{\quad=\quad} l@{}}
$C$ & molar concentration \\
$C_p$ & specific heat capacity at constant pressure \\
$E$ & total energy \\
$e$ & internal energy \\
$h$ & enthalpy \\
$h^0$ & enthalpy of formation \\
$\mathbf{I}$ & Kronecker tensor \\
$\mathbf{j}$ & diffusive flux of species \\
$k$ & turbulent kinetic energy \\
$N$ & total number of species \\
$n$ & index of time step \\
$\mathrm{Pr}$ & Prandtl number \\
$p$ & pressure \\
$\dot{Q}$ & source term in energy equation due to reaction \\
$\mathbf{q}$ & heat flux in energy equation \\
$R$ & gas constant \\
$R_0$ & universal gas constant, $R_0 = 8.3145 \, \mathrm{J/mol/K}$ \\
$\mathrm{Sc}$ & Schmidt number \\
$T$ & temperature \\
$t$ & time \\
$u$ & velocity in $x$ direction \\
$\mathbf{V}$ & velocity vector, $\mathbf{V} = (u, \, v)$ \\
$v$ & velocity in $y$ direction \\
$W$ & molecular weight \\
$x,y$ & Cartesian coordinates \\
$Y$ & mass fraction \\

$\Delta t$ & time step \\
$\mu$ & viscosity coefficient \\
$\rho$ & density \\
$\boldsymbol{\tau}$ & viscous stress tensor \\
$\omega$ & specific dissipation rate \\
$\dot{\omega}$ & production rate of species by reaction \\

$(\cdot)^T$ & transpose of quantity \\
$\bar{(\cdot)}$ & Reynolds-averaged quantity \\
$\tilde{(\cdot)}$ & Farve-averaged quantity \\

$(\cdot)_i$ & species \\
$(\cdot)_T$ & quantity of turbulence \\
$(\cdot)_\infty$ & quantity at the top far away from the mixing layer \\
\end{longtable*}}

\section{Introduction}
\lettrine{T}o reduce the weight and widen the range of operation, the designer continues to pursue compact design of the combustor and the turbomachinery of a gas turbine engine. In a compact combustor, the fuel residence time becomes shorter than the the time required for complete combustion. As a result, the combustion process would be extended into the downstream turbine passage. This increases the challenge of heat transfer in turbine at the first sight. However, the thermodynamic analysis by Sirignano and Liu \cite{sirignano1999performance, liu2001turbojet} showed that intentional augmented burning in the turbine passage, called turbine-burner, allows for significant benefits: 1) reduction in after-burner length and weight, 2) reduction in specific fuel consumption, and 3) increase in specific thrust. 

To take advantage of the turbine-burner design concept, it is necessary to address some fundamental issues of aerodynamics and combustion associated with it. In a turbine passage, the compressible turbulent flow is subjected to strong streamwise and transverse pressure gradients produced by the turbine blade profiles. 
The flow accelerates from subsonic to supersonic in a very short distance, creating a challenge for flameholding. Large gradients of temperature, velocity, and species concentration occur on the fuel-oxidizer interface due to mixing and combustion. This can result in hydrodynamic instabilities that might significantly affect the energy conversion, heat transfer, and aerodynamic loading on the turbine blades, and the character of turbulent flow \cite{sirignano2012turbine}. 

The high-accelerating transonic flow with mixing and chemical reaction is an important area of applied scientific research.
Sirignano and Kim \cite{sirignano1997diffusion} obtained the similarity solution for a diffusion flame in the two-dimensional, laminar, steady, compressible mixing layer with constant pressure gradient along the streamwise direction. 
Fang et al. \cite{fang2001ignition} extended the study to include cases with arbitrary pressure gradients by using a finite-difference method for the boundary-layer equations. The influence of pressure gradient, initial temperature, initial velocity, and transport properties on the ignition process and flame structure was studied.
Mehring et al. \cite{mehring2001ignition} further extended the laminar boundary-layer computation to include the effects of turbulence. 
Cai et al. \cite{cai2001ignition, cai2001combustion} investigated the turbulent, transonic, reacting flows in a mixing layer and curved duct by solving the two-dimensional Reynolds-Averaged Navier-Stokes (RANS) equations. The flame structures in transonic flows with large axial and transverse pressure gradients were examined.
Cheng et al. \cite{cheng2007nonpremixed, cheng2008nonpremixed, cheng2009reacting} simulated the development of reacting mixing layers in straight and curved ducts from laminar flow to transition regime by solving the two-dimensional Navier-Stokes equations. 
These numerical computations of the reacting flows with pressure gradients are based on the boundary-layer equations or the two-dimensional Navier-Stokes equations in which the pressure gradient is provided by the variation of flow passage width in the third direction. In order to simulate the reacting turbulent flow in a real turbine, numerical methods to solve the full three-dimensional Navier-Stokes equations with chemical reaction must be developed.

Since the turbine vane is downstream of the primary combustion chamber in a gas turbine engine, the gases at the turbine entrance are a mixture of unburned air and reaction products. However, a pure gas model, treating the medium as a single species with altered specific heat capacities, is usually used for flow simulation in a turbine when there is no chemical reaction and heating \cite{zhu2017numerical}. This simplification has little influence on the aerodynamic performance and heat transfer in a turbine. However, it significantly affects the heat release and flow characteristics if a combustion process is incorporated into a turbine due to the different chemically thermodynamic properties of each species. To accurately model the reacting flow and thus predict its influence on the aerodynamic and thermodynamic performance of a turbine, the vitiated air composed of unburned air and reaction products should be used at the turbine entrance instead \cite{walsh2024}. The differences between pure air and vitiated air are also necessary to be evaluated.

In the present paper, a code to solve the three-dimensional compressible RANS equations with chemical reaction and turbulence models by using finite-volume method is developed and implemented. The code is then applied to study the laminar and turbulent reacting flows in an accelerating mixing layer and to compare the vitiated air and pure air in the same mixing layer and a real turbine cascade. 
The governing equations and numerical methods are presented in Sec. \ref{sec:governing_equations} and Sec. \ref{sec:numerical_methods}, respectively. 
The nonreacting laminar flow in a mixing layer is presented in Sec. \ref{sec:nonreaction_laminar}. The reacting laminar case is discussed in Sec. \ref{sec:reaction_laminar}. The reacting turbulent case is given in Sec. \ref{sec:reaction_turbulent}. The differences between vitiated air and pure air for the turbulent mixing layer are discussed in Sec. \ref{sec:vitiated_air}. The reacting turbulent flow in a turbine cascade is analyzed in Sec. \ref{sec:turbine_cascade}. 
The concluding remarks are given in Sec. \ref{sec:conclusion}.

\section{Governing Equations} \label{sec:governing_equations}

\subsection{Reynolds-Averaged Navier-Stokes Equations}
The Reynolds-Averaged Navier-Stokes (RANS) equations for compressible flows are expressed by the following transport equations for mass, momentum and energy
\begin{subequations} \label{eq:navier-stokes}
    \begin{align}
        \frac{\partial\bar{\rho}}{\partial t} + \nabla \cdot (\bar{\rho}\widetilde{\mathbf{V}}) & 
            = 0 \label{eq:continuity} \\
        \frac{\partial(\bar{\rho} \widetilde{\mathbf{V}})}{\partial t} + \nabla \cdot (\bar{\rho}\widetilde{\mathbf{V}} \widetilde{\mathbf{V}}) &
            = -\nabla \bar{p} + \nabla \cdot \boldsymbol{\tau} \label{eq:momentum} \\
        \frac{\partial(\bar{\rho} \widetilde{E})}{\partial t} + \nabla \cdot (\bar{\rho} \widetilde{E}\widetilde{\mathbf{V}}) & = -\nabla \cdot (\bar{p} \widetilde{\mathbf{V}}) + \nabla \cdot (\widetilde{\mathbf{V}} \cdot \boldsymbol{\tau}) - \nabla \cdot \mathbf{q} + \dot Q \label{eq:energy} 
    \end{align}
\end{subequations}
The chemical reaction in the flow is taken into consideration by the mass fraction transport equation for each species in a mixture with $N$ species
\begin{equation} \label{eq:species}
    \frac{\partial{\bar{\rho} \widetilde{Y}_i}}{\partial t} + \nabla \cdot (\bar{\rho} \widetilde{Y}_i \widetilde{\mathbf{V}}) = -\nabla \cdot \mathbf{j}_i + \dot{\omega}_i, \ i = 1, \ 2, \ ..., \ N  
\end{equation}

The energy equation is expressed in terms of the total energy, $\widetilde{E}$, which consists of the internal energy and the kinetic energy, i.e.
\begin{equation}
    \widetilde{E} = \tilde{e} + \frac{1}{2}\widetilde{\mathbf{V}} \cdot \widetilde{\mathbf{V}}
\end{equation}
where the internal energy $\tilde{e}$ is related to the enthalpy $\tilde{h}$ by 
\begin{equation}
    \tilde{e} = \tilde{h} - \frac{\bar{p}}{\bar{\rho}}
\end{equation}
The enthalpy is the summation of the sensible enthalpy weighted by the mass fraction
\begin{equation}
    \tilde{h} = \sum_{i=1}^{N}{\widetilde{Y}_i \tilde{h}_i}
\end{equation}
with 
\begin{equation}
    \tilde{h}_i = \int_{T_0}^{\widetilde{T}}{C_{p,i}\mathrm{d}T}
\end{equation}
where the specific heat capacity at constant pressure $C_{p,i}$ is a function of temperature given by the empirical polynomial formula of NASA \cite{mcbride1993coefficients} for each species. An additional heat source term $\dot Q$ appears on the right-hand side of the energy 
equation
\begin{equation}
    \dot Q = -\sum_{i=1}^{N}{\dot\omega_i h_i^0}
\end{equation}
where $h_i^0$ is the enthalpy of formation of species $i$ at the reference temperature $T_0$.

The viscous stress tensor $\boldsymbol{\tau}$ is the sum of the molecular stress tensor $\boldsymbol{\tau}_L$ and turbulent stress tensor $\boldsymbol{\tau}_T$ with
\begin{subequations}
    \begin{align}
    \boldsymbol{\tau}_L &= \mu \left[\nabla \widetilde{\mathbf{V}} + (\nabla \widetilde{\mathbf{V}})^T \right] -\frac{2}{3}\mu \left(\nabla \cdot \widetilde{\mathbf{V}} \right) \mathbf{I} \\
    \boldsymbol{\tau}_T &= \mu_T \left[\nabla \widetilde{\mathbf{V}} + (\nabla \widetilde{\mathbf{V}})^T \right] -\frac{2}{3}\mu_T \left(\nabla \cdot \widetilde{\mathbf{V}} \right) \mathbf{I}
    \end{align}
\end{subequations}
where $\mu$ is the molecular viscosity computed by the mass-weighted summation of molecular viscosity of each species given by the Sutherland's law \cite{white2006viscous}, and $\mu_T$ is the turbulent viscosity determined by the turbulence model in the next subsection.

The diffusive flux of species $i$ is given by 
\begin{equation}
    \mathbf{j}_i = -\left( \frac{\mu}{\mathrm{Sc}_i} + \frac{\mu_T}{\mathrm{Sc}_T} \right) \nabla \widetilde{Y}_i
\end{equation}
where $\mathrm{Sc}_i$ and $\mathrm{Sc}_T$ are the Schmidt number of species $i$ and the turbulent Schmidt number, respectively. In the present study, we set $\mathrm{Sc}_i = 1.0$ and $\mathrm{Sc}_T = 1.0$.

The heat flux in the energy equation is computed by 
\begin{equation}
    \mathbf{q} = -\left( \frac{\mu}{\mathrm{Pr}} + \frac{\mu_T}{\mathrm{Pr}_T} \right) \nabla \tilde{h} + \sum_{i=1}^{N}{\tilde{h}_i\mathbf{j}_i}
\end{equation}
where the last term stands for the energy transport due to mass diffusion of each species with different enthalpy, and $\mathrm{Pr}$ and $\mathrm{Pr}_T$ are the Prandtl number and the turbulent Prandtl number, respectively. In the present study, we set $\mathrm{Pr} = 1.0$ and $\mathrm{Pr}_T = 1.0$.

A perfect gas is assumed in this study, in which the pressure, density and temperature are related by the equation of state
\begin{equation}
    \bar{p} = \bar{\rho}R\widetilde{T}
\end{equation}
where $R$ is the gas constant of the mixture, computed by the mass-weighted summation of the gas constant of each species $R_i$ with $R_i = R_0/W_i$.

\subsection{Turbulence Model}
The improved $k\mbox{-}\omega$ Shear-Stress Transport (SST) model presented by Menter et al. \cite{menter2003ten} in 2003 is used to evaluate the turbulent viscosity.  
This model combines the advantages of the $k\mbox{-}\varepsilon$ model and the $k\mbox{-}\omega$ model.  
In the inner zone of a boundary layer, the SST model degenerates into the $k\mbox{-}\omega$ model, which avoids the stiff source term in the $k\mbox{-}\varepsilon$ model due to the damping function in the viscous sublayer and the defect in capturing the proper behaviors of turbulent flows with adverse pressure gradients up to separation.
In the outer zone of a boundary layer or in a free-shear flow, the SST model switches to the $k\mbox{-}\varepsilon$ model, avoiding the strong sensitivity of the $k\mbox{-}\omega$ model to freestream turbulence. 
In addition, the SST model takes into account the transport of principal turbulent shear stress to enhance its ability to predict turbulent flows with adverse pressure gradient and separation.

The SST model is established by the transport equations for turbulent kinetic energy $k$ and specific dissipation rate $\omega$
\begin{subequations} \label{eq:k-omega}
    \begin{align}
        \dfrac{\partial \bar{\rho} k}{\partial t} + \nabla\cdot(\bar{\rho} k \widetilde{\mathbf{V}}) &= P - \beta^*\bar{\rho} k\omega + \nabla \cdot [(\mu+\sigma_k\mu_T) \nabla k] \\
        \dfrac{\partial \bar{\rho} \omega}{\partial t} + \nabla\cdot(\bar{\rho}\omega \widetilde{\mathbf{V}}) &= \dfrac{\gamma\bar{\rho}}{\mu_T}P - \beta\bar{\rho}\omega^2 + \nabla\cdot\left[(\mu+\sigma_\omega\mu_T)\nabla\omega\right] + 2(1-F_1)\dfrac{\bar{\rho}\sigma_{\omega2}}{\omega}\nabla k\cdot \nabla\omega
    \end{align}
\end{subequations}
where the production term in the $k$ equation is
\begin{equation} \label{eq:turbulent_Pk}
    P = \mathrm{min}(\mu_TS^2, 10\beta^*\bar{\rho} k\omega)
\end{equation}

The turbulent viscosity is then computed by 
\begin{equation} 
    \mu_T = \dfrac{a_1\bar{\rho} k}{\mathrm{max}\left(a_1 \omega, F_2S\right)}
\end{equation}
where $S$ is the second invariant of strain rate tensor
\begin{equation}
    S = \sqrt{2S_{ij}S_{ij}}, \ S_{ij} = \frac{1}{2}({\tilde{u}_{j,i} + \tilde{u}_{i,j}})
\end{equation}

The blending functions $F_1$ and $F_2$ are defined by, respectively
\begin{gather}
    \begin{split}
        F_1 &= \mathrm{tanh}(\Gamma^4), \ \Gamma = \mathrm{min}\left[\mathrm{max}\left(\dfrac{500\mu}{\bar{\rho}\omega d^2}, \; \dfrac{\sqrt{k}}{\beta^*\omega d}\right), \; \dfrac{4\sigma_{\omega 2} \bar{\rho} k}{C\!D_{k\omega} d^2}\right], \ C\!D_{k\omega} = \mathrm{max}\left(\frac{2\bar{\rho}\sigma_{\omega 2}}{\omega}\nabla k\cdot \nabla\omega, \; 10^{-10}\right) \\
        F_2 &= \mathrm{tanh}(\Pi^2), \ \Pi = \mathrm{max}\left(\dfrac{500\mu}{\bar{\rho}\omega d^2}, \; \dfrac{2\sqrt{k}}{\beta^*\omega d}\right)
    \end{split}
\end{gather} 
where $d$ is the distance to the nearest wall. $F_1$ is set to zero away from the wall ($k\mbox{-}\varepsilon$ model), and switched to one inside the boundary layer ($k\mbox{-}\omega$ model). $F_2$ is one for boundary-layer flows and zero for free-shear layers. Both of them are artificially set to zero in the turbulent mixing-layer case below.

Each of the constants in the SST model is computed by a blend of the corresponding constants of the $k\mbox{-}\omega$ and $k\mbox{-}\varepsilon$ models via $\phi = F_1 \phi_1 + (1-F_1) \phi_2$, where $\phi = \sigma_k$, $\sigma_\omega$, $\beta$, $\gamma$. The other constants are: $a_1 = 0.31$, $\beta^* = 0.09$, $\sigma_{k1} = 0.85$, $\sigma_{k2} = 1.0$, $\sigma_{\omega_1} = 0.5$, $\sigma_{\omega_2} = 0.856$, $\beta_1 = 3/40$, $\beta_2 = 0.0828$, $\gamma_1 = 5/9$, $\gamma_2 = 0.44$.

\subsection{Chemistry Model}
The combustion of methane ($\rm{CH_4}$) in air is considered in the current computations. The production rate $\dot\omega_i$ of each species due to chemical reaction is calculated by the Westbrook and Dryer's one-step reaction mechanism \cite{westbrook1984chemical}
\begin{equation} \label{eq:methane_reaction}
    {\rm CH}_4 + 2 {\rm O}_2 + 7.52 {\rm N}_2 \rightarrow {\rm CO}_2 + 2 {\rm H}_2{\rm O} + 7.52 {\rm N}_2 
\end{equation}
where only four species, i.e. methane ($\rm{CH_4}$), oxygen ($\rm{O_2}$), carbon dioxide ($\rm{CO_2}$), and water vapor ($\rm{H_2O}$), are tracked besides nitrogen ($\rm{N_2}$) in air. Thus, the number of species $N = 5$ in the present work. Note that although this work only focuses on one-step reaction of one type of fuel, it is straightforward to extend the present method to different fuels and oxidizers with more complex chemical reaction mechanisms if the increased computational costs are acceptable.

The reaction rate by the laminar kinetics is given by the modified Arrhenius expression combined with the law of mass action
\begin{equation} \label{eq:Arrhenius_expression}
    {\varepsilon} = A \widetilde{T}^\beta e^{-E_a/(R_0 \widetilde{T})} C_{\rm{fuel}}^a C_{\rm{ox}}^b 
\end{equation}
where "$\rm{fuel}$" and "$\rm{ox}$" stand for $\rm{CH_4}$ and $\rm{O_2}$ in this study, respectively, and $C_i$ is the molar concentration of species $i$ defined by $C_i = \bar{\rho} \widetilde{Y}_i/W_i$. According to Westbrook and Dryer \cite{westbrook1984chemical}, $A = 1.3 \times 10^{9} \, \rm{s^{-1}}$, $\beta = 0$, $E_a = 202.506 \, \rm{kJ/mol}$, $a = -0.3$, and $b = 1.3$ for methane. 
To be consistent with the setups of Fang et al. \cite{fang2001ignition} and Mehring et al. \cite{mehring2001ignition}, the values of $A$ are adjusted to be $2.8 \times 10^{9} \, \rm{s^{-1}}$ and $1.3 \times 10^{10} \, \rm{s^{-1}}$ in the laminar and turbulent cases below, respectively. Computations show that this change has little influence on flame development except ignition length. It is found that ignition may not happen if the original value of $A$ is used in the turbulent mixing-layer case.
Note that the influence of turbulence on the reaction rate has been neglected in this analysis. Since the turbulent length scale is orders of magnitude smaller than the scale across which the largest temperature gradient occurs in the ignition region of mixing-layer flow as analyzed by Mehring et al. \cite{mehring2001ignition}, the averaged reaction rate by Eq. (\ref{eq:Arrhenius_expression}) is considered to be reasonable.
The net rate of production of species $i$ by the chemical reaction is thus calculated by
\begin{equation}
    \dot\omega_i = W_i(v_i^{\prime\prime} - v_i^\prime) \varepsilon
\end{equation}
where $v_i^\prime$ is the stoichiometric coefficient for reactant $i$ in Eq. (\ref{eq:methane_reaction}), and $v_i^{\prime\prime}$ is the stoichiometric coefficient for product $i$. Obviously, the net production rate of nitrogen is zero.

\section{Numerical Methods} \label{sec:numerical_methods}

\subsection{Numerical Solver}
An in-house three-dimensional code of simulating the steady and unsteady transonic flows for single species within turbomachinery blade rows has been developed, validated, and applied by Refs. \cite{liu1993multigrid, yao2001development, sadeghi2005computation, zhu2018flow, zhu2018influence}. The code solves the Navier-Stokes equations together with various turbulence models by using the second-order cell-centred finite-volume method based on multi-block structured grid. The central schemes with artificial viscosity, flux difference splitting schemes, and advection upstream splitting methods with various options to reconstruct the left and right states have been developed and implemented in the code. In this study, the present code is extended to the case of multiple species with varying specific heat capacities and to include appropriate chemistry models.

The species transport equations (\ref{eq:species}) have the same form as the basic conservation equations (\ref{eq:navier-stokes}) and (\ref{eq:k-omega}). Consequently, the numerical methods for the Navier-Stokes equations are still applicable, except that the chemical source terms in both species equations and energy equation need to be treated separately to avoid stiffness. 
The convective and viscous fluxes are discretized by the JST scheme \cite{jameson1981numerical} and the second-order central scheme, respectively. 
The local time-stepping method is introduced to accelerate the convergence to steady state. Thus, the time $t$ in the governing equations (\ref{eq:navier-stokes}), (\ref{eq:species}), and (\ref{eq:k-omega}) is interpreted as a pseudo time, and a large enough pseudo-time step determined by the local flow field can be used in each grid cell since time accuracy is not required for steady-state solutions. 
The Lower-Upper Symmetric-Gauss-Seidel (LU-SGS) method \cite{yoon1988lower} is applied to the pseudo-time stepping to obtain the converged steady-state solutions. The parallel technique based on Message Passing Interface (MPI) is adopted to further accelerate the computation by distributing grid blocks among CPU processors.

Note that the continuity equation (\ref{eq:continuity}) and the $N$ species transport equations (\ref{eq:species}) are not independent of each other. The summation of the $N$ species equations should reduce to the continuity equation, which gives the following restrictions on the terms in the species transport equations
\begin{subequations}
    \begin{align}
        \sum_{i=1}^{N} \widetilde{Y}_i &= 1, \ 0 \leq \widetilde{Y}_i \leq 1 \label{eq:restrict_Yi} \\
        \sum_{i=1}^{N} \mathbf{j}_i &= 0 \label{eq:restrict_Ji} \\
        \sum_{i=1}^{N} \dot\omega_i &= 0 \label{eq:restrict_omega_i}
    \end{align}
\end{subequations}
These restrictions should be satisfied during the computation to maintain consistency of the final converged solution. Equation (\ref{eq:restrict_omega_i}) is automatically satisfied due to the balanced stoichiometric coefficients in Eq. (\ref{eq:methane_reaction}). Additional treatments should be adopted to guarantee the other two conditions. To ensure Eq. (\ref{eq:restrict_Yi}), after each iteration for Eq. (\ref{eq:species}), the mass fraction of each species is corrected as 
\begin{equation}
    \widetilde{Y}_i^{\rm corr} = \frac{\widetilde{Y}_i}{\sum_{k=1}^{N} \widetilde{Y}_k}
\end{equation}
To ensure Eq. (\ref{eq:restrict_Ji}), the diffusive flux of each species is corrected as
\begin{equation}
    \mathbf{j}_i^{\rm corr} = \mathbf{j}_i - \widetilde{Y}_i\sum_{k=1}^{N} \mathbf{j}_k
\end{equation}

\subsection{Splitting Scheme}
In the species and energy equations, the source terms exhibit fundamentally different physical properties from the terms of convection and diffusion due to significantly smaller time scales for chemical reactions than for the flow, resulting in strong stiffness in solving the governing equations. The operator-splitting method is a natural choice to achieve efficient integration in time for unsteady problem. 

Equations (\ref{eq:species}) and (\ref{eq:energy}) can be rewritten as
\begin{equation} \label{eq:stiff_integration}
    \frac{\mathrm{d}W}{\mathrm{d}t} = {\cal{T}}(W)+ {\cal{S}}(W)
\end{equation}
where $W = [\bar\rho \widetilde{Y}_i, \, \bar\rho \widetilde{E}]^T$, with size of $N+1$, is the state variables. ${\cal{T}}$ and ${\cal{S}}$ represent the transport term (convection and diffusion) and the reacting source term, respectively. The operator-splitting method integrates the two terms sequentially. Consider the integration from time step $n$ to $n+1$ over the time interval of $\Delta t$. A natural splitting scheme first integrates the Ordinary Differential Equation (ODE) of ${\cal{S}}$ over the time interval $\Delta t$ with the solution at the time step $n$ as initial value, and then solves the Partial Differential Equation (PDE) of ${\cal{T}}$ over $\Delta t$ with the solution at the end of the first step as initial value. It is easy to prove that this simplest splitting is of first-order accuracy. The accuracy can be improved to second order by using the Strang splitting scheme \cite{strang1968construction}, in which the integration proceeds in a symmetric way: first a half interval $\Delta t/2$ is taken with the ${\cal{S}}$ operator, then a full interval $\Delta t$ with the ${\cal{T}}$ operator, and finally another half interval $\Delta t/2$ with the ${\cal{S}}$ operator. Both splitting schemes are strongly stable and applicable to unsteady problems. However, they are not steady-state preserving \cite{wu2019efficient}. A numerical integration method is called steady-state preserving, if given an initial solution $W_0$ satisfying ${\cal{T}}(W_0)+ {\cal{S}}(W_0) = 0$, the solution of the next step remains $W_0$, regardless of the step size.

We propose here a steady-state preserving splitting scheme for solving stiff reacting flow. The integration of Eq. (\ref{eq:stiff_integration}) from pseudo-time step $n$ to $n+1$ over the time interval of $\Delta t$ is split into two separate sub-integrations
\begin{subequations} \label{eq:splitting_scheme}
    \begin{align}
        \frac{\mathrm{d}W^*}{\mathrm{d}t} &= {\cal{T}}(W^n)+ {\cal{S}}(W^*), \ \ (W^*)^n = W^n   \label{eq:splitting_scheme_ode} \\ 
        \frac{\mathrm{d}W}{\mathrm{d}t} &= {\cal{R}}, \ \ {\cal{R}} = \frac{(W^*)^{n+1} - W^n}{\Delta t}  \label{eq:splitting_scheme_pde} 
    \end{align}
\end{subequations}

In the first chemical sub-integration, Eq. (\ref{eq:splitting_scheme_ode}) is integrated by a stiff ODE solver over the time interval $\Delta t$ with the initial value $W^n$, giving an intermediate value $(W^*)^{n+1}$ at the end of the sub-integration. Since the time scale of chemical reaction is several orders of magnitude smaller than that of the flow, the transport term ${\cal{T}}$ is evaluated at time step $n$ and is kept unchanged within the chemical sub-integration. 
Although advanced full-implicit method or Quasi-Steady-State (QSS) method \cite{mott2000quasi} is usually chosen as the stiff ODE solver, the simplest Euler explicit integration method is applied to solve Eq. (\ref{eq:splitting_scheme_ode}) in the present study due to the extremely stiff source terms introduced by the empirical global reaction mechanism in the absence of reverse reaction. The chemical time step in the explicit integration is determined by the spectral radius of the Jacobian matrix ${\partial{\cal{S}}}/{\partial W}$. 

Once the intermediate solution $(W^*)^{n+1}$ is obtained by the chemical sub-integration, the LU-SGS method is then applied to the flow sub-integration (\ref{eq:splitting_scheme_pde}) over the time interval $\Delta t$ to obtain the solution $W^{n+1}$ at time step $n+1$. In the original LU-SGS method for unsplit problems, the residual on the right-hand side is computed by the solution at the time step $n$, i.e., ${\cal{R}} = {\cal{T}}(W^n) + {\cal{S}}(W^n)$. However, in Eq. (\ref{eq:splitting_scheme_pde}), the residual is replaced by the difference of solutions at the intermediate time step and the time step $n$. In consideration of Eq. (\ref{eq:splitting_scheme_ode}), this residual can be regarded as the weighted average of ${\cal{T}}(W) + {\cal{S}}(W)$ over the time interval $\Delta t$
\begin{equation}
    {\cal{R}} = \frac{(W^*)^{n+1} - W^n}{\Delta t} = \frac{1}{\Delta t} \int_{t_n}^{t_{n+1}}{\frac{\mathrm{d}W^*}{\mathrm{d}t} \mathrm{d}t} = \frac{1}{\Delta t} \int_{t_n}^{t_{n+1}}{[{\cal{T}}(W^n)+ {\cal{S}}(W^*)] \mathrm{d}t} = {\cal{T}}(W^n) + \frac{1}{\Delta t} \int_{t_n}^{t_{n+1}}{{\cal{S}}(W^*) \mathrm{d}t}
\end{equation}
It is considered to be more reasonable to maintain the stability of the integration than that computed at time step $n$.

The contributions of transport term and reacting source term are incorporated into both the chemical sub-integration and the flow sub-integration, which guarantees that the right-hand sides are always consistent with the original differential governing equations for steady problem. In other words, the present splitting scheme is steady-state preserving. However, since the contribution of chemical source term is not included into the Jacobian matrix of LU-SGS iteration, this splitting scheme may degrade the convergence as the chemical time scale becomes much smaller than the flow time scale. Even so, this splitting scheme is attractive since it is easy to be implemented in the existing LU-SGS method. We need only reset residuals of species and energy equations before performing LU-SGS iteration. Thus, in the second flow sub-integration, the LU-SGS iteration for species and energy equations can be performed together with the other non-stiff equations, which avoids the separate solving of governing equations. It is especially important for the energy equation since it is closely coupled with the continuity and momentum equations. 
In addition, in contrast to the Strang splitting scheme, the computational cost of the presented scheme is smaller since only one chemical sub-integration and one flow sub-integration are necessary within one time step determined by the flow time scale.

\subsection{Computational Configuration}
To model the combustion flow in the turbine burner, the diffusion flame in a two-dimensional steady transonic mixing layer with strong favorable pressure gradient is considered in this study. The flow condition is from the cases of Fang et al. \cite{fang2001ignition}, Mehring et al. \cite{mehring2001ignition}, and Cai et al. \cite{cai2001ignition}. To produce the prescribed streamwise pressure gradient in the mixing layer, a configuration of converging-diverging nozzle is created in this paper, which was not necessary in Fang's and Mehring's work since the boundary-layer approximation was made. It was also avoided in Cai's work by introducing a streamtube thickness function into the two-dimensional equations.
At the inlet of the nozzle, the hot air mixed with burned gases flows into the upper side and comes into contact with the fuel vapor from the lower side. To achieve the prescribed pressure levels in the nozzle passage, given the flow conditions and nozzle height at the inlet, the downstream profiles of the upper and lower surfaces can be determined by the isentropic relations of quasi-one-dimensional flow for air and fuel, respectively. Since this is based on the assumption of quasi-one-dimensional flow without mixing and reaction, there exists a difference between the computed pressure in the diffusion flame and the prescribed one. This is eliminated by reshaping the nozzle profiles according to the pressure difference using the isentropic relations again. 

\begin{figure}[htb!]
    \centering
    \includegraphics[width=0.495\linewidth]{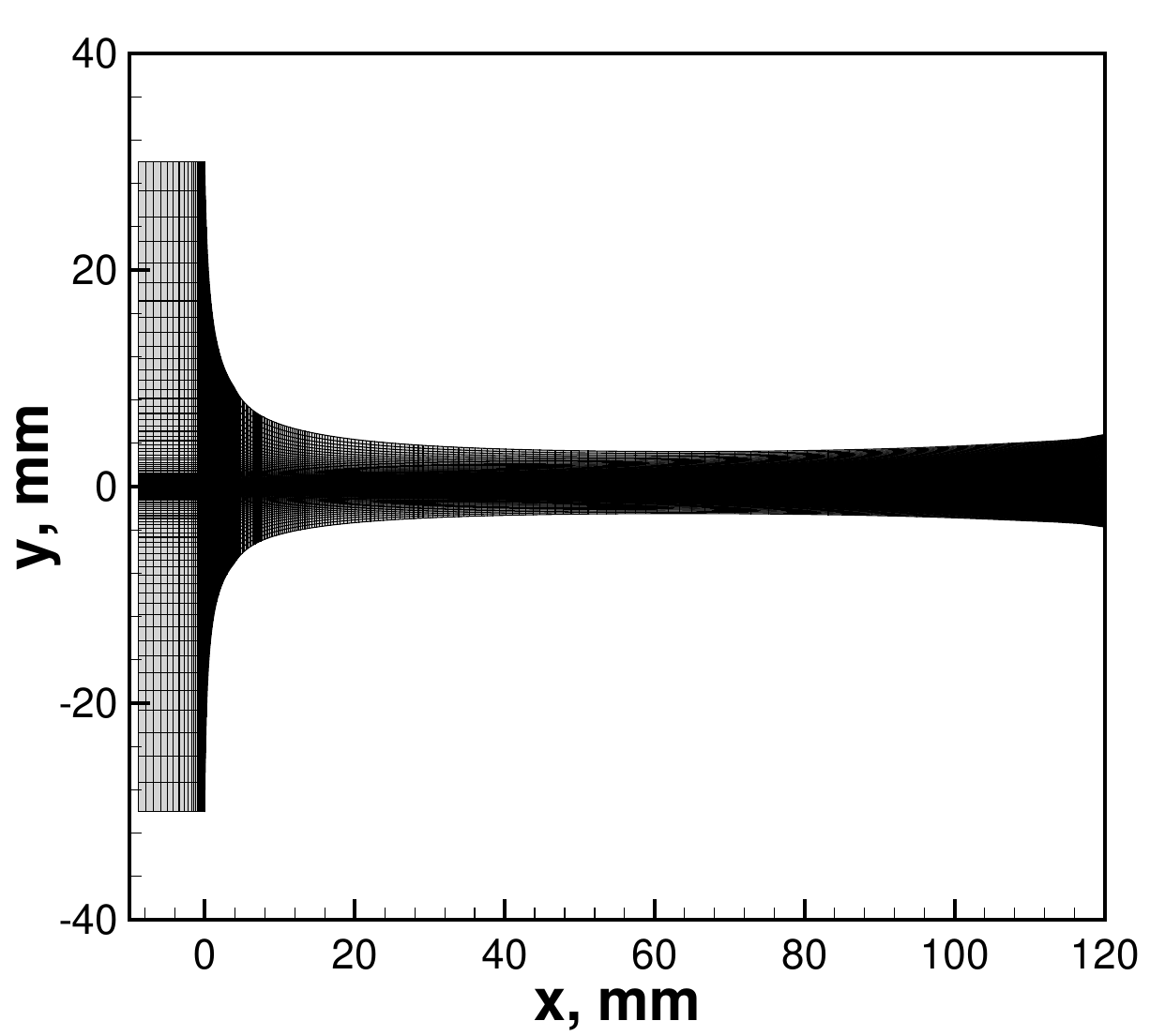}
    \caption{Nozzle configuration and grid for turbulent mixing layer}
    \label{fig:nozzle_grid}
\end{figure}

Figure \ref{fig:nozzle_grid} shows the converging-diverging nozzle configuration in the turbulent mixing-layer case. To reduce the disturbance of inlet boundary condition on the downstream flow, a uniform inlet section is added ahead of the mixing layer. The upper side and lower side of the nozzle are almost symmetric, both of which rapidly converge at the inlet, gradually slow down in the middle, and keep diverging after the throat at $x = 70 \, \rm{mm}$. To reduce the slope of the side surfaces near the inlet while keeping them away from the mixing layer at the throat, the nozzle height at the inlet should be carefully chosen. The half height at the inlet is $30 \, \rm{mm}$ for the turbulent case, whereas it is reduced to $3.5 \, \rm{mm}$ for the laminar case due to the thinner mixing layer. As a result, the half heights at the throats for the turbulent and laminar cases are about $3 \, \rm{mm}$ and $0.4 \, \rm{mm}$, respectively.

\begin{figure}[htb!]
    \centering
    \includegraphics[width=0.495\linewidth]{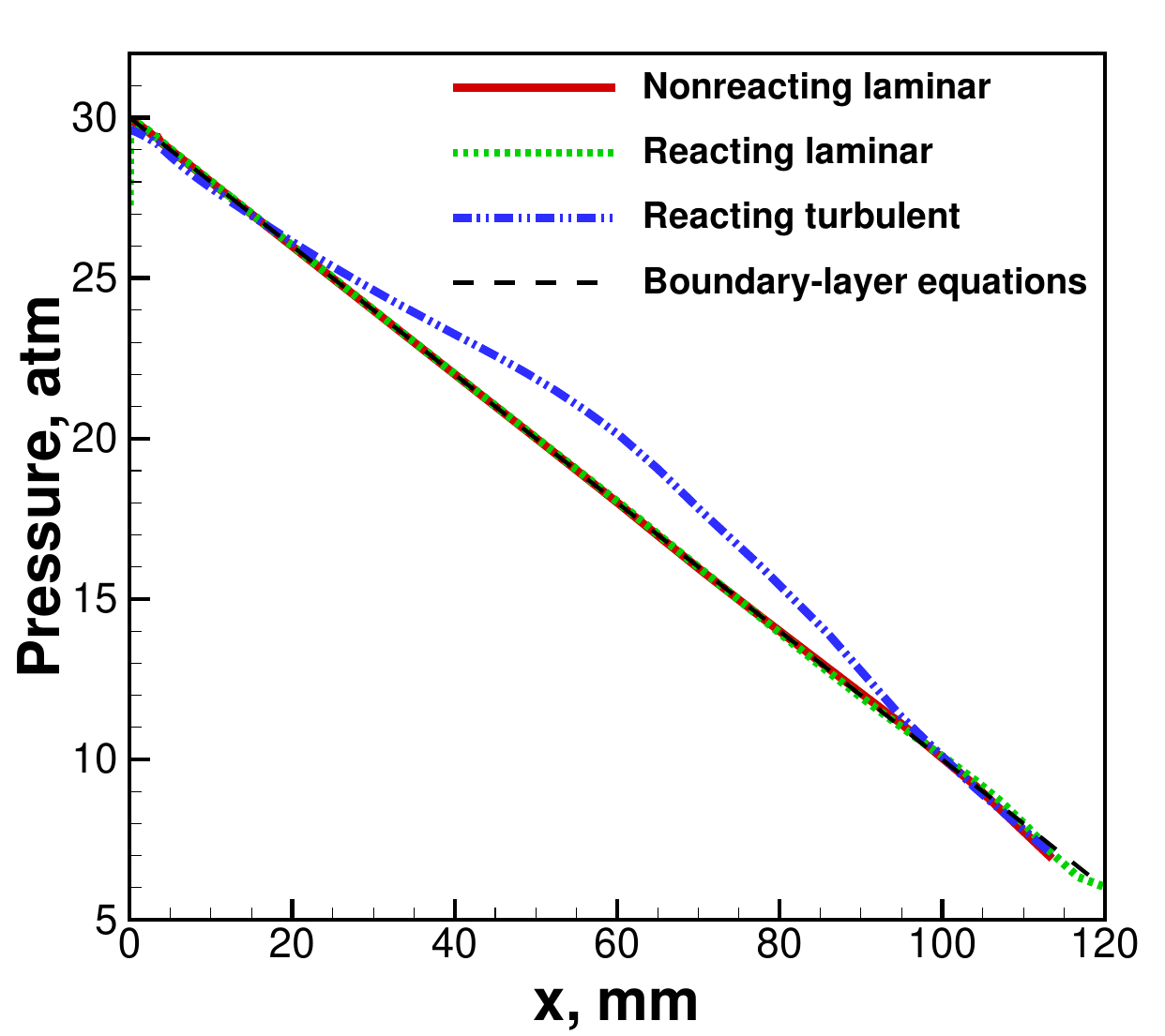}
    \caption{Pressure variations along center line of nozzle}
    \label{fig:pressure_centerline}
\end{figure}

Figure \ref{fig:pressure_centerline} presents the pressure variations along the center line of the nozzle for three cases to be analyzed in Secs. \ref{sec:nonreaction_laminar}, \ref{sec:reaction_laminar} and \ref{sec:reaction_turbulent}, along with the linearly varying pressure imposed in the boundary-layer approximation. The pressure variations in the two laminar cases are sufficiently close to the boundary-layer case. However, the pressure levels for the reacting turbulent case deviate from the linear values, especially in the middle of the nozzle. This is because the thick turbulent shear layer approaches the side surfaces and is evidently perturbed by them in the downstream nozzle. This pressure difference in the reacting turbulent case is believed to have some influence on the development of the mixing layer and combustion process in it, which will be discussed later.

Refs. \cite{fang2001ignition} and \cite{mehring2001ignition} studied the diffusion flame in the mixing layer under 13 different flow conditions by solving the boundary-layer equations. One of those cases, identified as the base case, is chosen to be simulated by using the full Navier-Stokes equations in this paper. The base case, corresponding to a constant streamwise pressure gradient of $-200 \, \rm{atm/m}$, has a static pressure of $30 \, \rm{atm}$ at the inlet. The temperature and velocity of the air at the inlet are $T_{\rm{air}} = 1650 \, \rm{K}$ and $u_{\rm{air}} = 50 \, \rm{m/s}$, respectively. Those of the fuel are $T_{\rm{fuel}} = 400 \, \rm{K}$ and $u_{\rm{fuel}} = 25 \, \rm{m/s}$.

The inviscid slip and adiabatic wall boundary conditions with zero normal pressure gradient are specified on the two side surfaces of the nozzle. At the inlet, the total pressure and total temperature of air and methane are fixed on the upper stream and lower stream, respectively. For turbulent cases, the turbulent intensity and ratio of turbulent to molecular viscosity for air are set as $5\%$ and 10.0 at the inlet, respectively. Those for the fuel are $10\%$ and 100.0. All flow quantities are extrapolated (zero streamwise gradient) at the exit flow boundary since the flow is supersonic there. This ensures that no backward waves propagate into the computational domain at the exit. The segment of the center line between the inlet and the starting point of convergence is set as a symmetric plate to avoid pre-mixing of air and fuel.

\subsection{Grid and Grid Independence}
Multi-block grids with matched interfaces between neighboring blocks are generated in the nozzle. Figure \ref{fig:nozzle_grid} shows the grid for the turbulent case, which is rescaled along the vertical direction to obtain the grid for the laminar case. The vertical grid lines cluster near the inlet, whereas the transverse grid lines cluster near the center line and scatter towards the two side surfaces. The height of the first cell off the center line is less than $0.006 \, \rm{mm}$ for the laminar case, and less than $0.05 \, \rm{mm}$ for the turbulent case. The total number of grid cells is 38272. 

\begin{figure}[htb!]
    \centering
    \begin{subfigure}[b]{0.495\linewidth}
        \centering
        \includegraphics[width=1\linewidth]{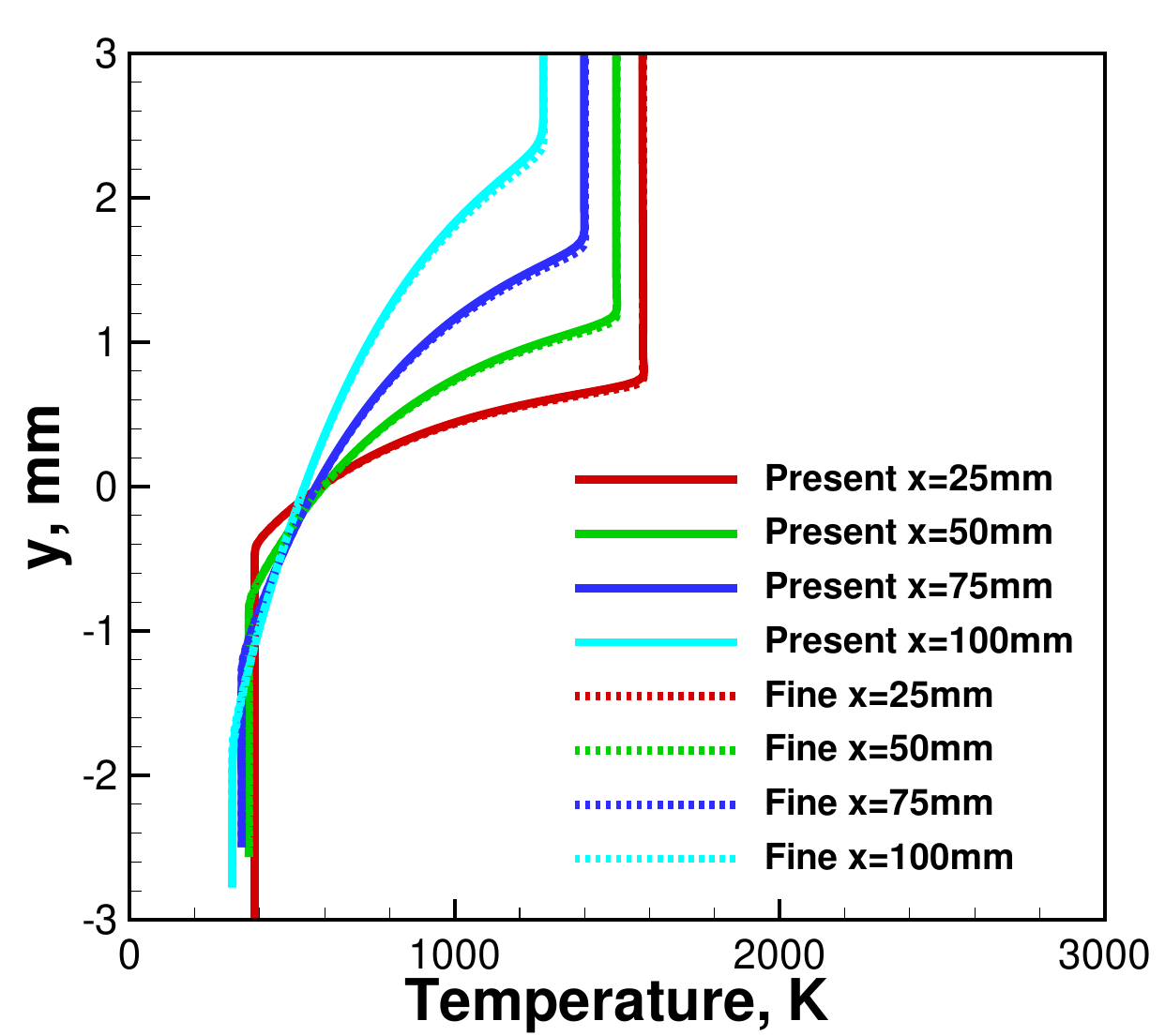}
        \caption{Nonreacting case}
        \label{fig:profiles_mesh_independence_T_noreaction}
    \end{subfigure}
    \begin{subfigure}[b]{0.495\linewidth}
        \centering
        \includegraphics[width=1\linewidth]{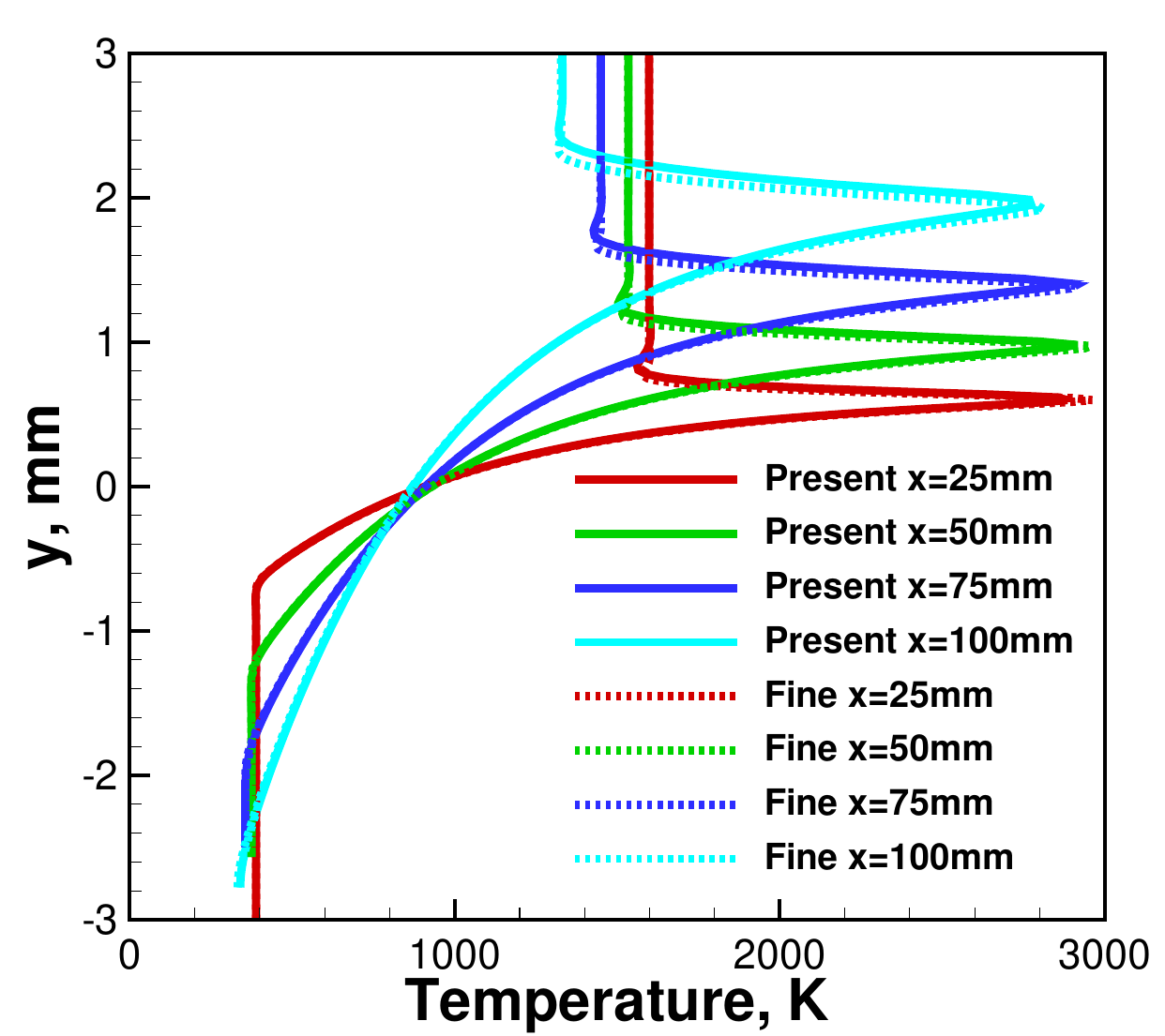}
        \caption{Reacting case}
        \label{fig:profiles_mesh_independence_T_reaction}
    \end{subfigure}
    \caption{Profiles of temperature at different streamwise positions for turbulent mixing layer on different grid resolutions}
    \label{fig:profiles_mesh_independence_T}
\end{figure}

To check the grid independence, the nonreacting case and reacting case in a turbulent mixing layer are performed on the current grid and another fine grid with 83776 cells, respectively. Figure \ref{fig:profiles_mesh_independence_T} compares the profiles of temperature at four streamwise positions for both cases. The solutions on the two grid levels are almost indistinguishable. The profiles of other flow variables (not shown here for brevity) show similar behavior, indicating achievement of grid independence on the coarser 38272 grid already for the turbulent case. For the laminar case, grid independence is also reached on the coarse grid since the grid size (length of cell size) in the vertical direction is smaller than the grid for the turbulent case because of the smaller throat height. Therefore, the results in the following section are based on the 38272 grid.

\begin{figure}[htb!]
    \centering
    \includegraphics[width=0.495\linewidth]{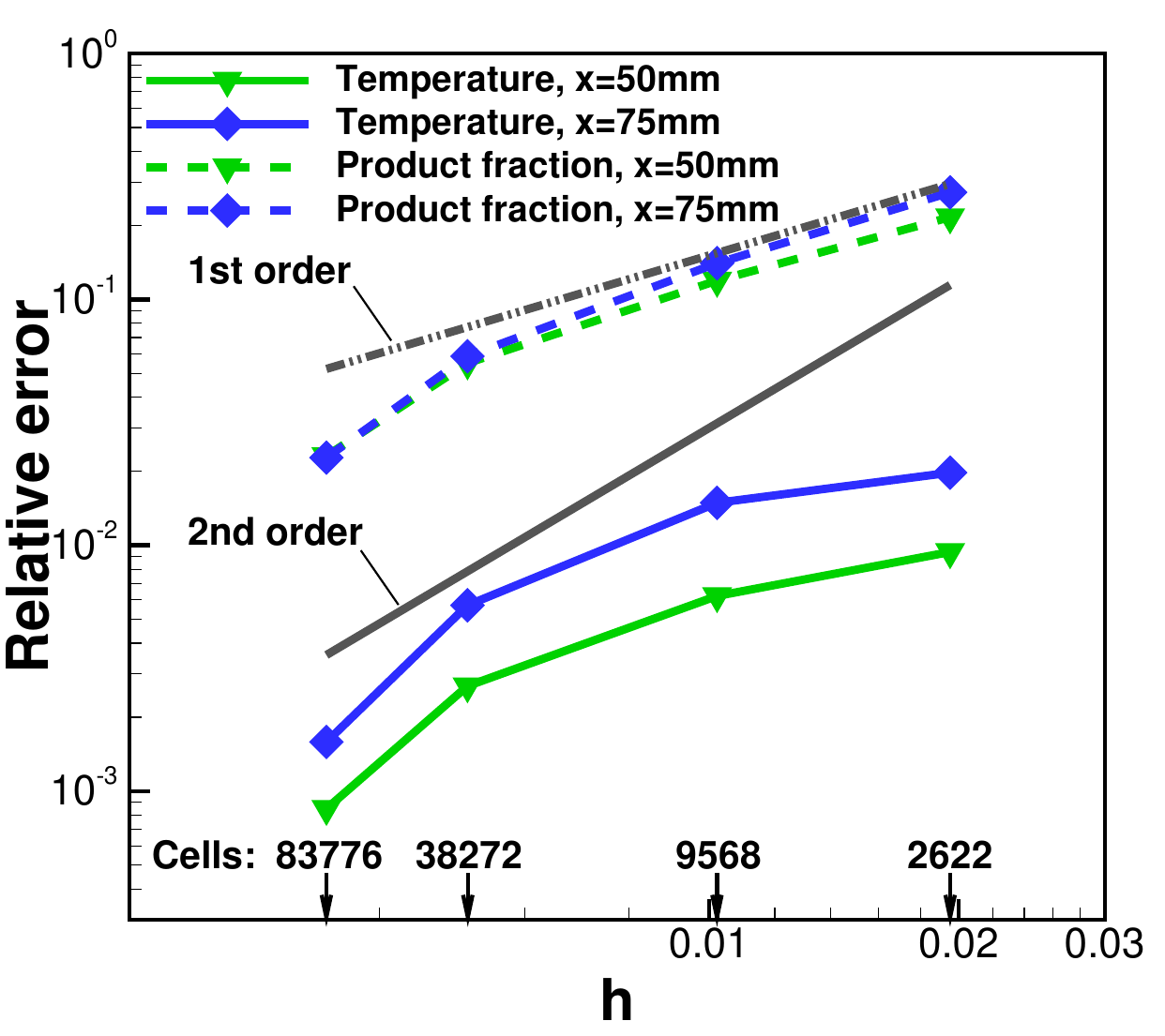}
    \caption{Variations of relative errors with grid spacing for reacting turbulent mixing layer}
    \label{fig:error_temperature_Yproduct}
\end{figure}

To further verify the grid convergence of the present code in chemically reacting flows, Fig. \ref{fig:error_temperature_Yproduct} shows the variations of relative errors of temperature and product mass fraction with the grid spacing for the reacting case in the turbulent mixing layer, along with the scales of first order and second order. The grid spacing $h$ is represented by $h = \sqrt{1/N_c}$ for the two-dimensional problem, where $N_c$ is the total number of cells on each grid level. The error on each grid level is defined as the relative difference of a mass-averaged property over a streamwise cross section with respect to that on an extra fine grid of 153824 cells. The relative errors of all properties, including those not shown here, reduce as the grid is refined. The accuracy of grid convergence between the two coarse grid levels is close to first order since the solutions on them, especially on the coarsest grid of 2622 cells, contain significant discretization errors. However, second-order accuracy, which is consistent with the spatial discretization accuracy of the present code, is achieved on the two fine grid levels. This also indicates that the two grids are in the asymptotic range of grid convergence for the present problem.

\section{Computational Results and Discussions} \label{sec:results_discussions}

\subsection{Nonreacting Laminar Mixing Layer} \label{sec:nonreaction_laminar}
\begin{figure}[htb!]
    \centering
    \begin{subfigure}[b]{0.495\linewidth}
        \centering
        \includegraphics[width=1\linewidth]{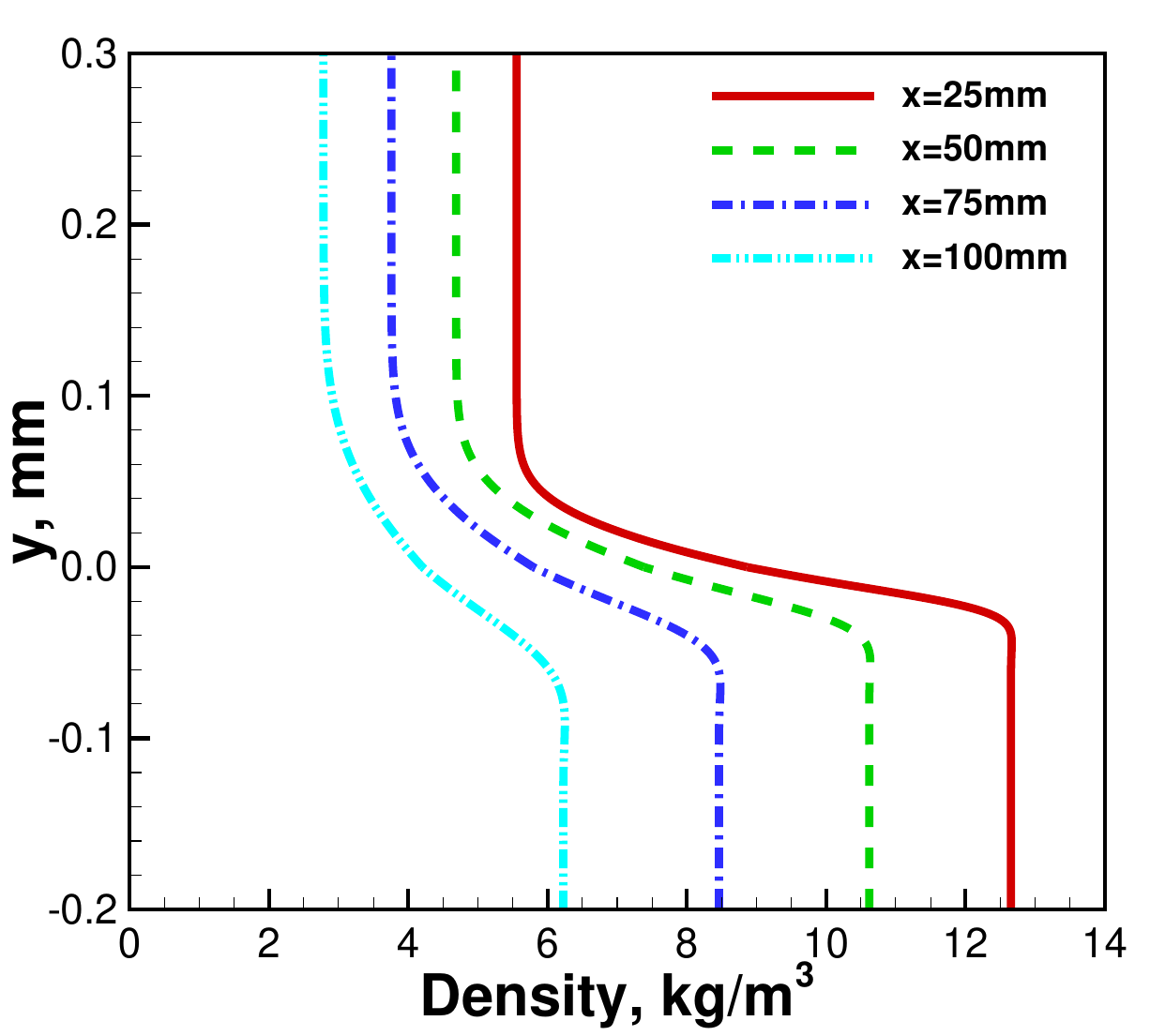}
        \caption{Density}
        \label{fig:profiles_mixing_laminar_noreaction_rho}
    \end{subfigure}
    \begin{subfigure}[b]{0.495\linewidth}
        \centering
        \includegraphics[width=1\linewidth]{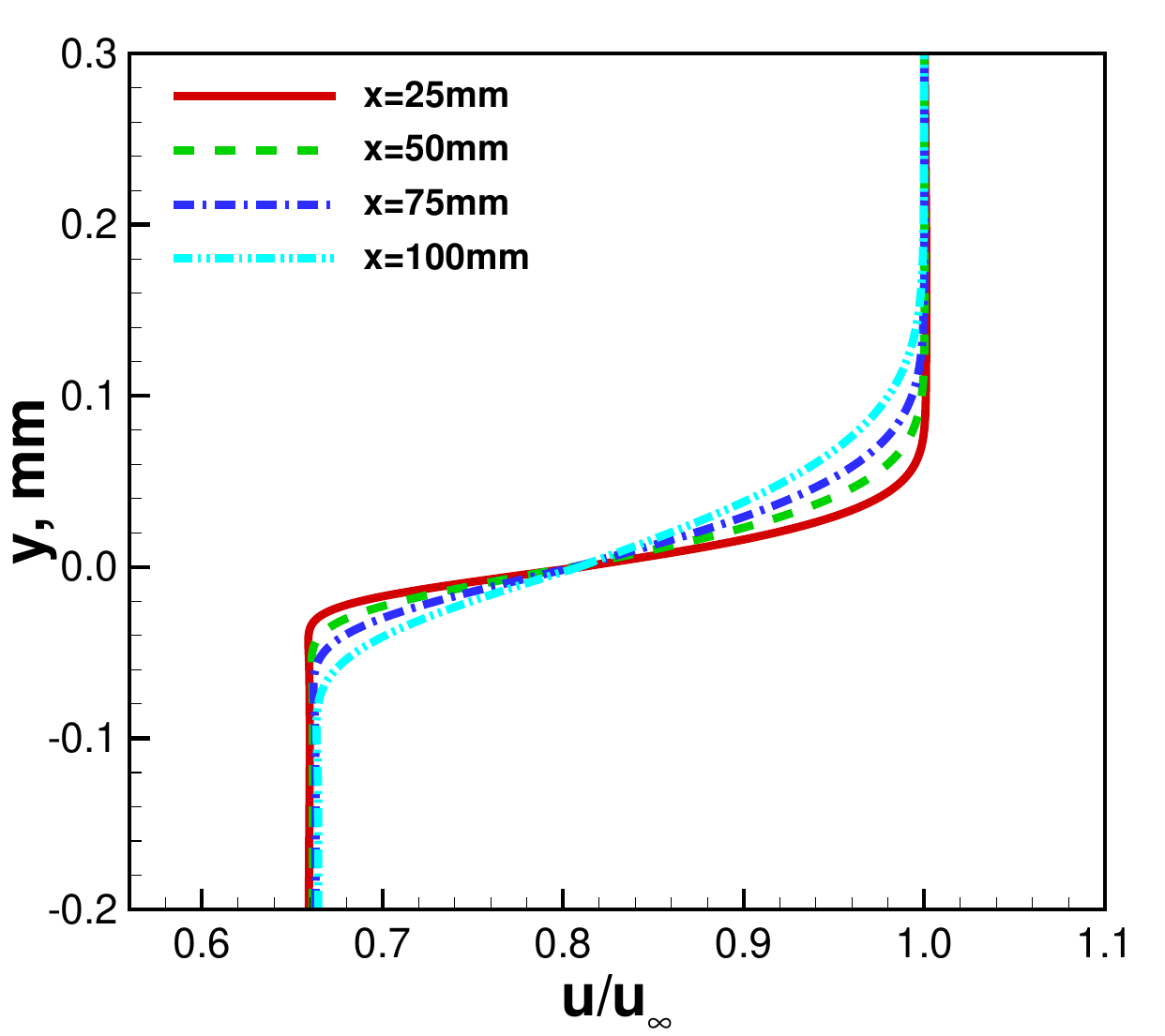}
        \caption{Velocity}
        \label{fig:profiles_mixing_laminar_noreaction_v}
    \end{subfigure}
    \caption{Profiles of density and velocity at different streamwise positions for nonreacting laminar mixing layer}
    \label{fig:profiles_mixing_laminar_noreaction}
\end{figure}

To demonstrate the abilities of the present code to deal with multi-component flows, we first examine the nonreacting laminar flow in the nozzle. At the trailing edge of the splitter plate ($x = 0$), the hot air on the upper stream starts to mix with the fuel vapor on the lower stream, producing velocity and thermal mixing layers in the middle of the nozzle. 
Figure \ref{fig:profiles_mixing_laminar_noreaction} shows the profiles of density and streamwise velocity at four different streamwise positions. The velocity is normalized by the corresponding freestream value in the air side at each position.
The mixing layer grows in thickness along the streamwise direction, and the freestream density decreases due to flow acceleration caused by the favorable pressure gradient. The thickness of the thermal mixing layer is almost the same as that of the velocity mixing layer since unity Prandtl number is used in the computation. In addition, the mixing layer on the upper side is thicker than that on the lower side. This is because the smaller density and larger molecular viscosity resulting from the higher temperature produce a smaller Reynolds number on the upper side although the velocity is higher there.

\subsection{Reacting Laminar Mixing Layer} \label{sec:reaction_laminar}
\begin{figure}[htb!]
    \centering
    \includegraphics[width=0.495\linewidth]{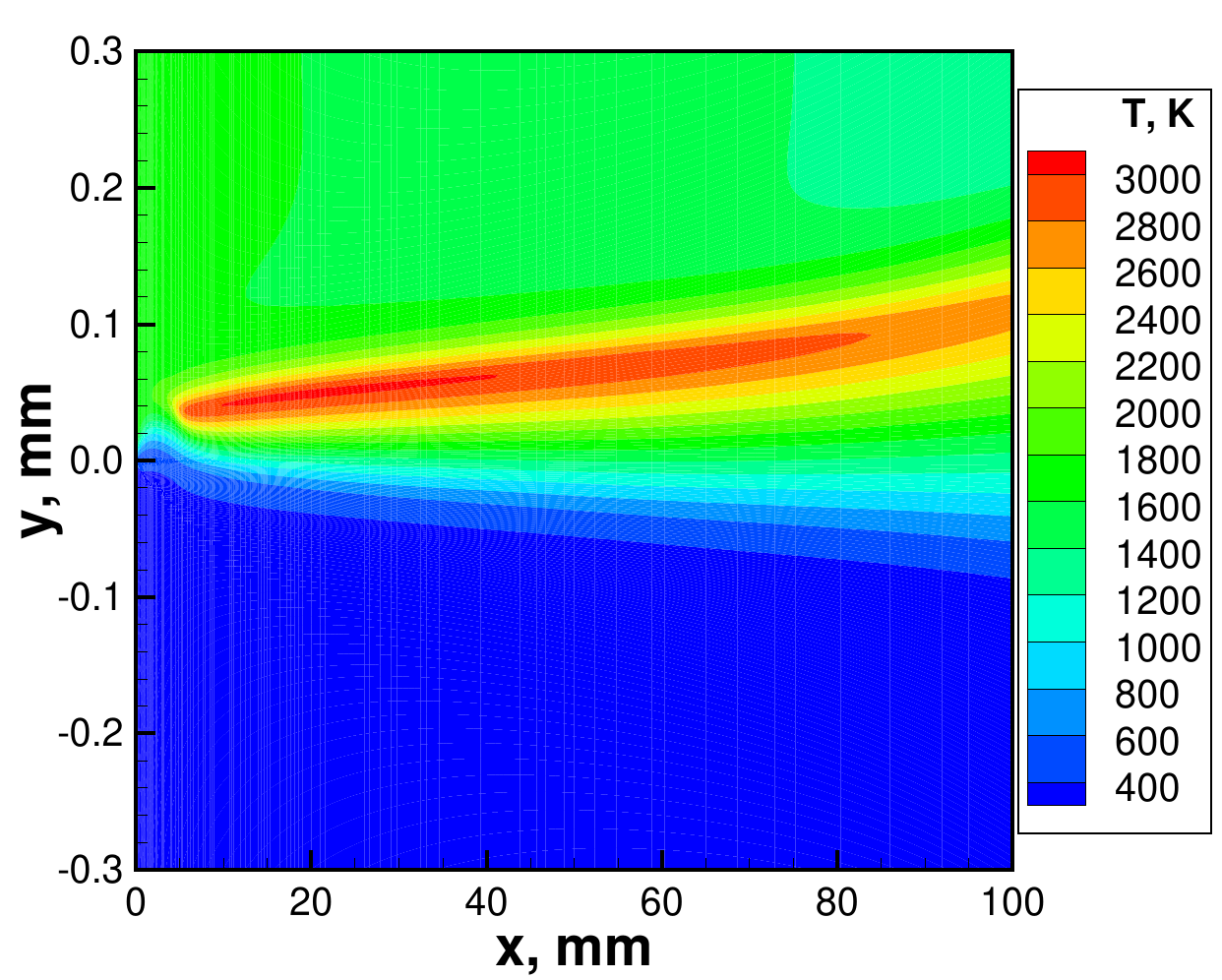}
    \caption{Contours of temperature for laminar mixing layer}
    \label{fig:contour_T_mixing_laminar}
\end{figure}

The contours of temperature for the laminar mixing layer are shown in Fig. \ref{fig:contour_T_mixing_laminar}. At the trailing edge of the splitter plate, the hot air on the upper stream starts to mix with the fuel vapor on the lower stream. Chemical reaction happens shortly downstream the trailing edge. A diffusion flame is established within the mixing layer as indicated by the high-temperature region slightly biased towards the air side. The reason for that is that the momentum of fuel on the lower stream is higher than that of air on the upper stream, resulting in an upward tilting of the mixing layer. The flame continues to move upward with the mixing layer along the streamwise direction. The freestream temperatures on the two sides decrease downstream because of the flow acceleration under the favorable pressure gradient. The peak temperature within the diffusion flame also decreases along the streamwise direction due to not only the decreasing freestream temperatures but also reduced reaction rate resulting from the reduced freestream temperatures and reactant concentrations. 

\begin{figure}[htb!]
    \centering
    \begin{subfigure}[b]{0.495\linewidth}
        \centering
        \includegraphics[width=1\linewidth]{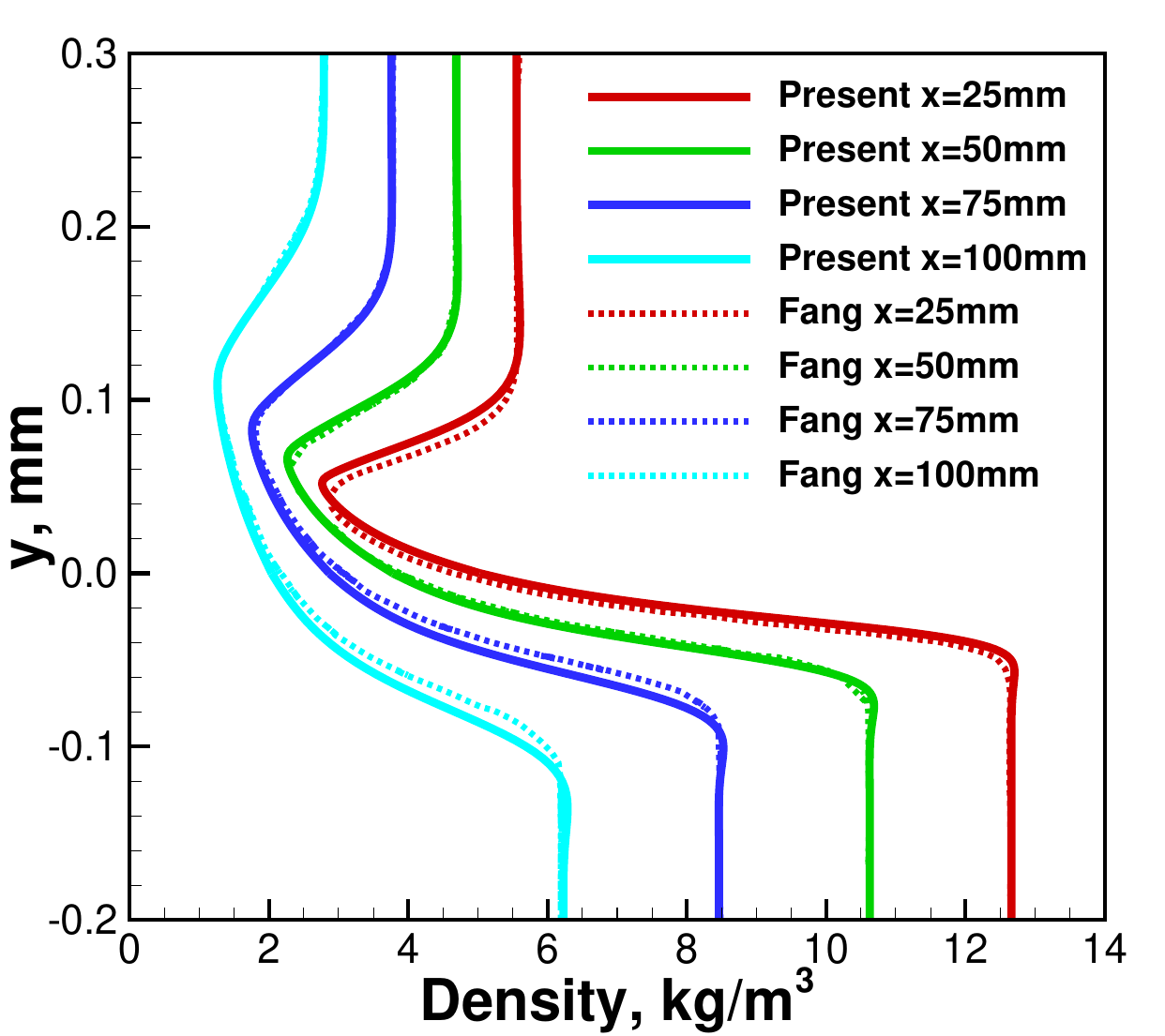}
        \caption{Density}
        \label{fig:profiles_mixing_laminar_rho}
    \end{subfigure}
    \begin{subfigure}[b]{0.495\linewidth}
        \centering
        \includegraphics[width=1\linewidth]{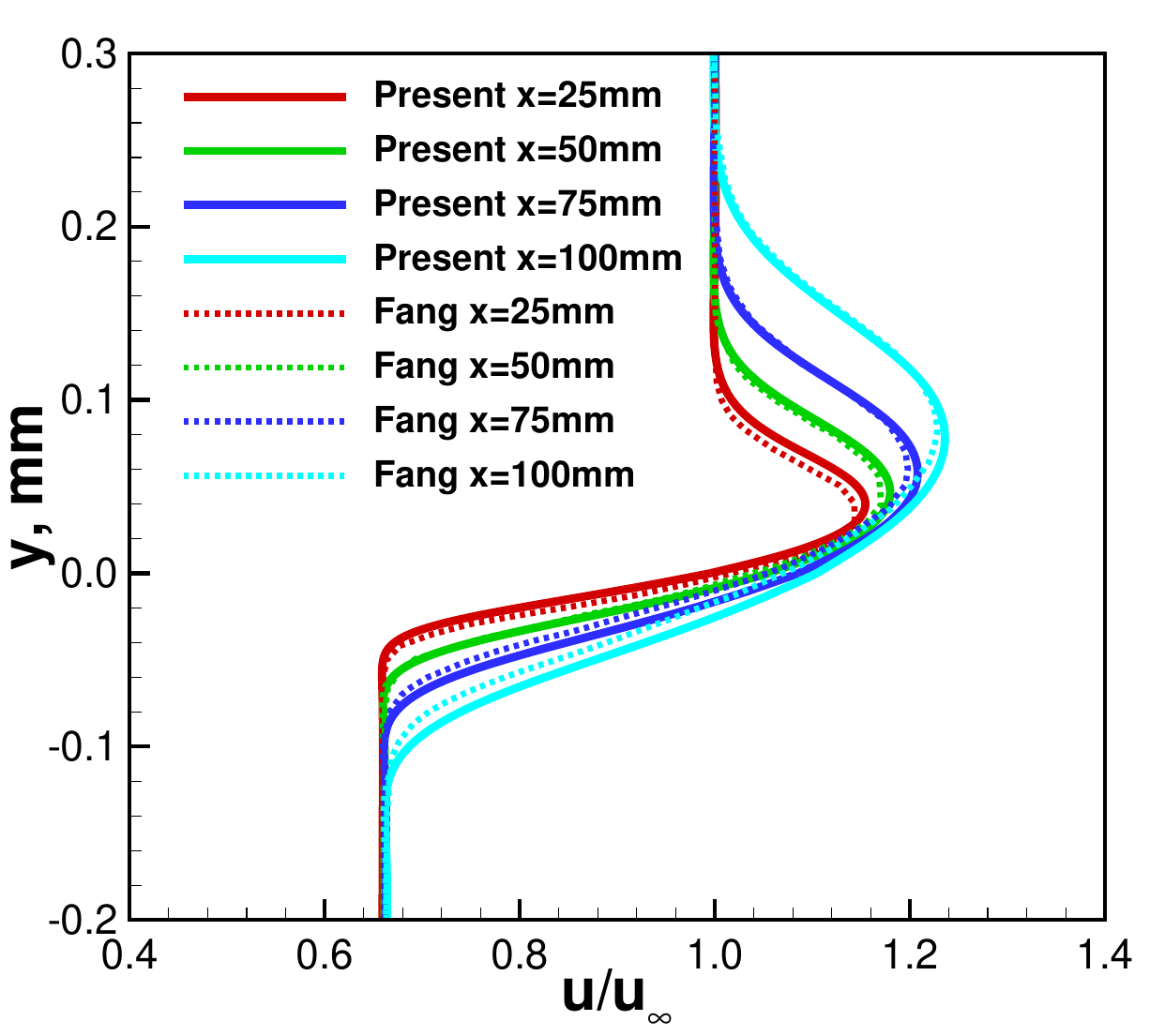}
        \caption{Velocity}
        \label{fig:profiles_mixing_laminar_v}
    \end{subfigure}
    \caption{Profiles of density and velocity at different streamwise positions for laminar mixing layer. Fang et al. used constant pressure gradient of -200 atm/m}
    \label{fig:profiles_mixing_laminar_rho_v}
\end{figure}

Figure \ref{fig:profiles_mixing_laminar_rho_v} shows the profiles of density and streamwise velocity at four different streamwise positions. The results computed by Fang et al. \cite{fang2001ignition} using the boundary-layer approximation are also shown for comparison. The present density and velocity profiles agree very well with those by Fang et al. Within the diffusion flame, however, the present density is slightly smaller and the velocity is slightly larger. This discrepancy is mainly because of the different governing equations used. In the flame, the density reaches a trough where the temperature peaks since the pressure is the same along the transverse direction. Along the streamwise direction, the density at the freestreams and in the flame decreases, which is consistent with the contours of temperature in Fig. \ref{fig:contour_T_mixing_laminar}. Although the normalized freestream velocity is unchanged along the streamwise direction, the peak velocity in the reaction region increases, which is different from the case in the nonreacting mixing layer in Fig. \ref{fig:profiles_mixing_laminar_noreaction}. This happens because the lighter gas in the reaction region gets accelerated more under the same pressure gradient.

\begin{figure}[htb!]
    \centering
    \includegraphics[width=0.495\linewidth]{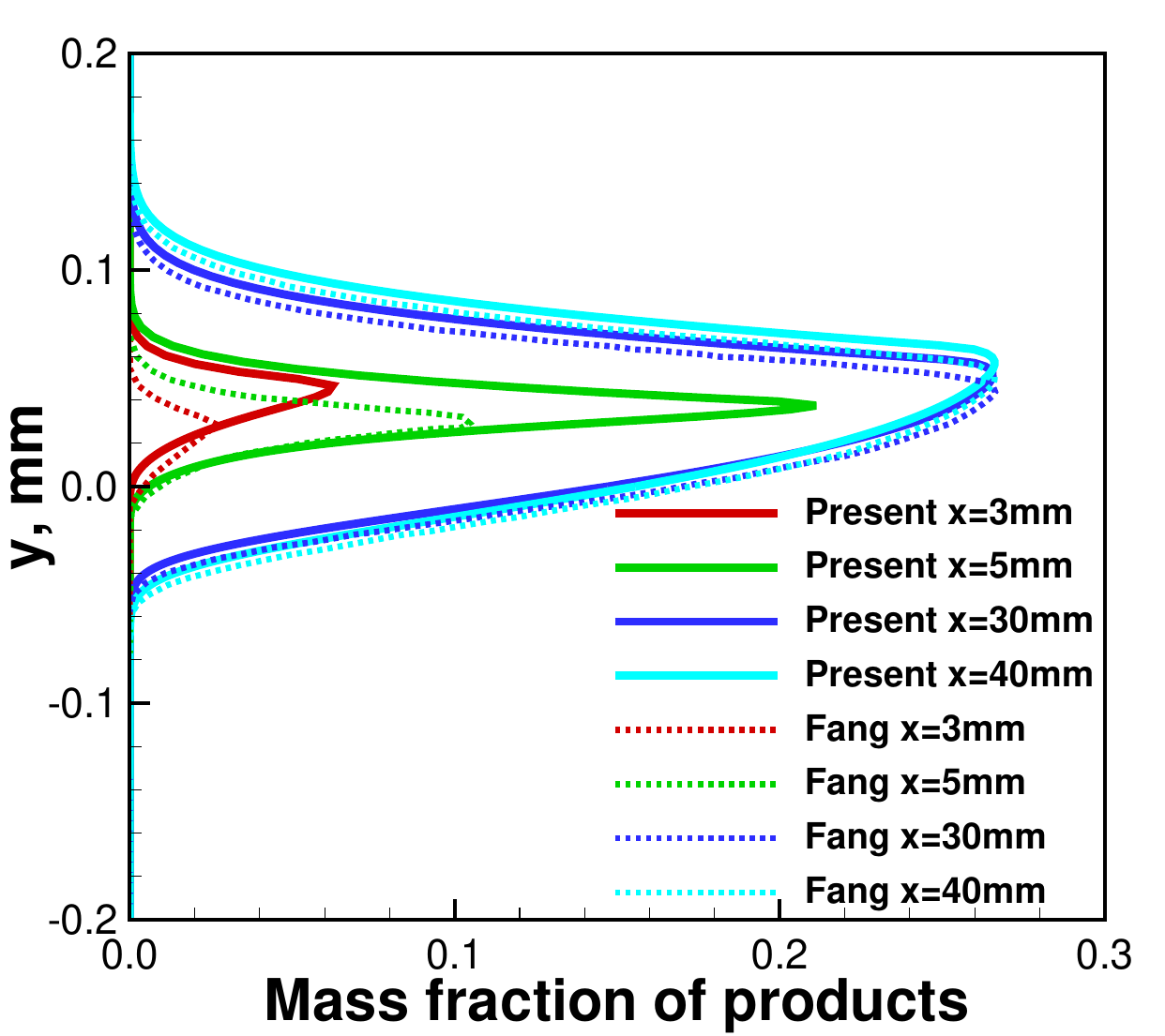}
    \caption{Profiles of mass fraction of products at different streamwise positions for laminar mixing layer. Fang et al. used constant pressure gradient of -200 atm/m}
    \label{fig:profiles_mixing_laminar_Yprod}
\end{figure}

Figure \ref{fig:profiles_mixing_laminar_Yprod} compares the profiles of mass fraction of products ($\rm{CO_2}$ and $\rm{H_2O}$) at four different streamwise positions. At the positions near the inlet ($x = 3 \, \rm{mm}$ and $x = 5 \, \rm{mm}$), both the thickness of reaction region and the peak value of mass fraction by the full Navier-Stokes equations differ from those by the boundary-layer equations. This is primarily attributed to the two-dimensional effects dominated at the initial stage of mixing-layer flow, which is neglected in the boundary-layer approximation. This is also due to the difficulty to exactly maintain a constant streamwise pressure gradient at the initial stage of the two-dimensional nozzle in the present computation. For both computations, the profiles of mass fractions vary sharply near their peaks, and the peak values obviously increase along the streamwise direction. This demonstrates that the chemical reaction dominates the flow over the molecular diffusion at the initial stage of the mixing layer. At the two downstream positions ($x = 30 \, \rm{mm}$ and $x = 40 \, \rm{mm}$), the two computational results are quite close to each other with the reaction region of the present simulation slightly biased to the upper side, consistent with the profiles of density and normalized velocity in Fig. \ref{fig:profiles_mixing_laminar_rho_v}. For both computations, the peak mass fraction of products, corresponding to the stoichiometric reaction, keeps unchanged at the two positions, while the thickness of reaction region continuously increases. This indicates that the diffusion begins to dominate the flow as the mixing layer further develops.

\subsection{Reacting Turbulent Mixing Layer} \label{sec:reaction_turbulent}
After validating the numerical method by the laminar case, the present method is extended to the reacting turbulent case. The flow condition is the same as that used for the laminar case, except that the nozzle height is enlarged about eight times to prevent the thicker turbulent mixing layer from touching the side surfaces downstream. 
To be consistent with the case in Mehring et al.'s work \cite{mehring2001ignition}, the standard $k\mbox{-}\omega$ model proposed by Wilcox \cite{wilcox1988reassessment} in 1988 rather than the SST model is used to evaluate the turbulent viscosity in this section.
The $k\mbox{-}\omega$ model exhibits a strong sensitivity to the freestream value of $\omega$ \cite{menter1992influence}. We set $k = 2.5 \times 10^{-4} \, \rm{m^2/s^2}$ and $\omega = 500 \, \rm{s^{-1}}$ at the inlet, same as the setup of Mehring et al. \cite{mehring2001ignition}.

\begin{figure}[htb!]
    \centering
    \begin{subfigure}[b]{0.495\linewidth}
        \centering
        \includegraphics[width=1\linewidth]{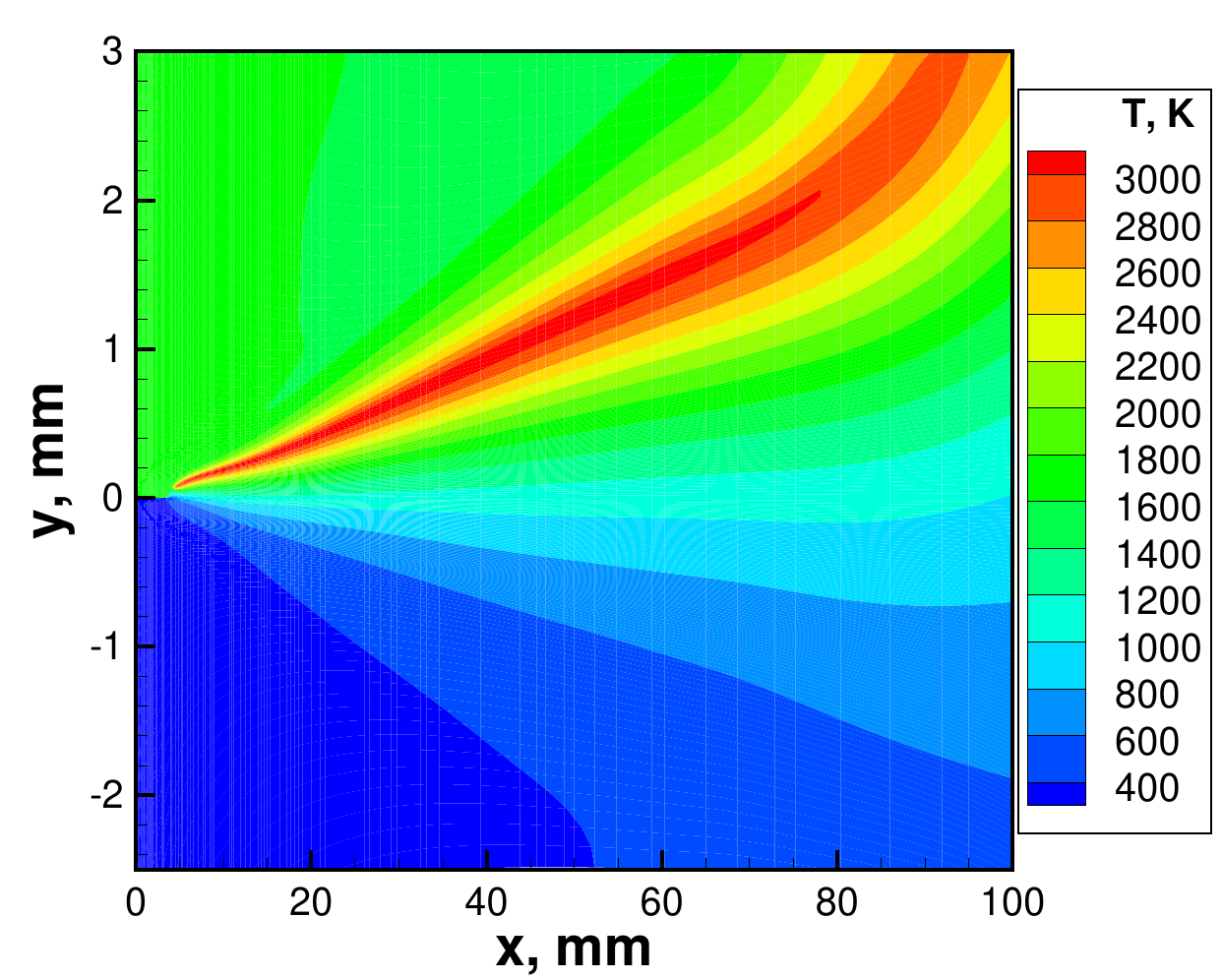}
        \caption{Present}
    \end{subfigure}
    \begin{subfigure}[b]{0.495\linewidth}
        \centering
        \includegraphics[width=1\linewidth]{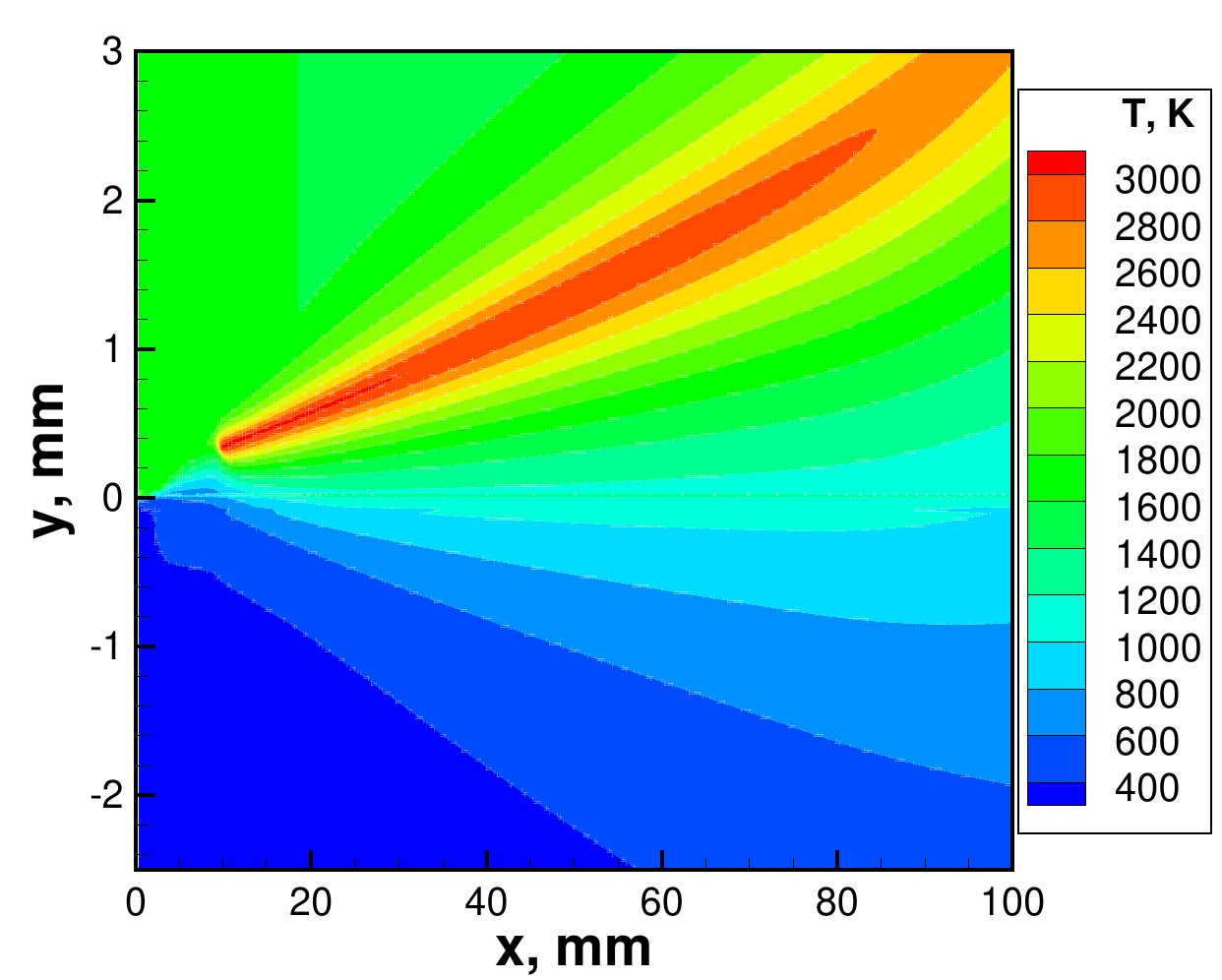}
        \caption{Mehring et al.}
    \end{subfigure}
    \caption{Contours of temperature for turbulent mixing layer. Mehring et al. used constant pressure gradient of -200 atm/m}
    \label{fig:contour_T_mixing_turbulent}
\end{figure}

The contours of temperature for the turbulent mixing layer are shown in Fig. \ref{fig:contour_T_mixing_turbulent}, along with the boundary-layer results by Mehring et al. \cite{mehring2001ignition}. Similar to the laminar case, a diffusion flame appears on the upper air side downstream of the splitter plate. The growths of both the mixing layer itself and flame are significantly larger than those in the laminar case shown in Fig. \ref{fig:contour_T_mixing_laminar} due to stronger diffusion of turbulence. In Mehring et al.'s results, the ignition occurs after a certain distance (about $10 \, \rm{mm}$) downstream from the trailing edge of the splitter plate, whereas it ignites much earlier (about $5 \, \rm{mm}$ after the splitter plate) in the present computation, as indicated by the high-temperature regions. This discrepancy is attributed to two reasons. On the one hand, the pressure gradient is exactly prescribed as $-200 \, \rm{atm/m}$ in the boundary-layer equations. However, it is produced by the converging-diverging side walls in the present full Navier-Stokes equations. The pressure levels near the nozzle inlet cannot stay exactly the same as the prescribed values due to the strong two-dimensional flow caused by the large slopes of the side walls. On the other hand, the boundary-layer approximation is not sufficiently accurate at the initial stage of mixing layer where the flow is fully two-dimensional in nature. The flame by the boundary-layer approximation spreads like a straight line along the streamwise direction. In the present computation, the flame keeps straight on the whole but with slight deflection near the inlet and after the throat. The deflection near the inlet is again due to the two-dimensional effects. The flame distorts after the throat since it is quite close to the top side surface due to the strong turbulent diffusion. We mainly focus on the streamwise ranges between $x = 10 \, \rm{mm}$ and $x = 70 \, \rm{mm}$ in the following.

\begin{figure}[htb!]
    \centering
    \includegraphics[width=0.495\linewidth]{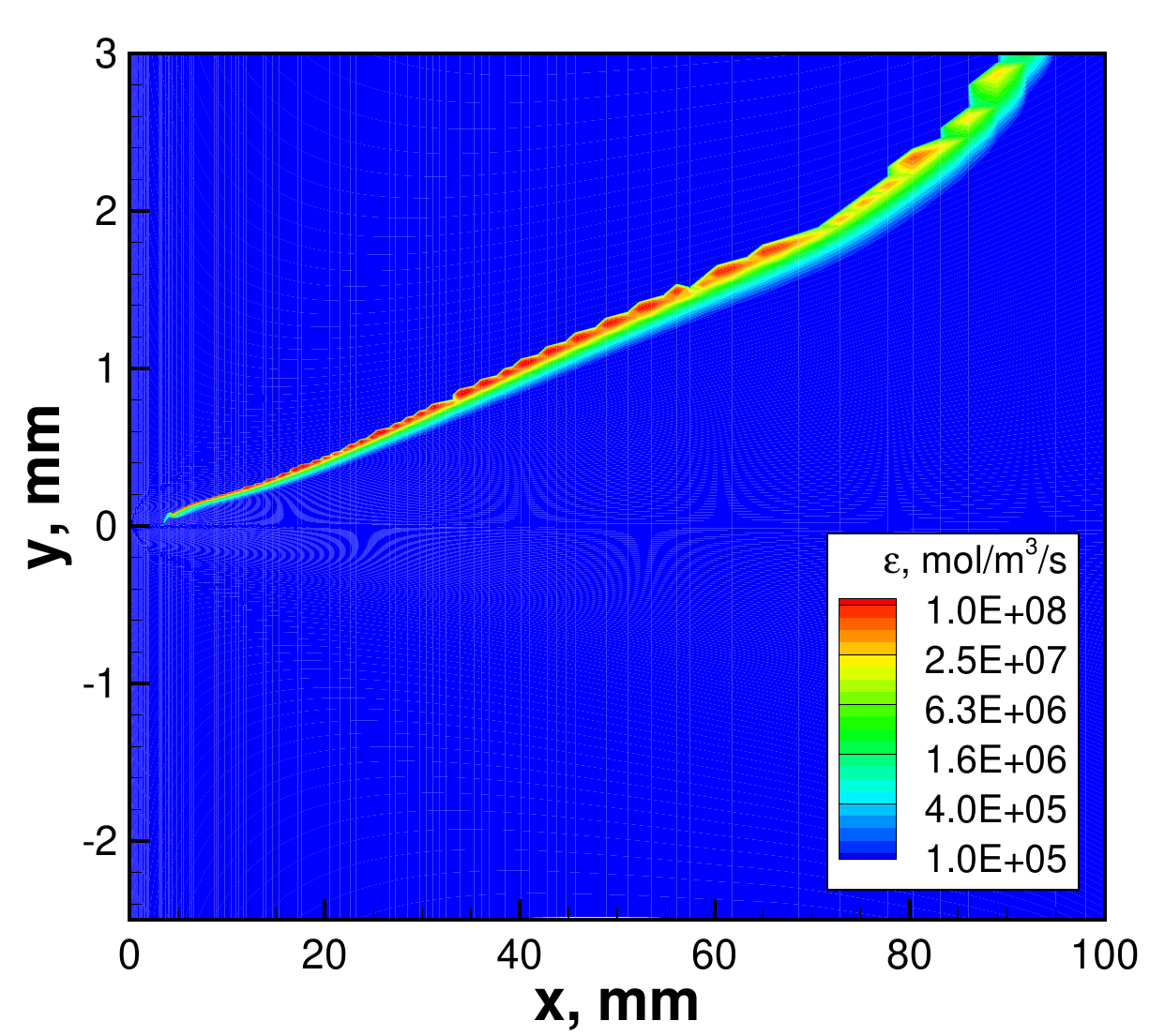}
    \caption{Contours of reaction rate for turbulent mixing layer}
    \label{fig:contour_reaction_rate_mixing_turbulent}
\end{figure}

\begin{figure}[htb!]
    \centering
    \begin{subfigure}[b]{0.495\linewidth}
        \centering
        \includegraphics[width=1\linewidth]{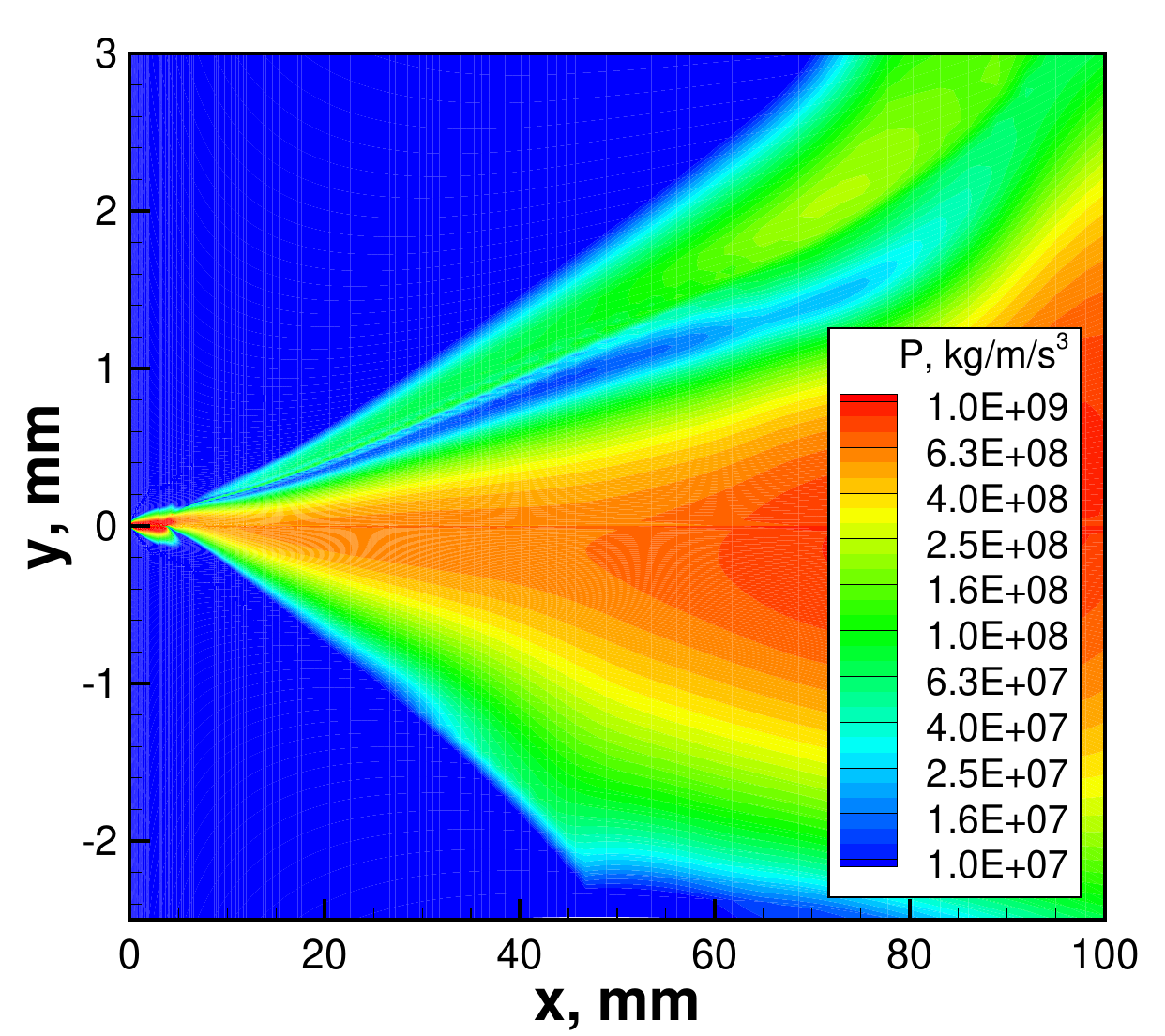}
        \caption{Production rate of turbulent kinetic energy}
        \label{fig:contour_Pk_mixing_turbulent}
    \end{subfigure}
    \begin{subfigure}[b]{0.495\linewidth}
        \centering
        \includegraphics[width=1\linewidth]{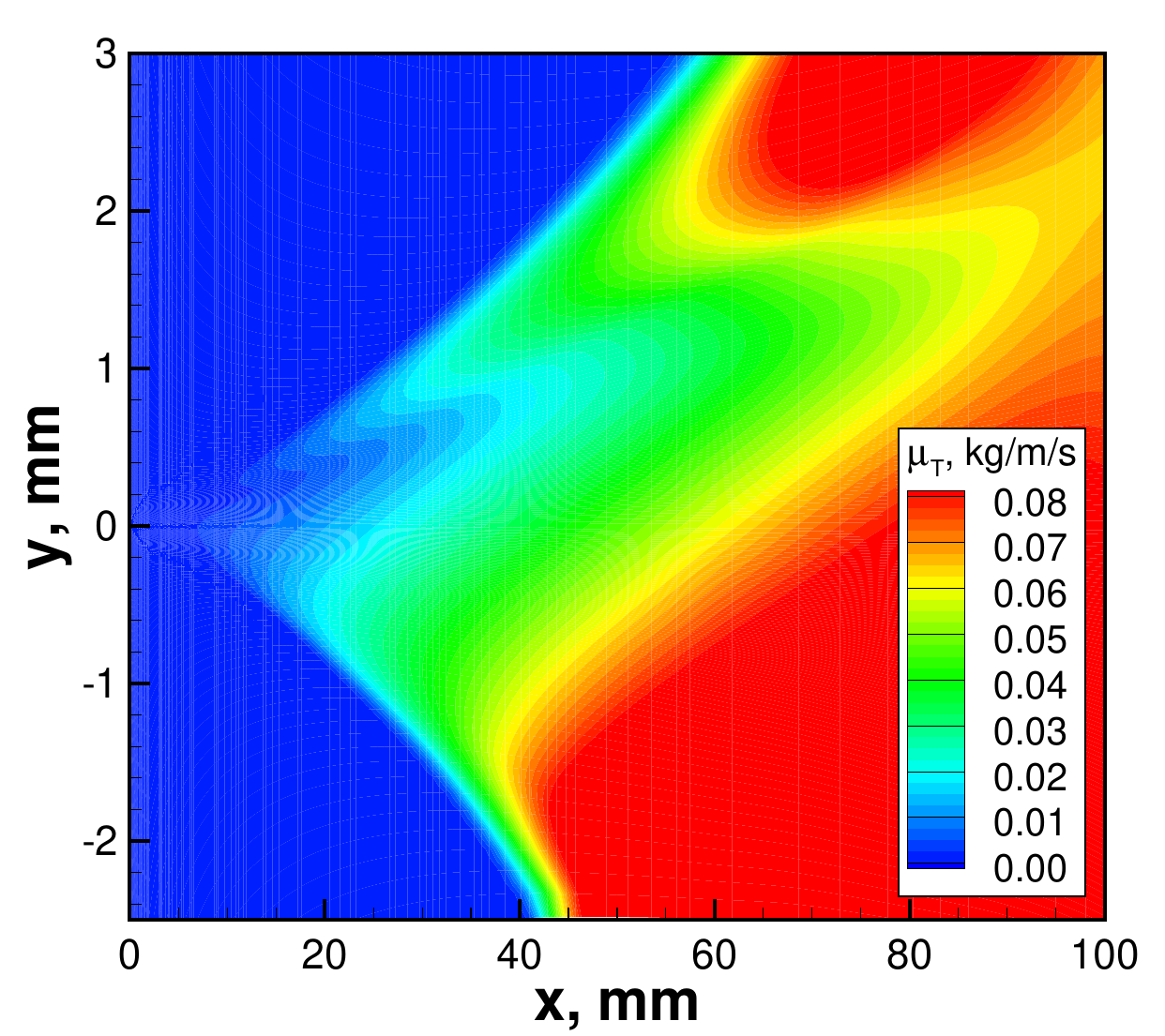}
        \caption{Turbulent viscosity}
        \label{fig:contour_mut_mixing_turbulent}
    \end{subfigure}
    \caption{Contours of turbulent properties for turbulent mixing layer}
    \label{fig:contour_mixing_turbulent}
\end{figure}

The present chemistry model does not explicitly include the influence of turbulent kinetics. However, the chemical reaction affects the turbulent transport of flame in the mixing layer. Figure \ref{fig:contour_reaction_rate_mixing_turbulent} shows the contours of the reaction rate computed by Eq. (\ref{eq:Arrhenius_expression}). The intense chemical reaction is concentrated around a narrow band on the upper side, where the fuel and oxidizer have been fully mixed. Note that the reaction rate on the air side is remarkably higher than that on the fuel side due to the negative exponent of fuel concentration in Eq. (\ref{eq:Arrhenius_expression}). As the mixing layer develops, the reaction region becomes thick due to diffusion, while its strength reduces due to the decreased temperature and concentrations of reactants. Similar to the laminar case, the mixing-layer flow is dominated by both reaction and diffusion in the beginning, while it becomes diffusion-dominated further downstream. High velocity gradient is generated in the reaction region under the combined actions of convection and diffusion. This induces significantly strong turbulent production in the reaction region, as indicated by the contours of production rate of turbulent kinetic energy in Fig. \ref{fig:contour_Pk_mixing_turbulent}. The production rate is computed by $P = \mu_TS^2$ in the $k\mbox{-}\omega$ model. Another stronger-production region near the center line originates from the shear in the turbulent mixing layer. The intense production of turbulence leads to high turbulent kinetic energy and thus large turbulent viscosity in the flame as shown in Fig. \ref{fig:contour_mut_mixing_turbulent}. As a result, not only the diffusion in the mixing layer itself is strengthened by the turbulence, but also the turbulent diffusion in the flame region is further enhanced by the chemical reaction.

\begin{figure}[htb!]
    \centering
    \begin{subfigure}[b]{0.495\linewidth}
        \centering
        \includegraphics[width=1\linewidth]{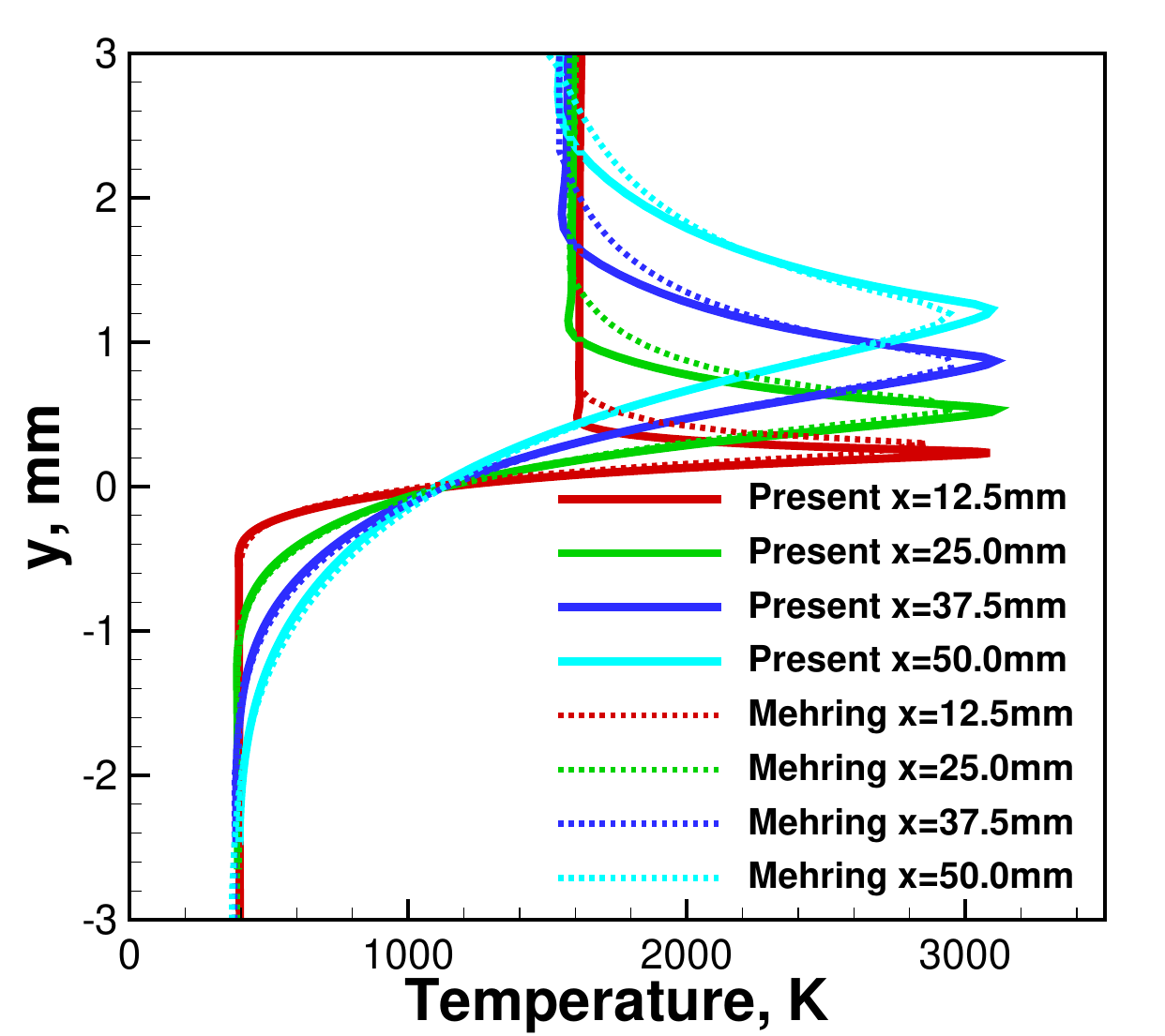}
        \caption{Temperature}
    \end{subfigure}
    \begin{subfigure}[b]{0.495\linewidth}
        \centering
        \includegraphics[width=1\linewidth]{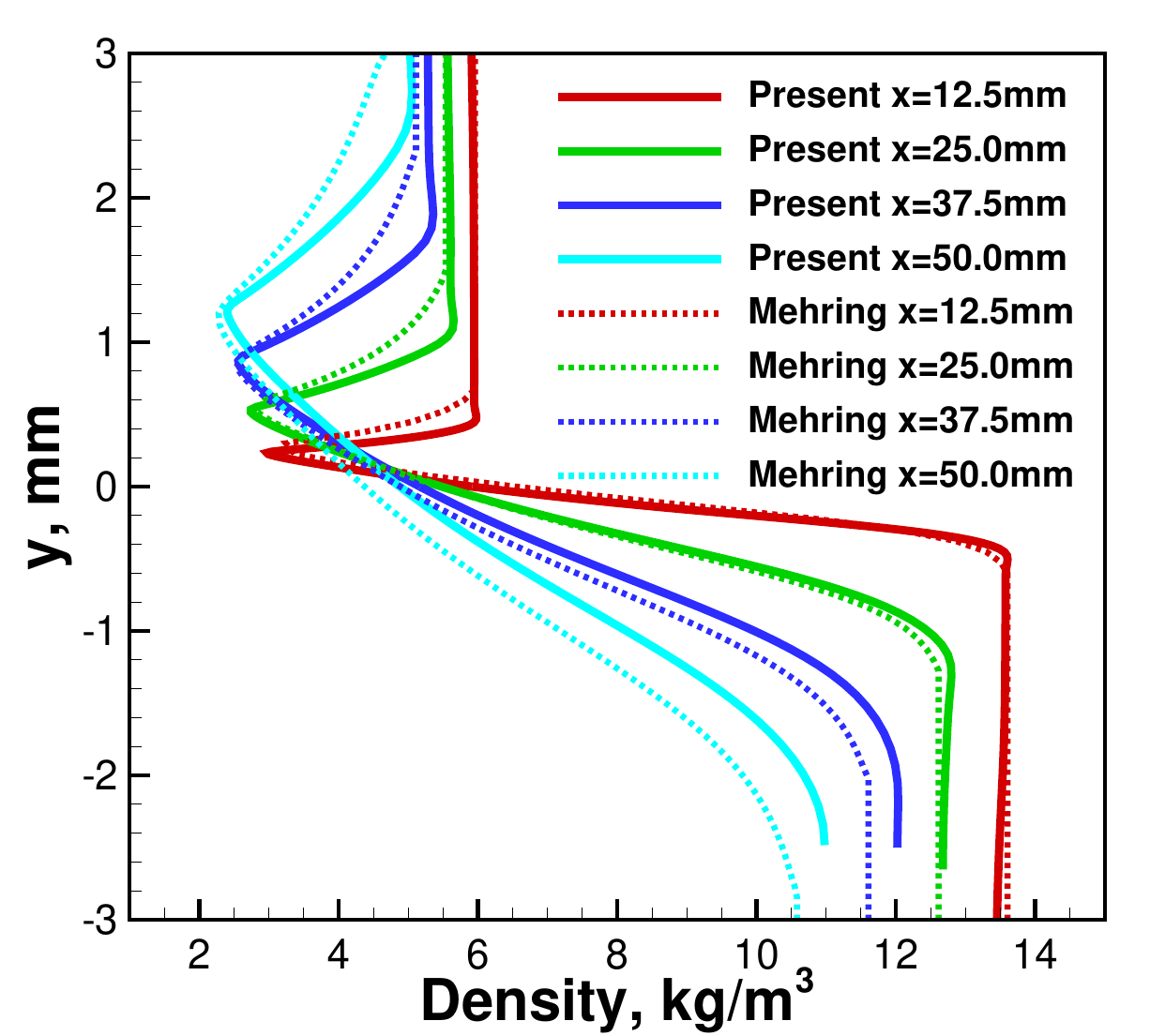}
        \caption{Density}
    \end{subfigure}
    \caption{Profiles at different streamwise positions for turbulent mixing layer. Mehring et al. used constant pressure gradient of -200 atm/m}
    \label{fig:profiles_mixing_turbulent}
\end{figure}

\begin{figure}[htb!]
    \centering
    \begin{subfigure}[b]{0.495\linewidth}
        \centering
        \includegraphics[width=1\linewidth]{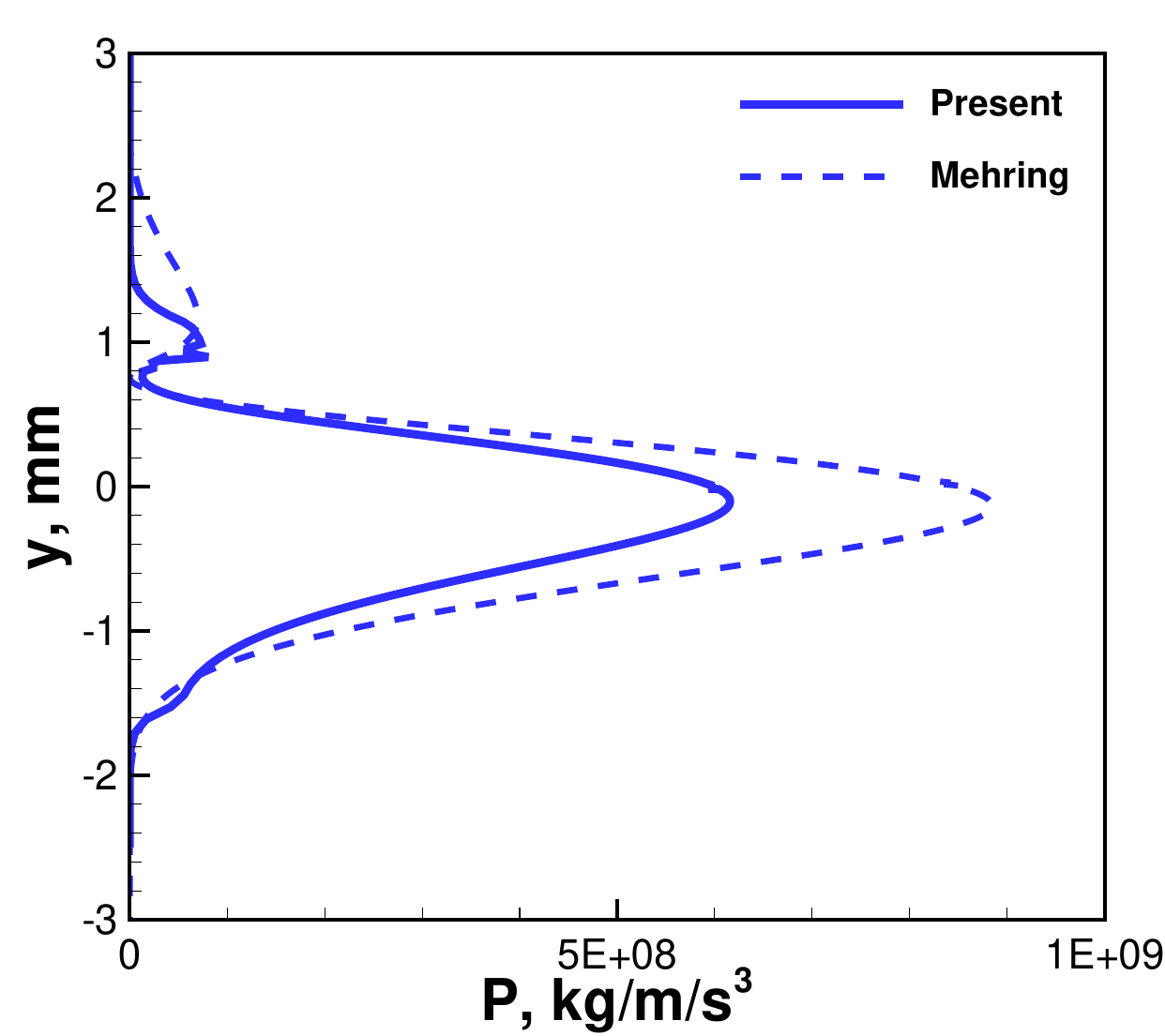}
        \caption{Production rate of turbulent kinetic energy}
    \end{subfigure}
    \begin{subfigure}[b]{0.495\linewidth}
        \centering
        \includegraphics[width=1\linewidth]{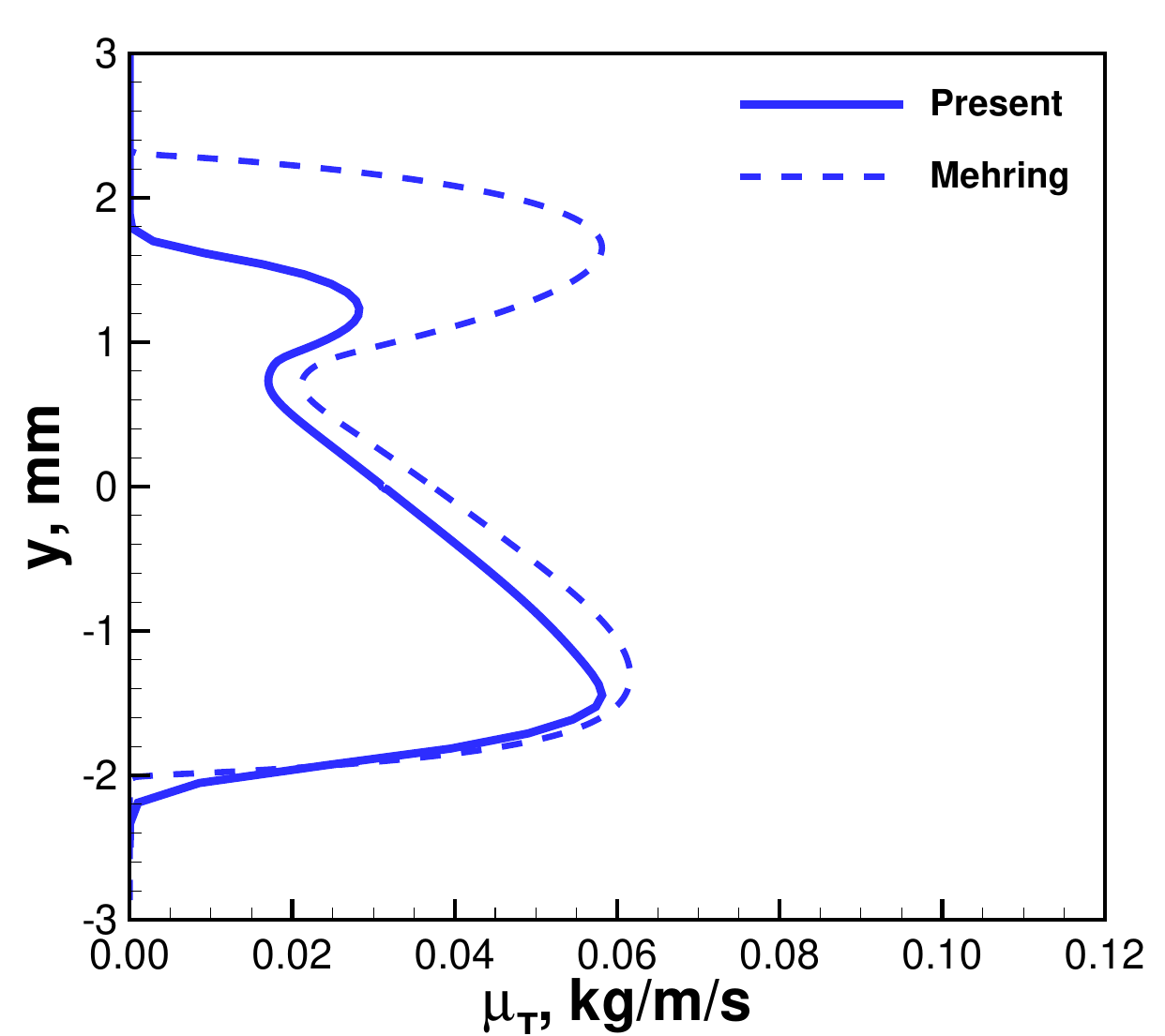}
        \caption{Turbulent viscosity}
    \end{subfigure}
    \caption{Profiles of turbulent properties at x = 37.5 mm for turbulent mixing layer. Mehring et al. used constant pressure gradient of -200 atm/m}
    \label{fig:profiles_mixing_turbulent_Pk_mut}
\end{figure}

Figure \ref{fig:profiles_mixing_turbulent} compares the profiles of temperature and density at four different streamwise positions. Compared to the laminar solutions in Fig. \ref{fig:profiles_mixing_laminar_rho_v}, the thickness of the turbulent mixing layer is one order of magnitude larger. However, the basic behavior remains the same. The present temperature and density agree with those by Mehring et al. in general. The boundary-layer solution is more diffusive as indicated by the thicker temperature and density shear layers. This is due to the higher production rate of turbulent kinetic energy, and thus the higher turbulent viscosity in the mixing layer and flame region for the boundary-layer approximation, as shown by the profiles at $x = 37.5 \, \rm{mm}$ in Fig. \ref{fig:profiles_mixing_turbulent_Pk_mut}. The stronger turbulent diffusion in the boundary-layer solution induces a lower peak temperature in the middle of the flame.

\begin{figure}[htb!]
    \centering
    \begin{subfigure}[b]{0.495\linewidth}
        \centering
        \includegraphics[width=1\linewidth]{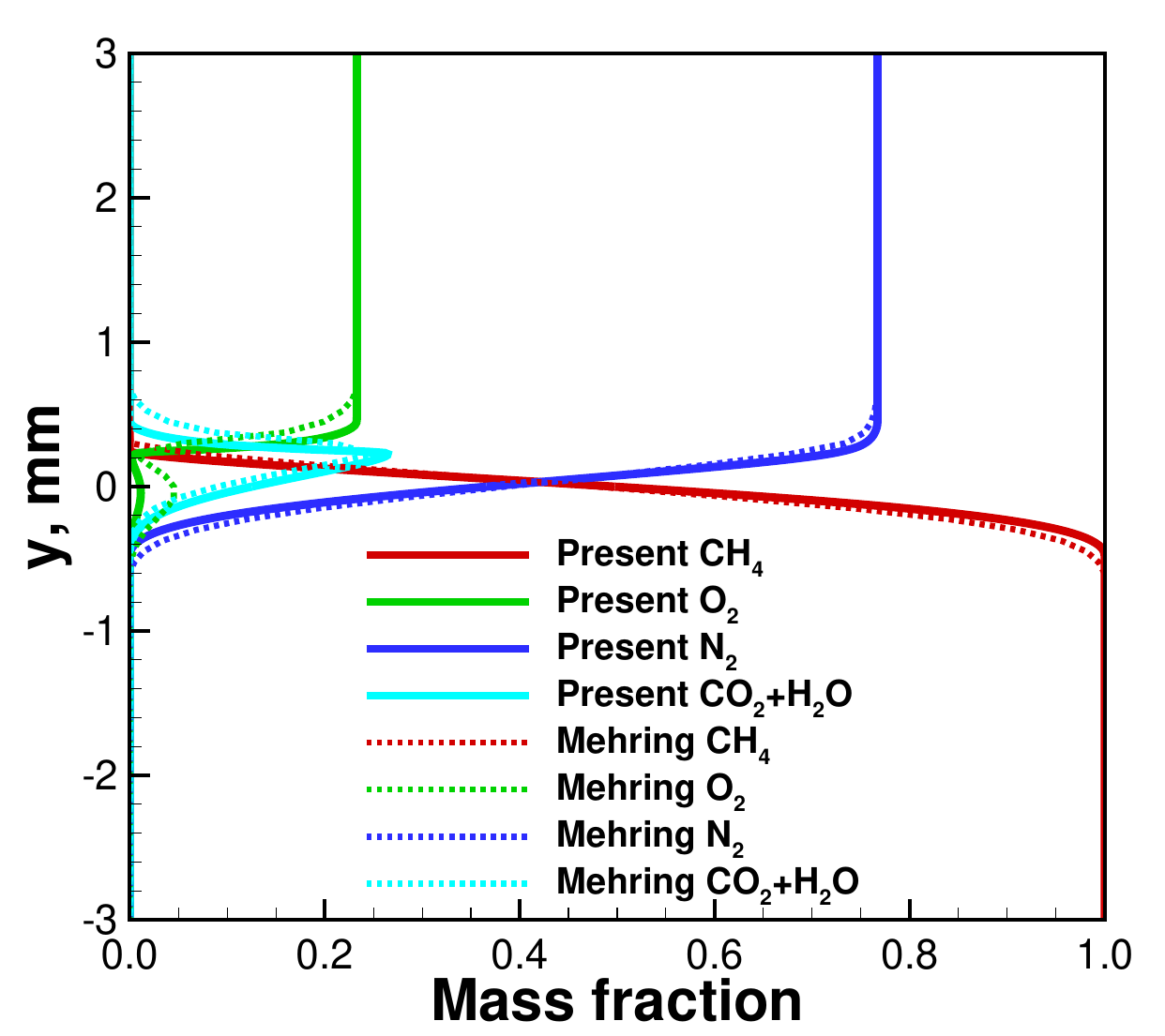}
        \caption{x=12.5 mm}
    \end{subfigure}
    \begin{subfigure}[b]{0.495\linewidth}
        \centering
        \includegraphics[width=1\linewidth]{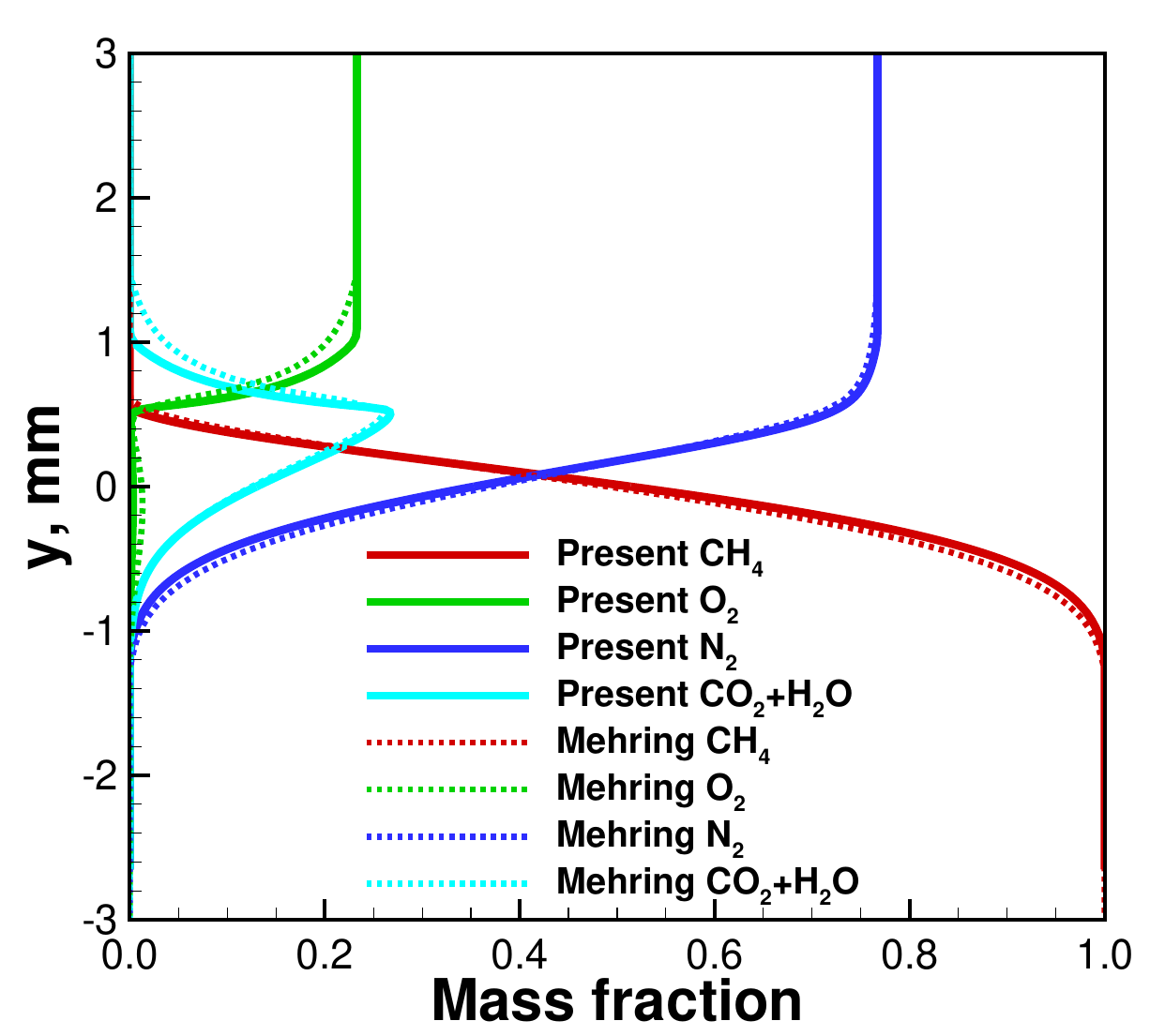}
        \caption{x=25.0 mm}
    \end{subfigure}
    \begin{subfigure}[b]{0.495\linewidth}
        \centering
        \includegraphics[width=1\linewidth]{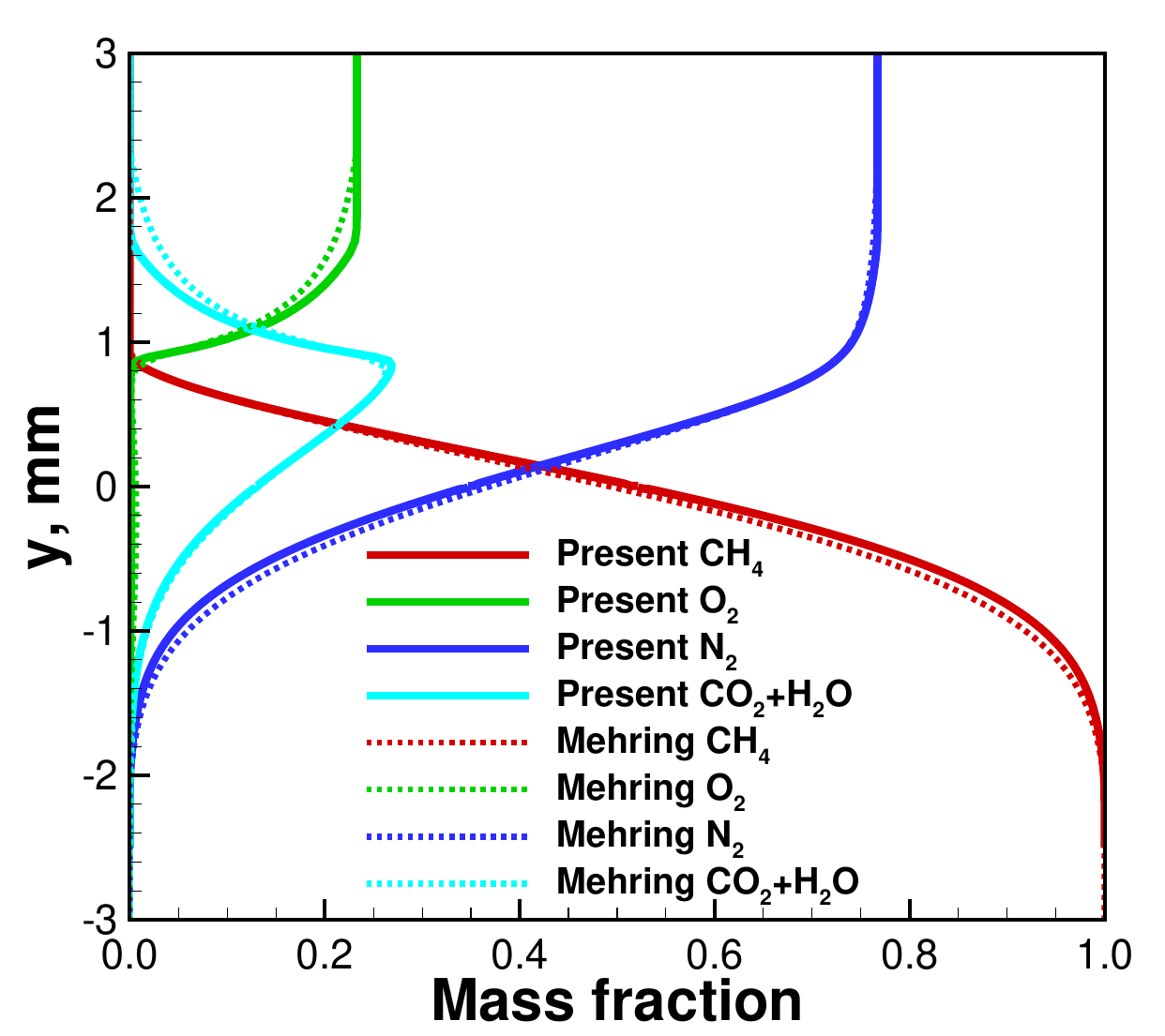}
        \caption{x=37.5 mm}
    \end{subfigure}
    \begin{subfigure}[b]{0.495\linewidth}
        \centering
        \includegraphics[width=1\linewidth]{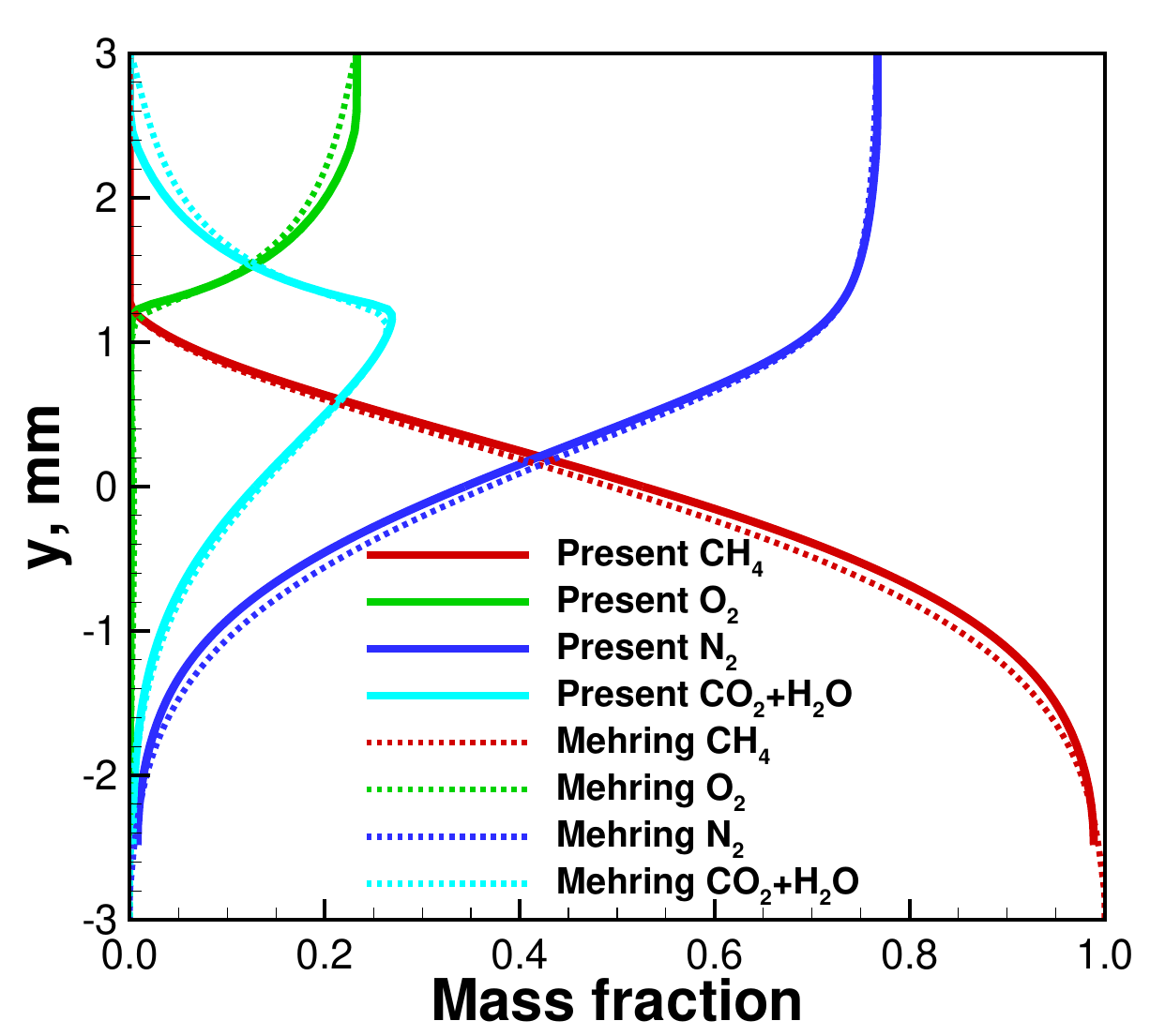}
        \caption{x=50.0 mm}
    \end{subfigure}
    \caption{Mass fraction profiles at different streamwise positions for turbulent mixing layer. Mehring et al. used constant pressure gradient of -200 atm/m}
    \label{fig:profiles_mixing_turbulent_Yi}
\end{figure}

Figure \ref{fig:profiles_mixing_turbulent_Yi} compares the profiles of mass fraction of each species at four different streamwise positions by the full Navier-Stokes and boundary-layer equations. At each streamwise position, the mass fraction of products reaches its peak at the transverse location where both methane and oxygen are simultaneously depleted. Nitrogen is inert in the present reaction model and is thus smoothly diffused from the air side to the fuel side. The peak product mass fraction stays almost constant along the streamwise direction. This is because the balance among convection, diffusion, and production in the species transport equations is independent of the pressure gradient after ignition. 
Similar to the profiles of density and temperature in Fig. \ref{fig:profiles_mixing_turbulent}, the two solutions for the mass fractions generally agree well with each other. The profiles of species mass fractions by the boundary-layer equations are slightly thicker than the full Navier-Stokes equations. The peak mass fractions of products by the full Navier-Stokes equations have slightly larger values. 

\begin{figure}[htb!]
    \centering
    \begin{subfigure}[b]{0.495\linewidth}
        \centering
        \includegraphics[width=1\linewidth]{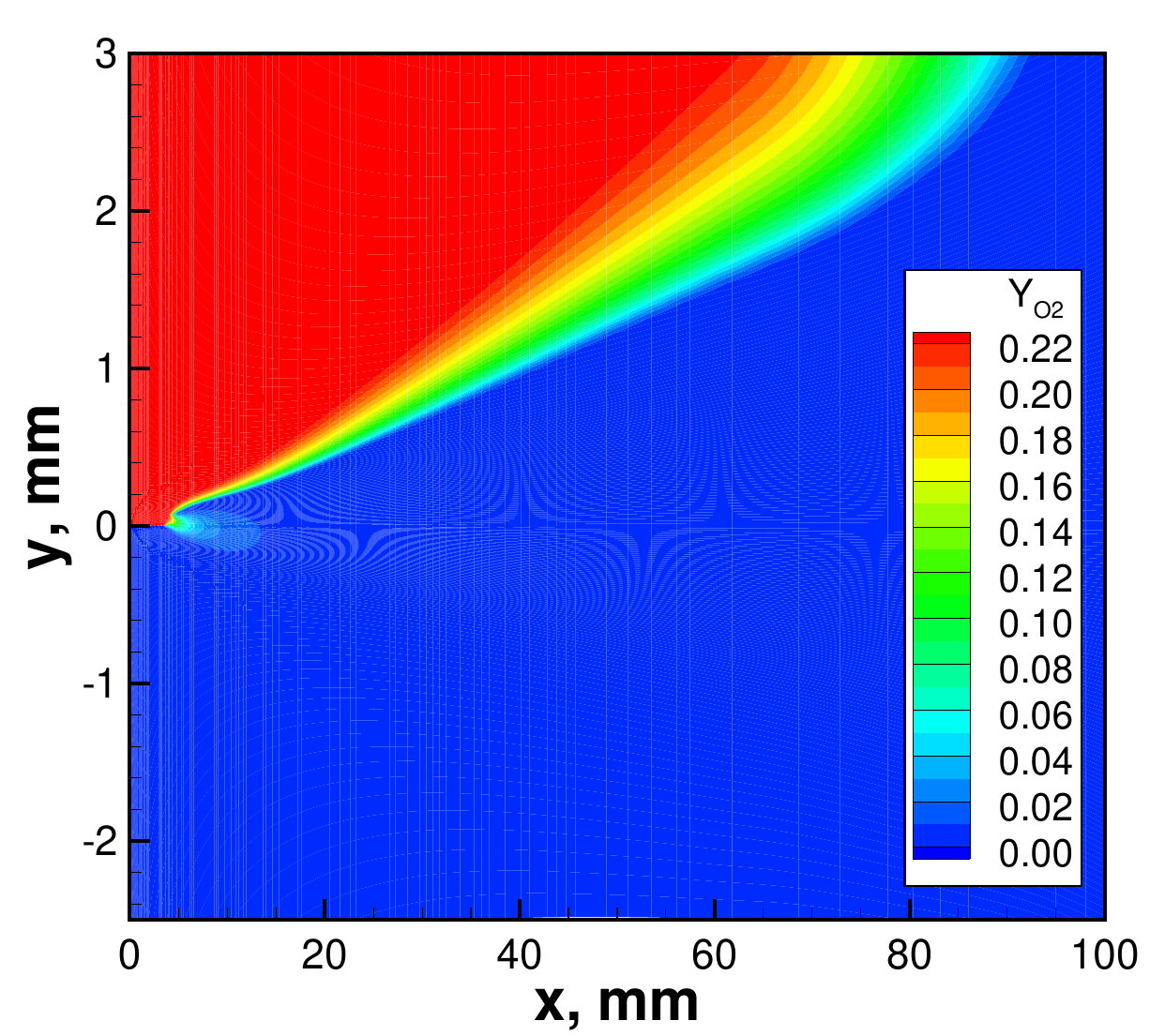}
        \caption{Present}
    \end{subfigure}
    \begin{subfigure}[b]{0.495\linewidth}
        \centering
        \includegraphics[width=1\linewidth]{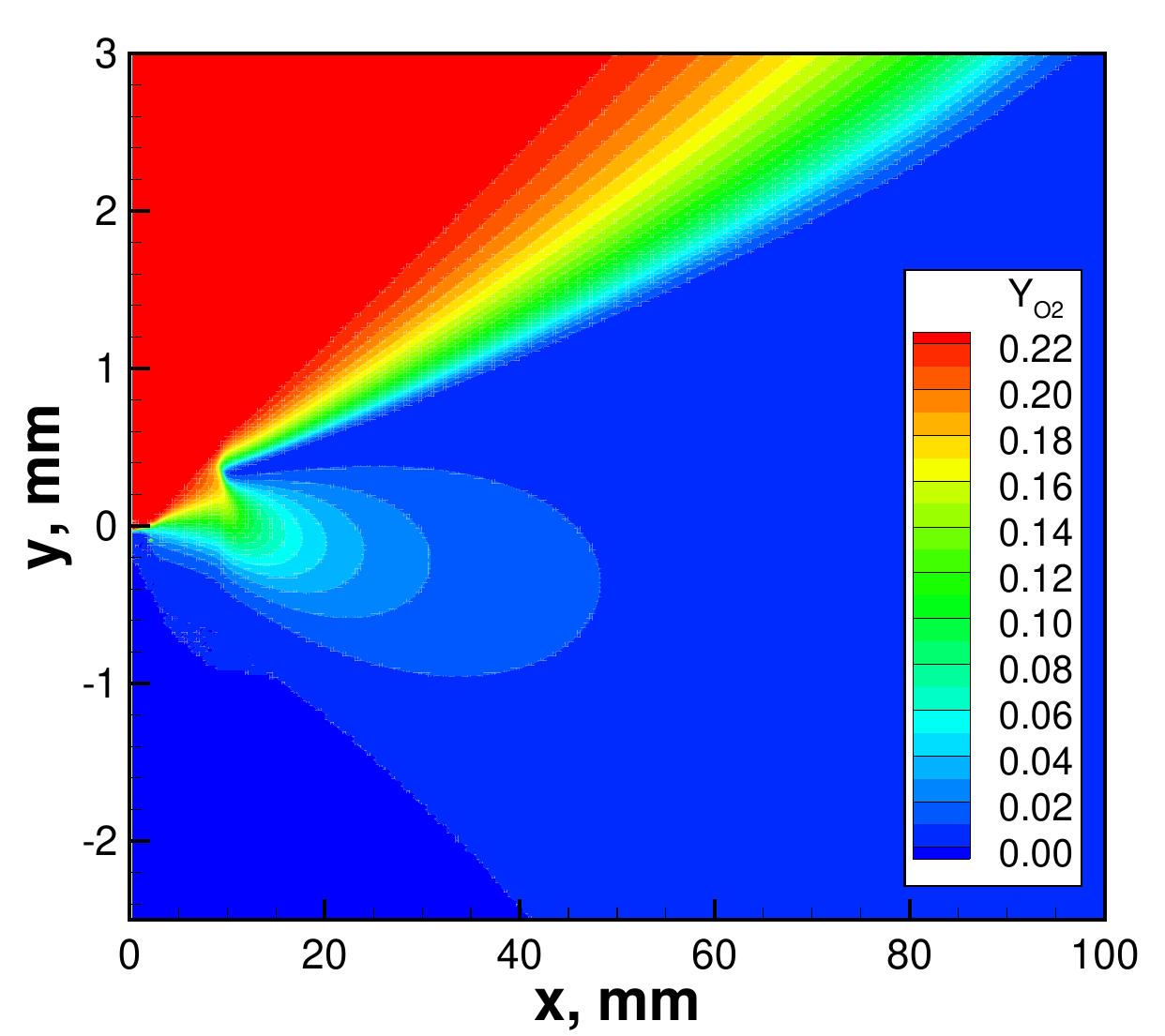}
        \caption{Mehring et al.}
    \end{subfigure}
    \caption{Contours of mass fraction of oxygen for turbulent mixing layer. Mehring et al. used constant pressure gradient of -200 atm/m}
    \label{fig:contour_O2_mixing_turbulent}
\end{figure}

At the first two streamwise positions ($x = 12.5 \, \rm{mm}$ and $x = 25.0 \, \rm{mm}$), there is a local peak in the mass fraction of oxygen slightly below the center line. In addition, the peak value in the boundary-layer solution is larger than those in the full Navier-Stokes solution. This can be explained by the contours of mass fraction of oxygen shown in Fig. \ref{fig:contour_O2_mixing_turbulent}. The flame is ignited after a certain distance downstream from the splitter plate. Before the ignition, the flow is dominated by convection and diffusion. Due to the higher velocity on the upper air side, a significant amount of oxygen is convected to and then diffused in the lower fuel side in front of the flame. Compared to the boundary-layer solution, a smaller quantity of oxygen is transported into the lower side in the present solution since it ignites much earlier. As a result, the peak value of oxygen mass fraction is smaller.

\subsection{Comparisons of Pure Air and Vitiated Air} \label{sec:vitiated_air}
To approximate the inlet flow conditions in a turbine burner, the case with vitiated air at the upper-stream inlet is considered. The vitiated air is used to simulate the exhaust gas from the upstream primary combustion chamber of a turbine engine. 
It is estimated that, to reach a turbine inlet temperature of $1650 \, \mathrm{K}$, a fuel-air mass ratio of 0.03 is needed for the stoichiometric combustion in the primary combustion chamber. As a result, the vitiated air at the turbine inlet consists of $73.77\%$ $\rm{N_2}$, $11.01\%$ $\rm{O_2}$, $8.04\%$ $\rm{CO_2}$, and $7.18\%$ $\rm{H_2O}$ by mass fraction. These compositions of vitiated air are specified as the inlet boundary conditions of species mass fractions at the upper stream of the nozzle. The other flow conditions are kept the same as the pure air case. To avoid the influence of nozzle geometry, the profiles without reshaping the side walls are applied in this section.

\begin{figure}[htb!]
    \centering
    \begin{subfigure}[b]{0.495\linewidth}
        \centering
        \includegraphics[width=1\linewidth]{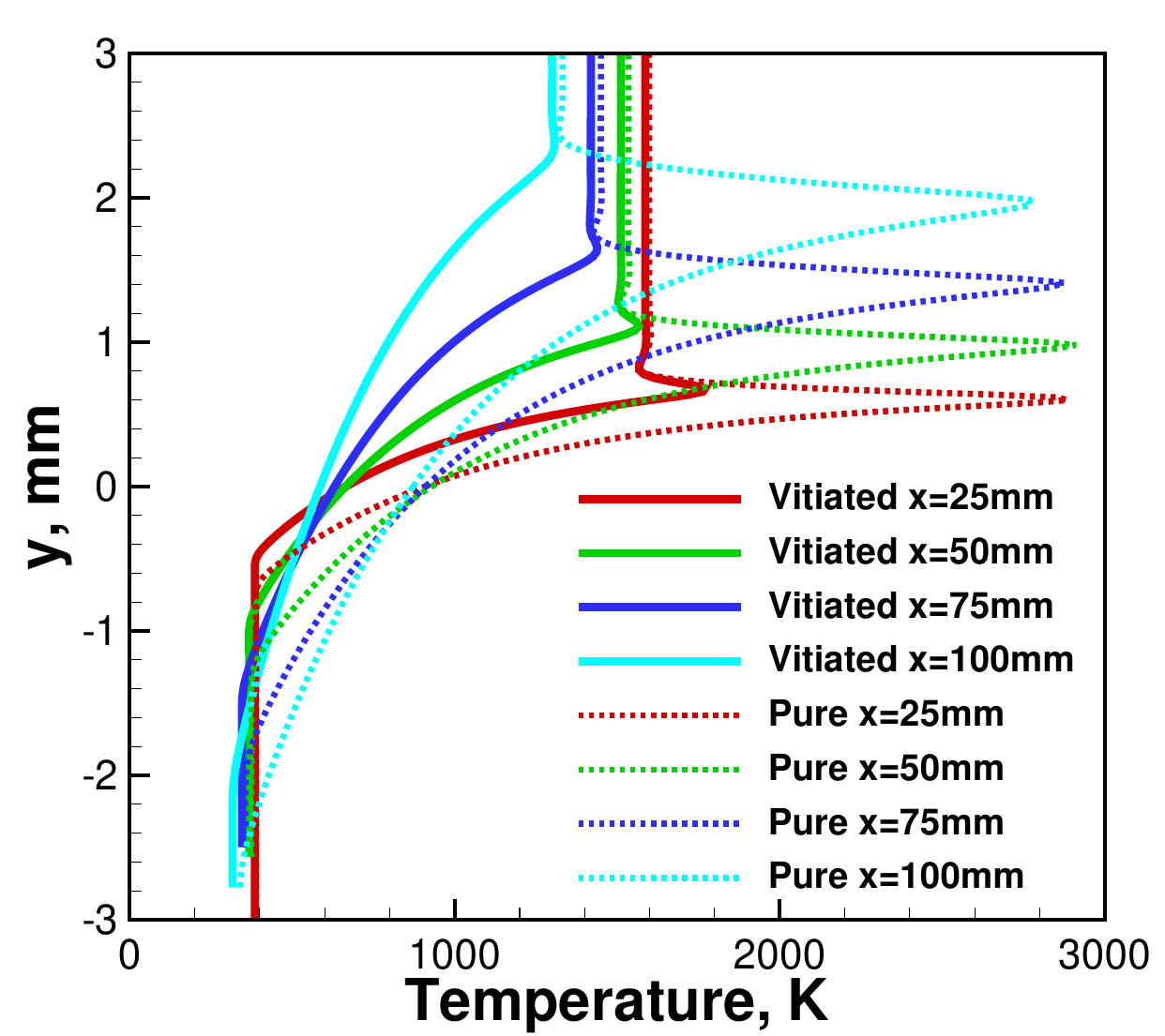}
        \caption{Temperature}
    \end{subfigure}
    \begin{subfigure}[b]{0.495\linewidth}
        \centering
        \includegraphics[width=1\linewidth]{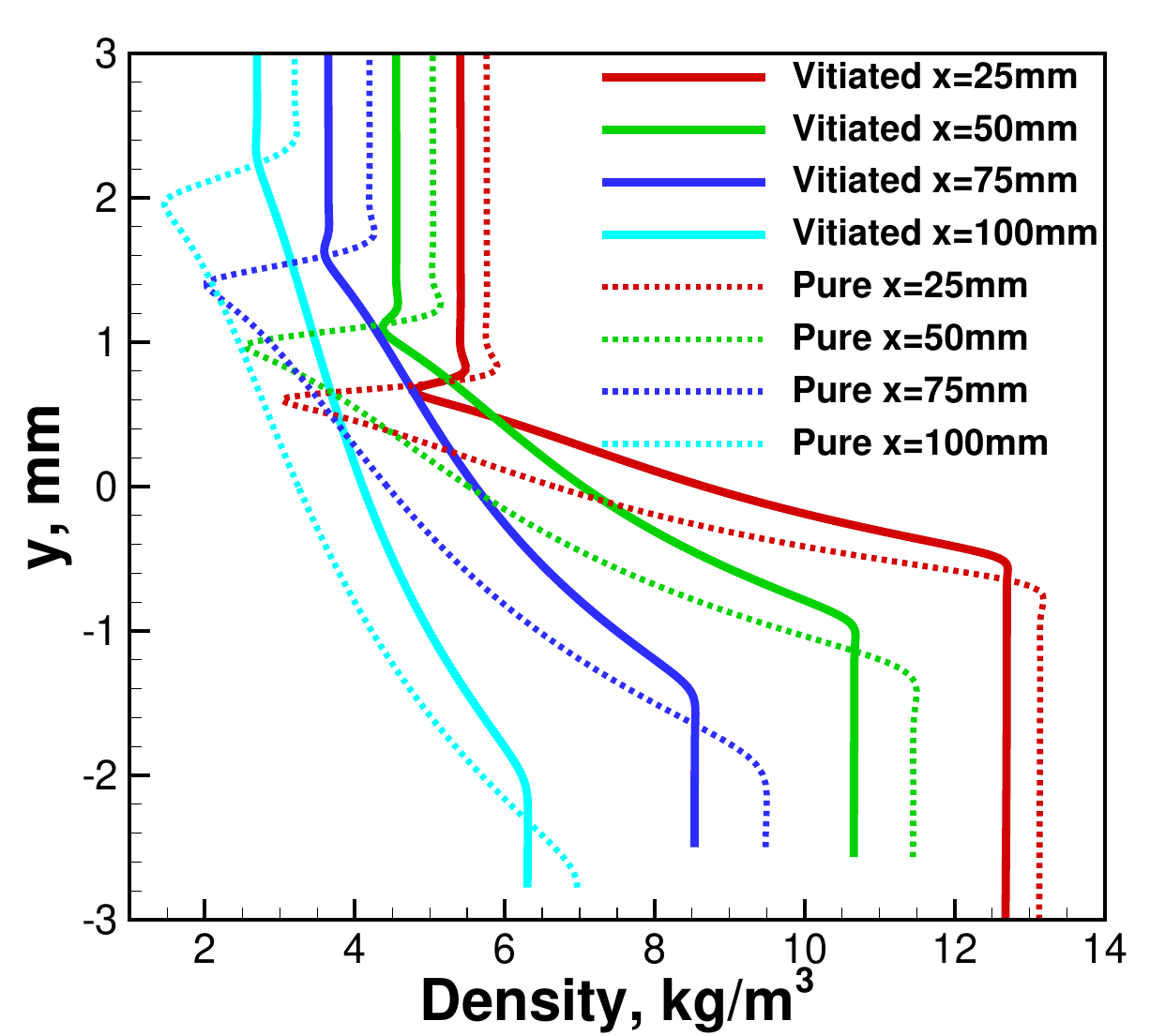}
        \caption{Density}
    \end{subfigure}
    \caption{Profiles at different streamwise positions for turbulent mixing layer with vitiated air inlet}
    \label{fig:profiles_mixing_turbulent_vitiated}
\end{figure}

Figure \ref{fig:profiles_mixing_turbulent_vitiated} compares the profiles of temperature and density at different streamwise positions by using the inlet conditions of pure air and vitiated air.  
Compared to the pure air case, the peak flame temperature in the vitiated air case is reduced since the lower oxygen concentration in the upper stream significantly weakens the chemical reaction. As a result, higher density levels are found in the flame for the vitiated air case under the same pressure levels as the pure air case. 
The transverse location of peak temperature and thus minimum density, corresponding to stoichiometric combustion, is sightly shifted upward for the vitiated air case. The oxygen is redistributed along the transverse direction under the actions of molecular and turbulent diffusion. However, the concentration of oxygen is globally lower in the vitiated air case. Thus, the amount of oxygen required for the stoichiometric combustion moves farther towards the upper freestream.
Compared to the pure air case, the thickness of the shear layer is decreased for the vitiated air case, as indicated by the profiles of temperature and density. This is because the reduced velocity gradients in the shear layer resulting from the weak chemical reaction produce a low production of turbulent kinetic energy. Consequently, the turbulent diffusion along the transverse direction is reduced. 

It is interesting to note that turbulence modeling has significant influence on the developments of flow and combustion in the mixing layer. Comparison between the pure-air results by the SST model in Fig. \ref{fig:profiles_mixing_turbulent_vitiated} and those by the $k\mbox{-}\omega$ model in Fig. \ref{fig:profiles_mixing_turbulent} shows that the mixing layer by the SST model develops slower. For example, the thickness of the shear layer on the air side is about $70\%$ of the $k\mbox{-}\omega$ model at $x = 25.0 \, \rm{mm}$, and it decreases to less than half at $x = 50.0 \, \rm{mm}$. This weaker turbulent diffusion of the SST model reduces the mixing between air and fuel, and thus the chemical reaction, as seen from the lower peak temperature at $x = 25.0 \, \rm{mm}$ and $x = 50.0 \, \rm{mm}$. This discrepancy between the two RANS turbulence models indicates that more elaborate turbulence modeling, such as large-eddy simulation or detached-eddy simulation, is necessary to accurately resolve the interaction between chemical reaction and turbulent flow.

\begin{figure}[htb!]
    \centering
    \includegraphics[width=0.495\linewidth]{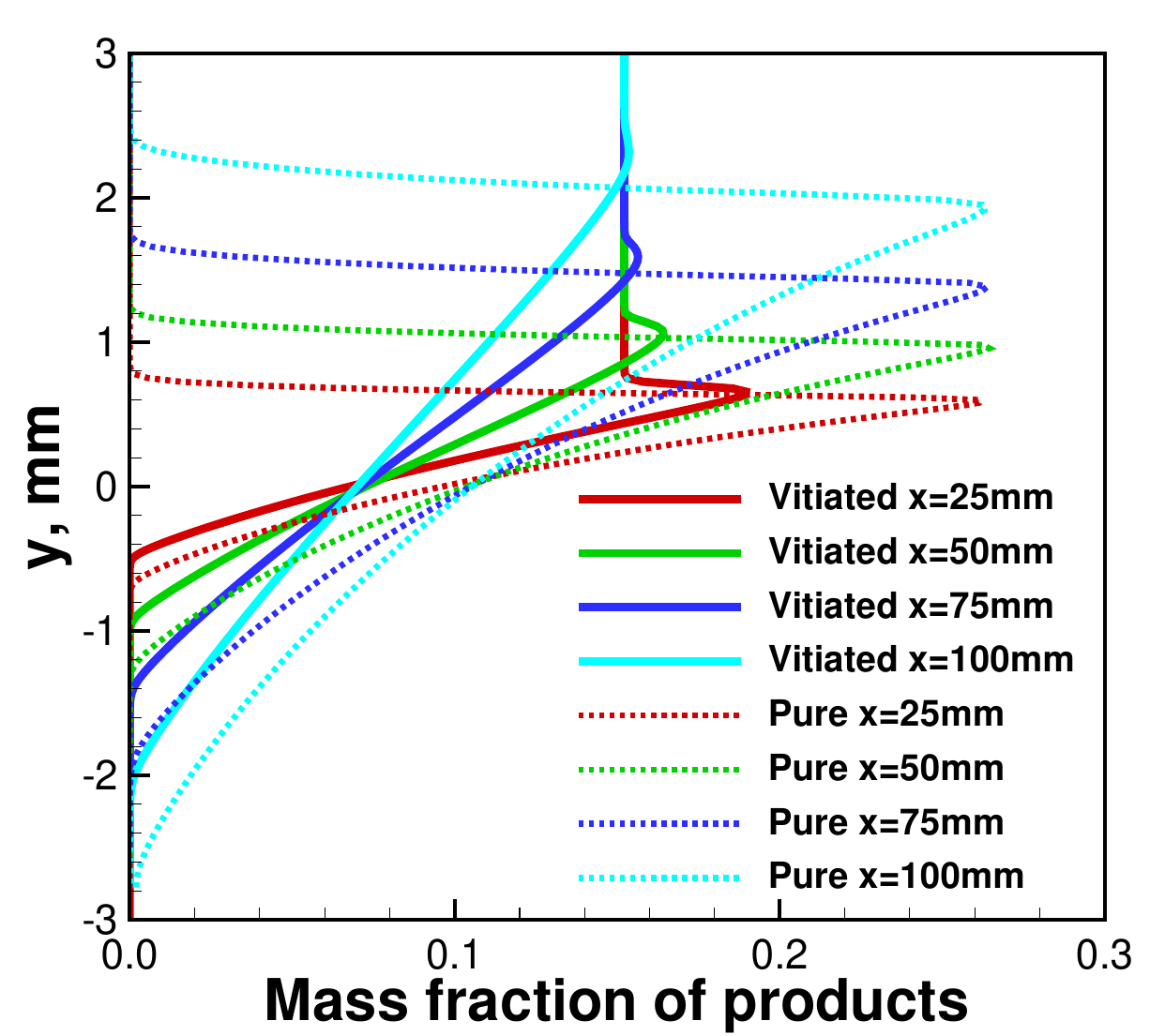}
    \caption{Profiles of mass fraction of products at different streamwise positions for turbulent mixing layer with vitiated air inlet}
    \label{fig:profiles_mixing_turbulent_vitiated_Yprod}
\end{figure}

The profiles of mass fraction of products at different streamwise positions are shown in Fig. \ref{fig:profiles_mixing_turbulent_vitiated_Yprod}. Similar to the temperature and density, the thickness of profile of mass fraction is obviously reduced in the vitiated air case. Although there already exists initial $\mathrm{CO_2}$ and $\mathrm{H_2O}$ in the freestream vitiated air, the peak mass fraction of products is still lower than that in the pure air case due to the significantly reduced chemical reaction. In fact, the flame almost becomes extinct after $x =75.0 \, \rm{mm}$, as indicated by the extremely low levels of product mass fraction and temperature at $x =75.0 \, \rm{mm}$ and $x = 100.0 \, \rm{mm}$. The upward shift of the peaks in the vitiated air case is also clearly observed. 

\subsection{Reacting Turbulent Flow in Turbine Cascade} \label{sec:turbine_cascade}
\begin{figure}[htb!]
    \centering
    \includegraphics[width=0.38\linewidth]{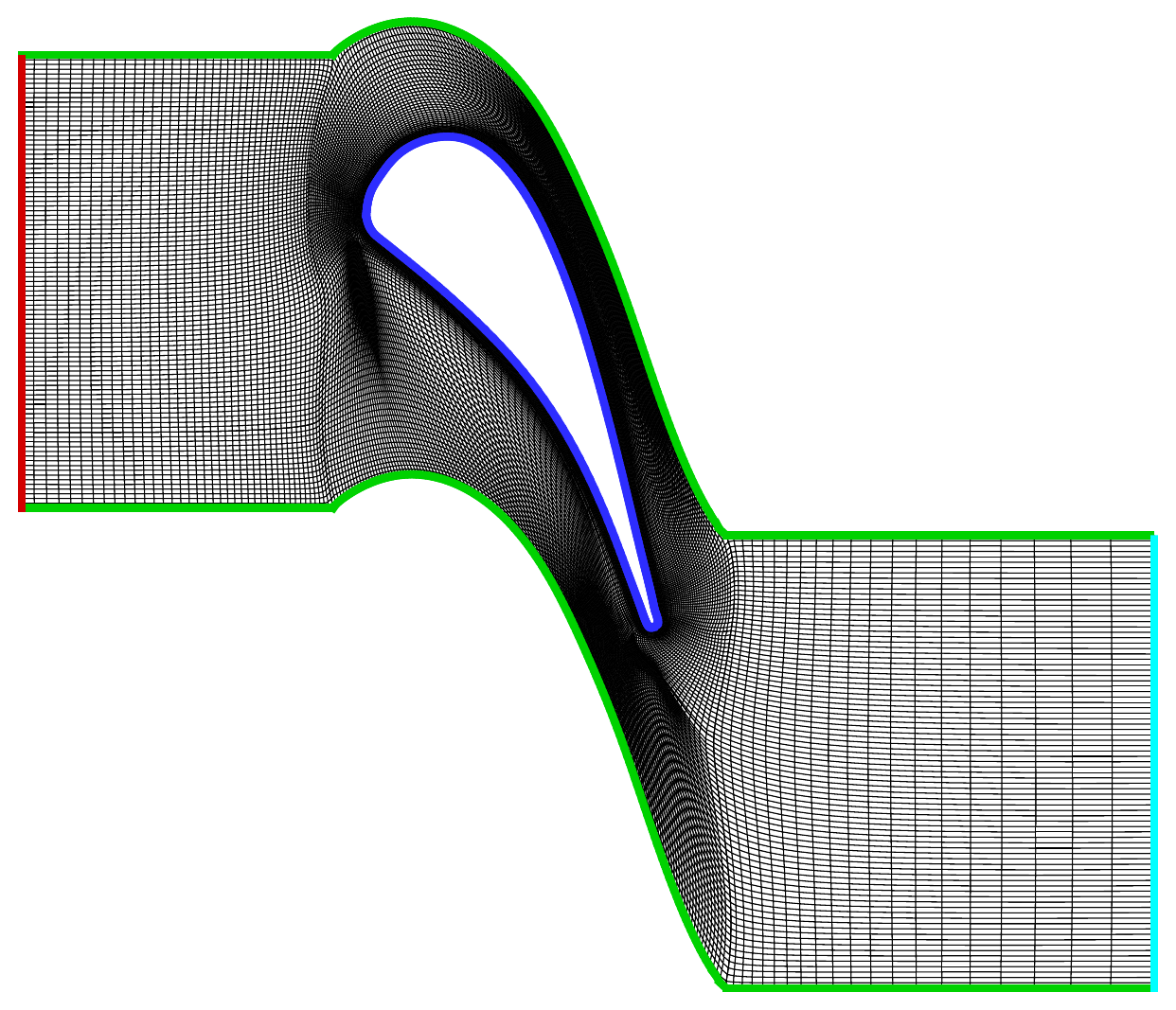}
    \caption{Configuration and grid of turbine cascade}
    \label{fig:grid_turbine_cascade}
\end{figure}

The reacting flow in a highly-loaded transonic turbine guide vane, named VKI LS89 \cite{arts1992aero}, is simulated and compared for the cases with pure air and vitiated air inlets. The chord of the vane is $76.674 \, \rm{mm}$, and the pitch-to-chord ratio is 0.85. The stagger angle of the blade is $55^\circ$. The multi-block structured grid, as shown in Fig. \ref{fig:grid_turbine_cascade}, is generated for a single cascade passage with translational periodicity on the pitchwise boundaries (green curves in Fig. \ref{fig:grid_turbine_cascade}). There are 317 grid points around the blade (blue curve in Fig. \ref{fig:grid_turbine_cascade}), with the points concentrated near the leading edge and trailing edge. The dimensionless distance $y^+$ of the first grid point away from the blade is less than one. The total grid has 26416 cells, which are divided into 7 blocks for the parallel computation.

At the inlet of the turbine cascade, methane with total temperature of $400 \, \rm{K}$ is injected over part of the middle section. Air with total temperature of $1650 \, \rm{K}$ is specified at the remaining part of the inlet section. The total pressure is uniform $166834 \, \rm{Pa}$, and the inlet flow angle is 0. The back pressure at the outlet is set as $101325 \, \rm{Pa}$. This produces an averaged streamwise pressure gradient of $-20 \, \rm{atm/m}$ within the cascade passage, which is an order of magnitude lower compared to the mixing-layer cases.

\begin{figure}[htb!]
    \centering
    \begin{subfigure}[b]{0.35\linewidth}
        \centering
            \includegraphics[width=1\linewidth]{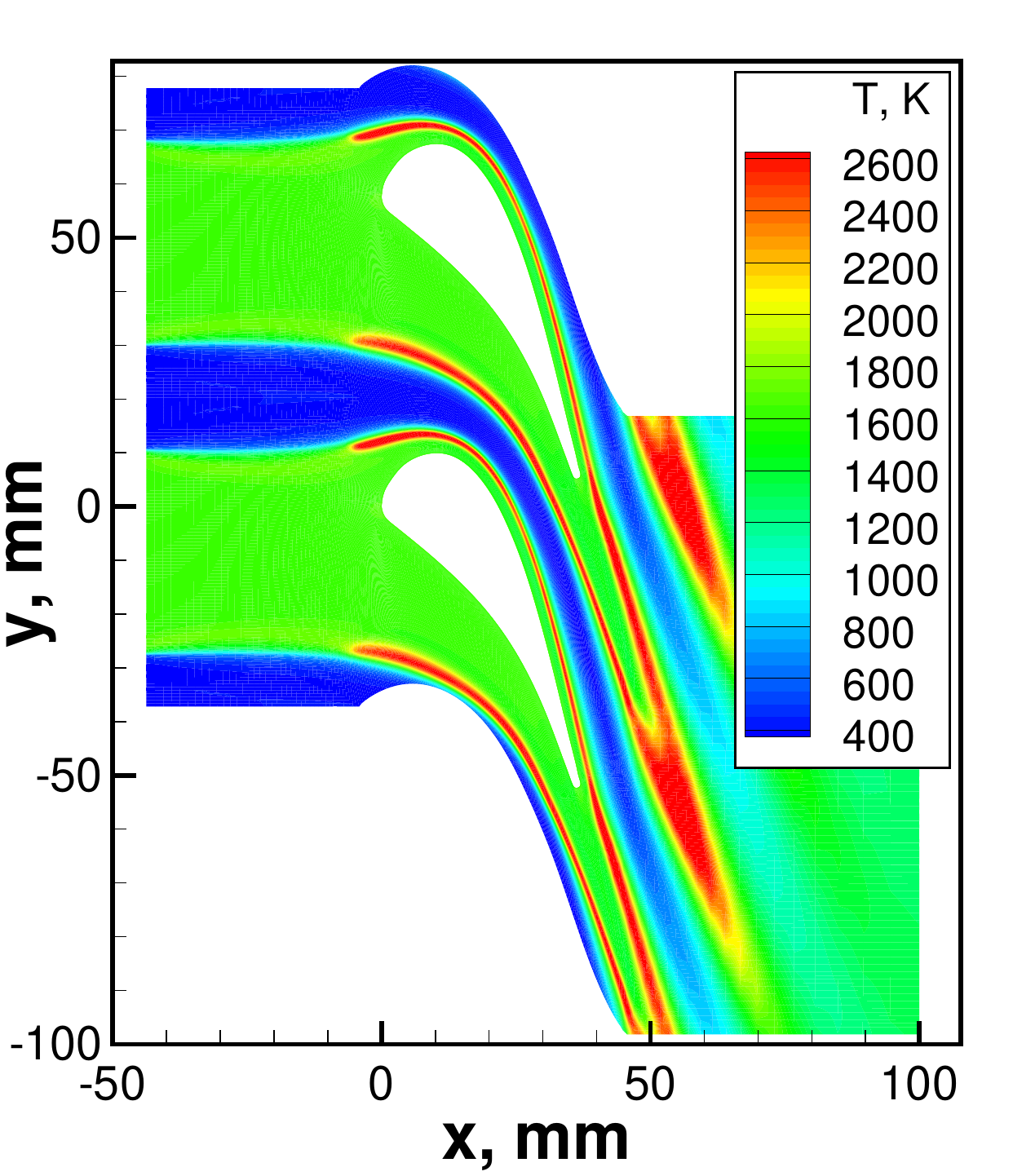}
        \caption{Pure air}
    \end{subfigure}
    \begin{subfigure}[b]{0.35\linewidth}
        \centering
        \includegraphics[width=1\linewidth]{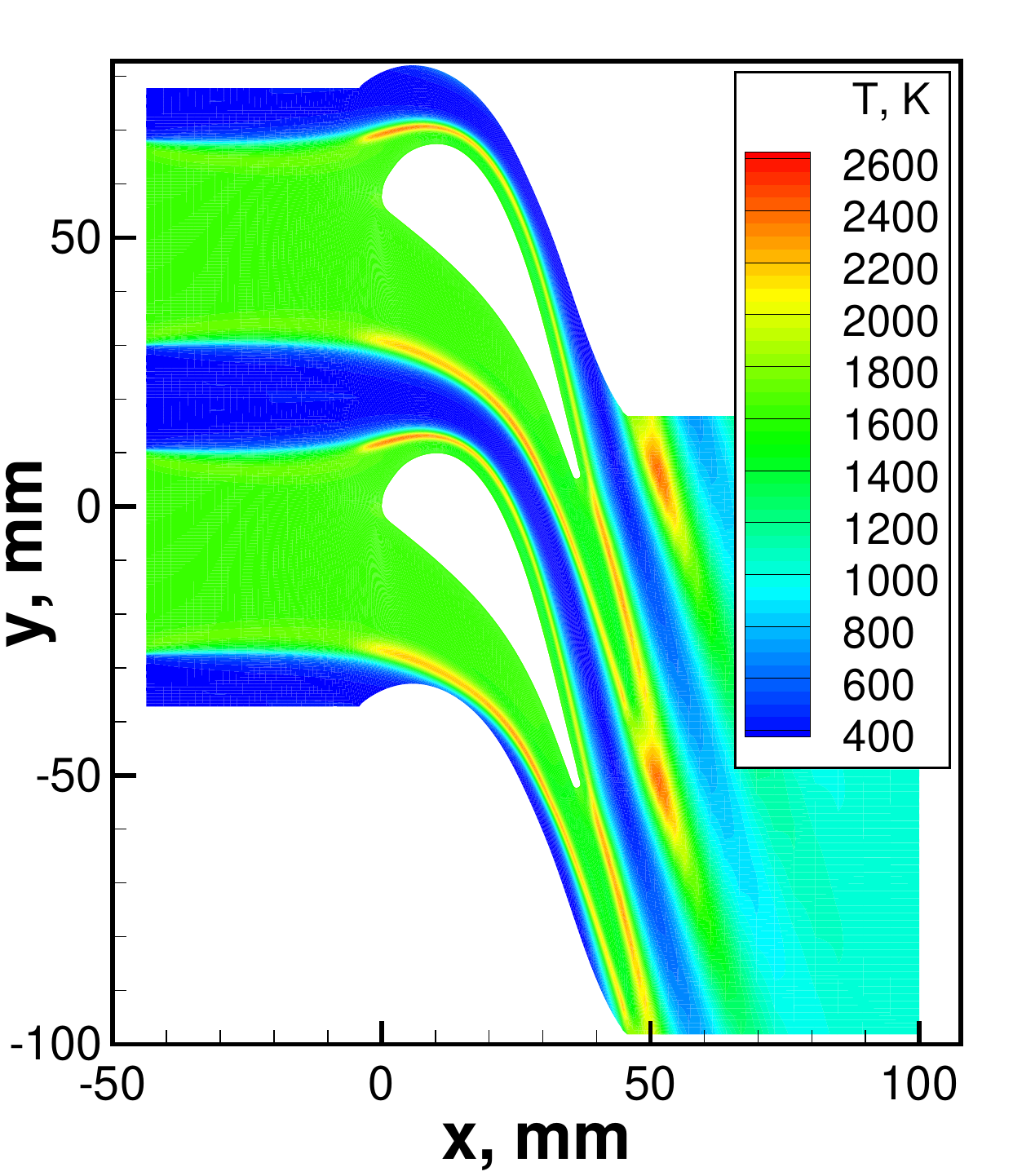}
        \caption{Vitiated air}
    \end{subfigure}
    \caption{Contours of temperature in turbine cascade}
    \label{fig:contour_T_turbine}
\end{figure}

Figure \ref{fig:contour_T_turbine} shows the contours of temperature for the pure air and vitiated air cases. Two diffusion flames, generated on the interfaces between the fuel and air, are transported downstream in the cascade passage. One of them is near the suction surface, and the other one goes through the middle of the passage. After the trailing edge, the middle branch merges with the suction-surface branch from the adjacent blade, and then they move downstream together. 
The variations of temperature for the vitiated air case are similar to the pure air case, but the flame temperature levels are obviously reduced. The maximum temperature within the flames is about $3200 \, \rm{K}$ for the pure air case, while it reduces to about $2400 \, \rm{K}$ for the vitiated air case.

\begin{figure}[htb!]
    \centering
    \begin{subfigure}[b]{0.35\linewidth}
        \centering
        \includegraphics[width=1\linewidth]{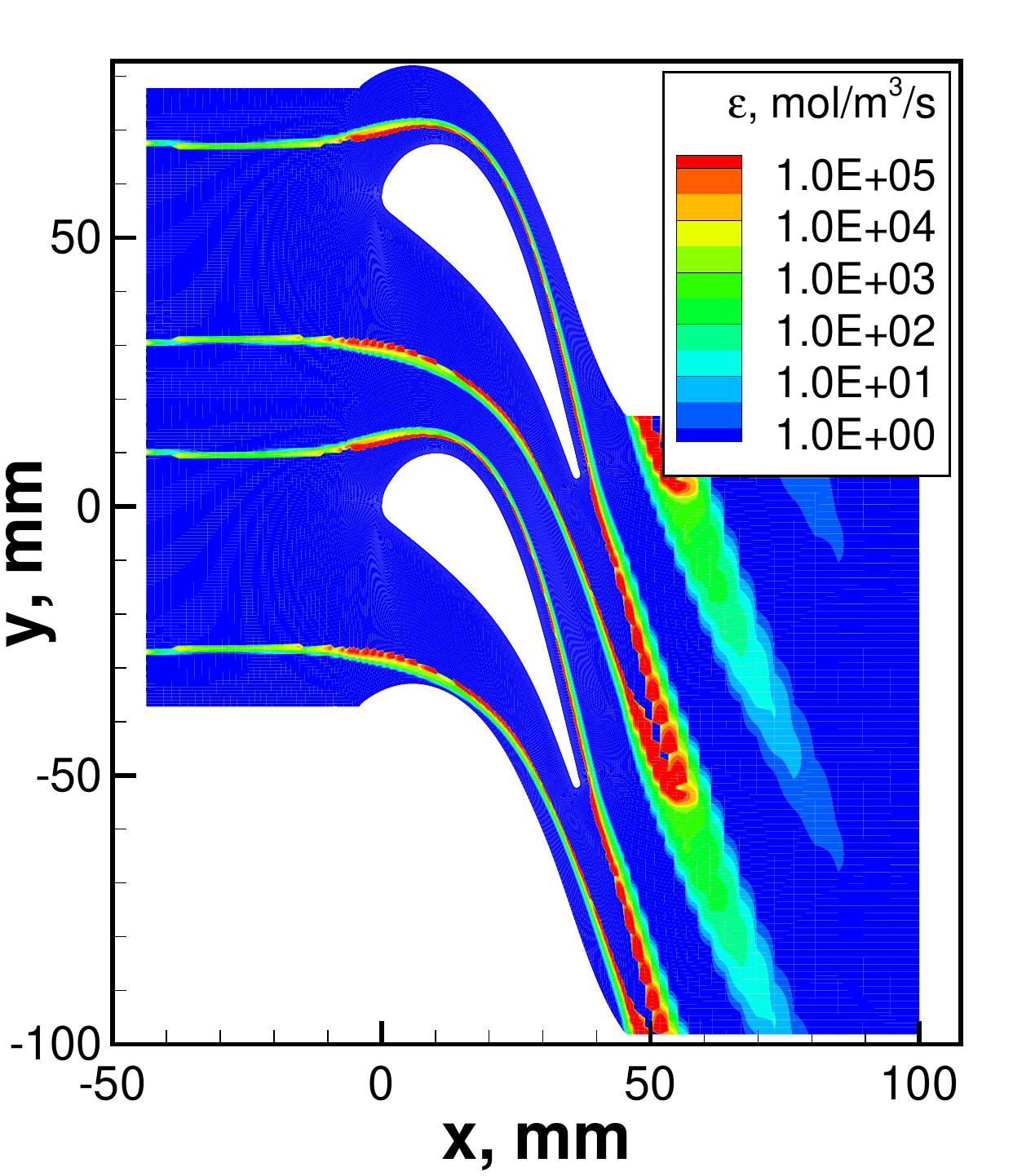}
        \caption{Pure air}
    \end{subfigure}
    \begin{subfigure}[b]{0.35\linewidth}
        \centering
        \includegraphics[width=1\linewidth]{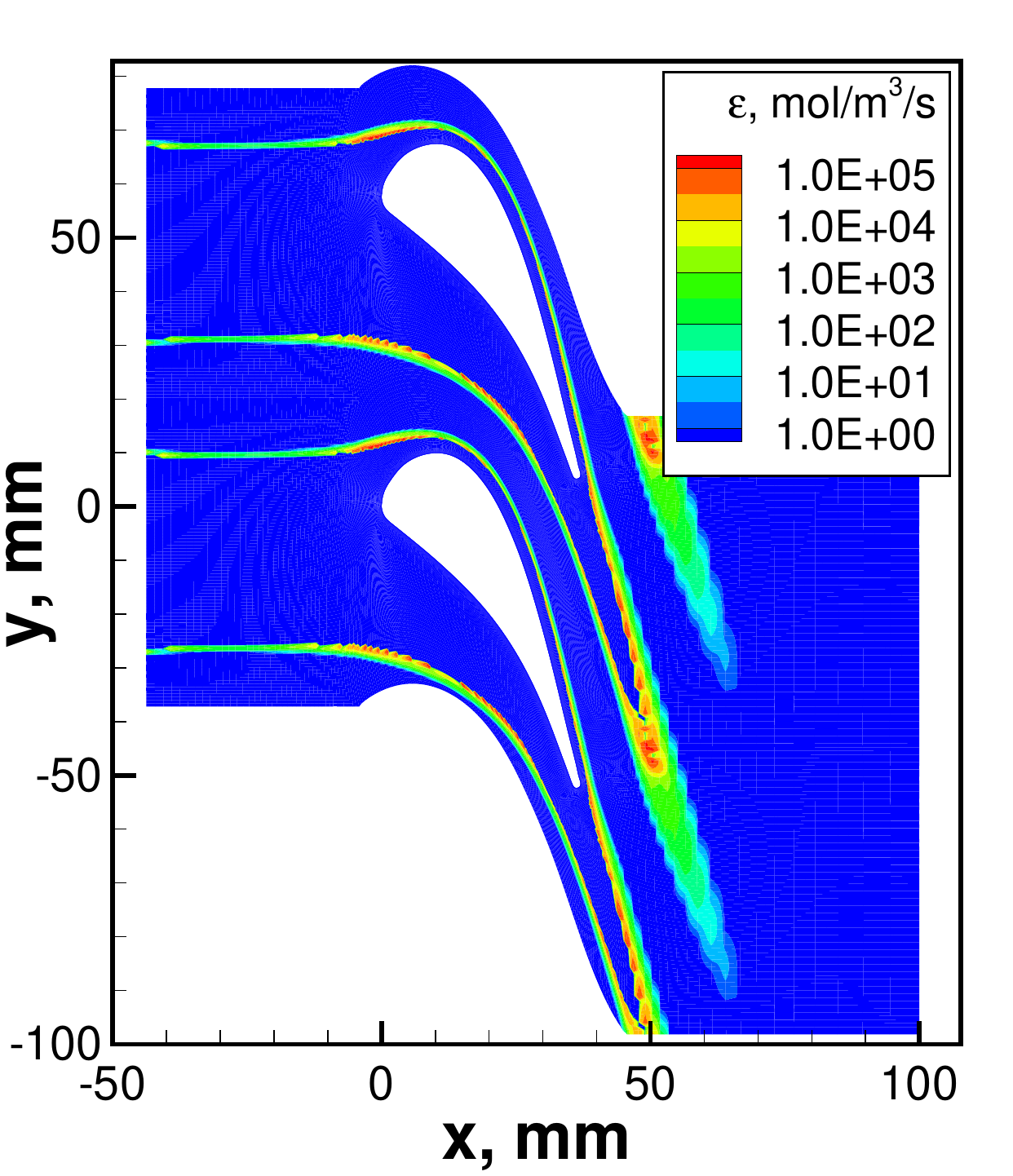}
        \caption{Vitiated air}
    \end{subfigure}
    \caption{Contours of chemical reaction rate in turbine cascade}
    \label{fig:contour_reaction_rate_turbine}
\end{figure}

Same as in the mixing-layer cases, the streamwise and transverse pressure gradients produced by the curved suction and pressure surfaces have significant effects on the flow and combustion process in the turbine cascade, which can be seen from the contours of chemical reaction rate for both pure air and vitiated air cases in Fig. \ref{fig:contour_reaction_rate_turbine}. The region with high reaction rate in the figure indicates the flame. 
Ignition immediately happens at the turbine inlet, even if the reaction rate is relatively low due to the insufficient mixing between fuel and air. 
Disturbed by the blade, the pressure starts to decrease along the streamwise direction near the leading edge. At first glance, the chemical reaction would be weakened by this favourable pressure gradient as in the mixing-layer cases above. However, the local velocity gradients resulting from the blade surface curvature significantly reinforce the molecular and turbulent diffusion and thus make the mixing between fuel and oxidizer sufficient enough, which consequently enhances the chemical reaction. This is why both the strength and thickness of the flames significantly increase near the leading edge. It turns out that the mixing dominates the chemical reaction over the pressure gradient at the initial stage of the turbine cascade.
However, after the suction peak ($x = 10 \, \mathrm{mm}$) on the suction surface, the two flames gradually become weak until reaching the trailing edge due to the strong favorable streamwise pressure gradient produced by the converged turbine passage. After the trailing edge, in the absence of constraints from blade surfaces, the gases within the two flames mix with the low-speed wakes from the suction and pressure surfaces. Hence, the two flames are enhanced again in both strength and thickness, and finally they merge together. 
The variations of temperature in the turbine passage in Fig. \ref{fig:contour_T_turbine} are clearly consistent with those for reaction rate. Note that the overall pressure gradient in the turbine passage is strongly lower than that in the mixing-layer case. Hence, extinction does not happen on any of the two flames within the accelerating turbine passage, although the reaction rate in the vitiated air case is evidently reduced. This is especially important for the flameholding in high-speed flow.

\begin{figure}[htb!]
    \centering
    \begin{subfigure}[b]{0.35\linewidth}
        \centering
        \includegraphics[width=1\linewidth]{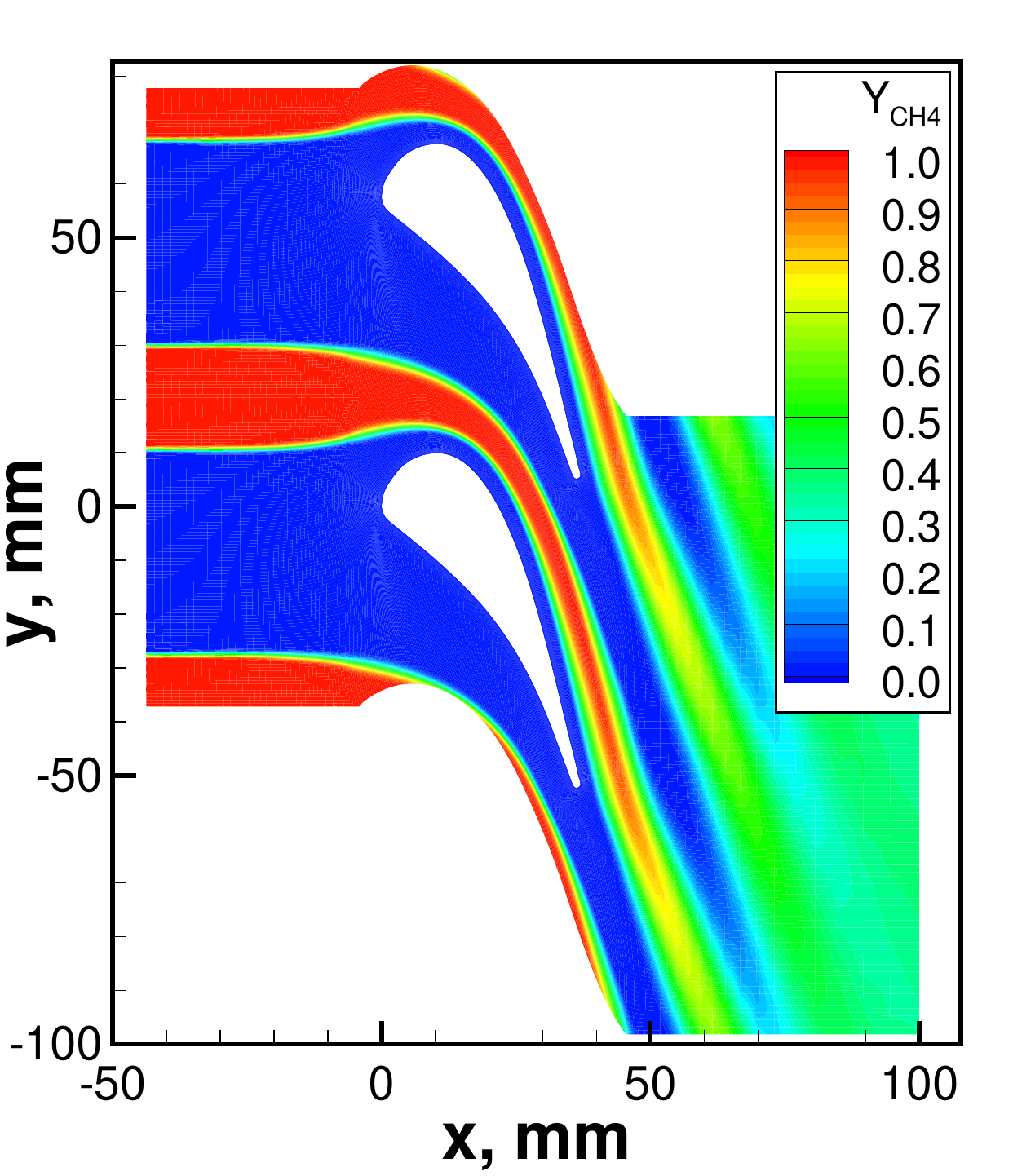}
        \caption{Pure air}
    \end{subfigure}
    \begin{subfigure}[b]{0.35\linewidth}
        \centering
        \includegraphics[width=1\linewidth]{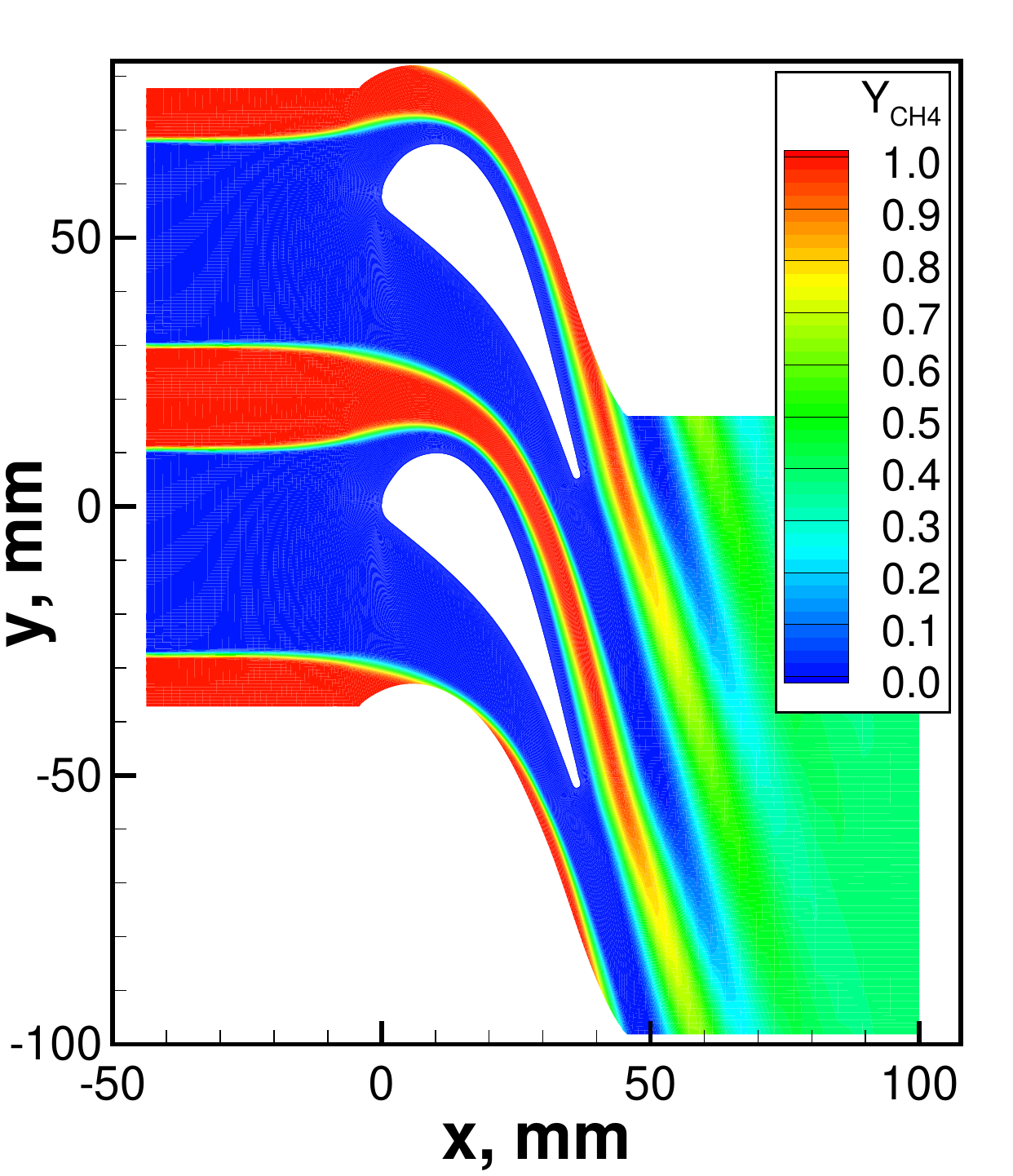}
        \caption{Vitiated air}
    \end{subfigure}
    \caption{Contours of mass fraction of methane in turbine cascade}
    \label{fig:contour_CH4_turbine}
\end{figure}

Figure \ref{fig:contour_CH4_turbine} shows the contours of mass fraction of methane for the pure air and vitiated air cases. The methane from the inlet is transported downstream and continuously consumed on the interfaces between fuel and oxidizer in the turbine passage. The mass fraction of methane is significantly reduced due to the increased reaction rate after the two flames merge. However, the methane is not depleted until it is transported out of the computational domain by the main stream. This is mainly due to the excessive fuel provided at the inlet. Fuel will be depleted if we further decrease the area portion it occupies at the inlet.

\begin{figure}[htb!]
    \centering
    \includegraphics[width=0.495\linewidth]{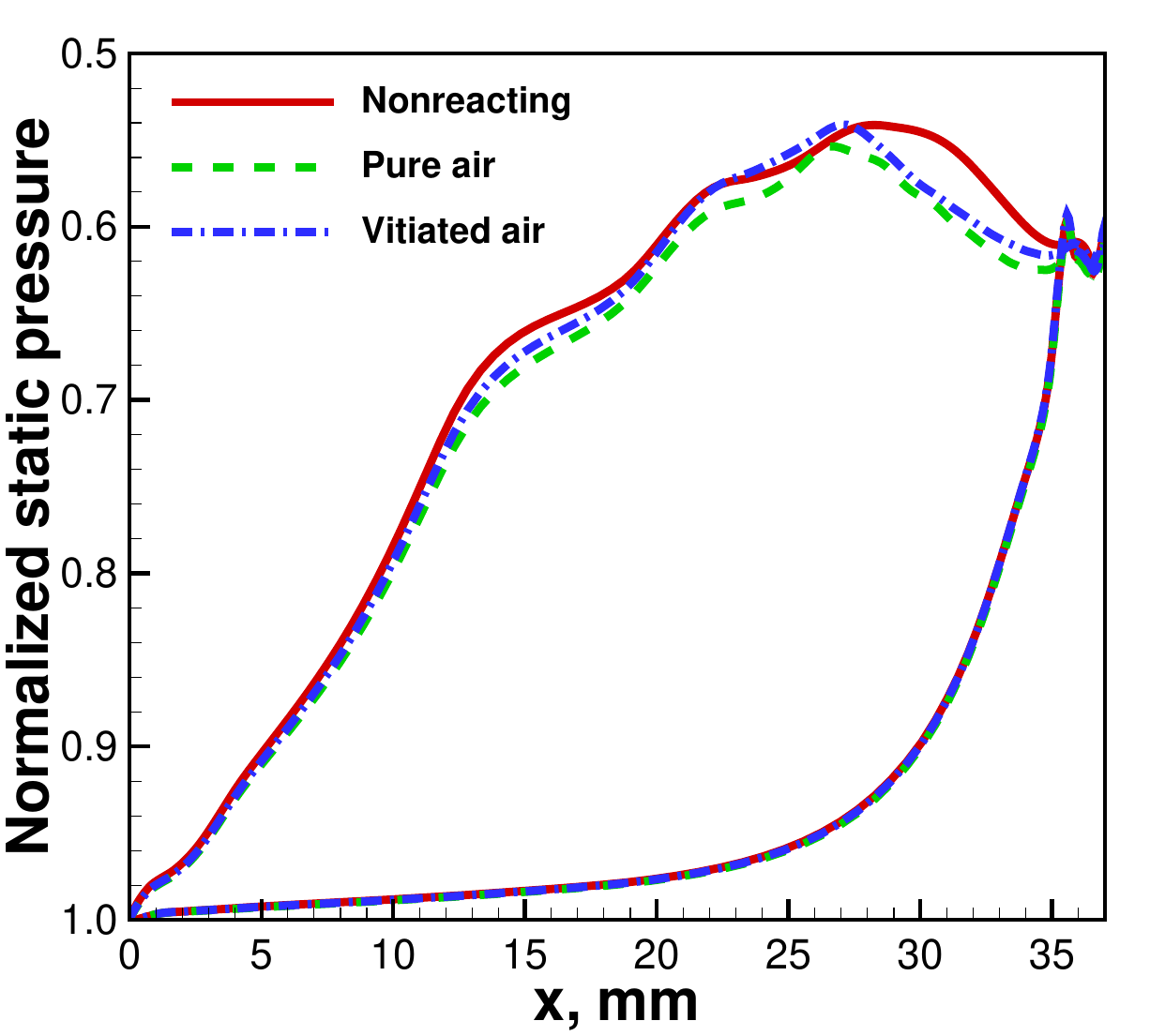}
    \caption{Distributions of pressure over turbine blade}
    \label{fig:p-x_turbine_wall}
\end{figure}

Chemical reaction in the turbine passage affects the aerodynamic performance of the blade. The distributions of static pressure over the turbine blade for the nonreacting case, pure air case, and vitiated air case are compared in Fig. \ref{fig:p-x_turbine_wall}, in which the static pressure is normalized by the total pressure at the inlet. The pressure distributions on the pressure surface are not affected by the combustion since the two flames in the turbine passage are far away from it. However, the pressure distributions on the suction surface are different for the three cases, especially after the suction peak. Compared to the nonreacting case, the pressure in the pure air case is higher on the suction surface since the increased temperature by the intense reaction in the passage and wake reduces the pressure diffusion in the turbine. This results in a lower pressure difference between the pressure and suction surfaces and thus reduces the aerodynamic loading of the blade. Resulting from the weaker chemical reaction, the pressure levels in the vitiated air case are between the levels in the nonreacting case and the pure air case.

\section{Conclusion} \label{sec:conclusion}
A finite-volume method for the compressible reacting Reynolds-averaged Navier-Stokes equations is developed by using a steady-state preserving splitting scheme to treat the stiff source terms. 
Laminar and turbulent reacting flows in an accelerating mixing layer are studied and compared to the boundary-layer solutions. The influence of vitiated air on the combustion process and aerodynamic performance is investigated for the cases of mixing layer and turbine cascade.

For the reacting flow in the accelerating mixing layer, a diffusion flame is established slightly biased towards the air side after the splitter plate and then transported downstream. The chemical reaction strongly enhances turbulent transport due to intensive production of turbulence by the increased velocity gradients and thus produces large turbulent viscosity in the reaction region. Compared to the laminar case, the turbulent shear-layer thickness is one order of magnitude larger. However, the basic behaviors of the two cases remain the same. 

Vitiated air has significant influence on the combustion process and aerodynamics. In the mixing layer, the peak temperature in the flame reduces while the minimum density increases compared to the pure air case. The location of peak reaction is slightly shifted upward to the air side. The thickness of shear layer is decreased due to the reduced turbulent diffusion by the weak chemical reaction. In the turbine cascade, the variations of temperature for the vitiated air case are similar to the pure air case, but the flame temperature levels are lower. Although the reaction rate for the vitiated air case is evidently reduced, the extinction does not happen within the accelerating passage.
The turbine cascade analysis indicates viability for the turbine-burner concept. Future three-dimensional large-eddy simulation with improved chemistry modeling will be pursued.


\section*{Acknowledgments}
The research was supported by the Office of Naval Research through Grant N00014-21-1-2467 with Dr. Steven Martens as program manager.

\bibliography{main}

\begin{thebibliography}{30}
\newcommand{\enquote}[1]{``#1''}
\providecommand{\natexlab}[1]{#1}
\providecommand{\url}[1]{\texttt{#1}}
\providecommand{\urlprefix}{URL }
\expandafter\ifx\csname urlstyle\endcsname\relax
  \providecommand{\doi}[1]{\discretionary{}{}{}https://doi.org/#1}\else
  \providecommand{\doi}[1]{\discretionary{}{}{}\urlstyle{rm}\url{https://doi.org/#1}}\fi

\bibitem[{Sirignano and Liu(1999)}]{sirignano1999performance}
Sirignano, W.~A., and Liu, F., \enquote{Performance Increases for Gas-Turbine
  Engines Through Combustion Inside the Turbine,} \emph{Journal of Propulsion
  and Power}, Vol.~15, No.~1, 1999, pp. 111--118.
\newblock \doi{10.2514/2.5398}.

\bibitem[{Liu and Sirignano(2001)}]{liu2001turbojet}
Liu, F., and Sirignano, W.~A., \enquote{Turbojet and Turbofan Engine
  Performance Increases Through Turbine Burners,} \emph{Journal of Propulsion
  and Power}, Vol.~17, No.~3, 2001, pp. 695--705.
\newblock \doi{10.2514/2.5797}.

\bibitem[{Sirignano et~al.(2012)Sirignano, Dunn-Rankin, Liu, Colcord, and
  Puranam}]{sirignano2012turbine}
Sirignano, W.~A., Dunn-Rankin, D., Liu, F., Colcord, B., and Puranam, S.,
  \enquote{Turbine Burners: Performance Improvement and Challenge of
  Flameholding,} \emph{AIAA Journal}, Vol.~50, No.~8, 2012, pp. 1645--1669.
\newblock \doi{10.2514/1.J051562}.

\bibitem[{Sirignano and Kim(1997)}]{sirignano1997diffusion}
Sirignano, W.~A., and Kim, I., \enquote{Diffusion Flame in a Two-Dimensional,
  Accelerating Mixing Layer,} \emph{Physics of Fluids}, Vol.~9, No.~9, 1997,
  pp. 2617--2630.
\newblock \doi{10.1063/1.869378}.

\bibitem[{Fang et~al.(2001)Fang, Liu, and Sirignano}]{fang2001ignition}
Fang, X., Liu, F., and Sirignano, W.~A., \enquote{Ignition and Flame Studies
  for an Accelerating Transonic Mixing Layer,} \emph{Journal of Propulsion and
  Power}, Vol.~17, No.~5, 2001, pp. 1058--1066.
\newblock \doi{10.2514/2.5844}.

\bibitem[{Mehring et~al.(January 2001)Mehring, Liu, and
  Sirignano}]{mehring2001ignition}
Mehring, C., Liu, F., and Sirignano, W.~A., \enquote{Ignition and Flame Studies
  for a Turbulent Accelerating Transonic Mixing Layer,} \emph{AIAA Paper
  2001-0190}, January 2001.
\newblock \doi{10.2514/6.2001-190}.

\bibitem[{Cai et~al.(January 2001)Cai, Icoz, Liu, and
  Sirignano}]{cai2001ignition}
Cai, J., Icoz, O., Liu, F., and Sirignano, W.~A., \enquote{Ignition and Flame
  Studies for Turbulent Transonic Mixing in a Curved Duct Flow,} \emph{AIAA
  Paper 2001-0189}, January 2001.
\newblock \doi{10.2514/6.2001-189}.

\bibitem[{Cai et~al.(March 2001)Cai, Icoz, Liu, and
  Sirignano}]{cai2001combustion}
Cai, J., Icoz, O., Liu, F., and Sirignano, W.~A., \enquote{Combustion in a
  Transonic Flow with Large Axial and Transverse Pressure Gradients,}
  \emph{Proceedings of the 2nd Joint Meeting of the US Sections of the
  Combustion Institute}, Oakland, March 2001.

\bibitem[{Cheng et~al.(2007)Cheng, Liu, and Sirignano}]{cheng2007nonpremixed}
Cheng, F., Liu, F., and Sirignano, W.~A., \enquote{Nonpremixed Combustion in an
  Accelerating Transonic Flow Undergoing Transition,} \emph{AIAA Journal},
  Vol.~45, No.~12, 2007, pp. 2935--2946.
\newblock \doi{10.2514/1.31146}.

\bibitem[{Cheng et~al.(2008)Cheng, Liu, and Sirignano}]{cheng2008nonpremixed}
Cheng, F., Liu, F., and Sirignano, W.~A., \enquote{Nonpremixed Combustion in an
  Accelerating Turning Transonic Flow Undergoing Transition,} \emph{AIAA
  Journal}, Vol.~46, No.~5, 2008, pp. 1204--1215.
\newblock \doi{10.2514/1.35209}.

\bibitem[{Cheng et~al.(2009)Cheng, Liu, and Sirignano}]{cheng2009reacting}
Cheng, F., Liu, F., and Sirignano, W.~A., \enquote{Reacting Mixing-Layer
  Computations in a Simulated Turbine-Stator Passage,} \emph{Journal of
  Propulsion and Power}, Vol.~25, No.~2, 2009, pp. 322--334.
\newblock \doi{10.2514/1.37739}.

\bibitem[{Zhu et~al.(June 2017)Zhu, Luo, and Liu}]{zhu2017numerical}
Zhu, Y., Luo, J., and Liu, F., \enquote{Numerical Investigation of Stator
  Clocking Effects on the Downstream Stator in a 1.5-Stage Axial Turbine,}
  \emph{Turbo Expo: Power for Land, Sea, and Air, GT2017-63273}, American
  Society of Mechanical Engineers, June 2017.
\newblock \doi{10.1115/GT2017-63273}.

\bibitem[{Walsh et~al.(January 2024 (To be available))Walsh, Zhan, Mehring,
  Liu, and Sirignano}]{walsh2024}
Walsh, S.~L., Zhan, L., Mehring, C., Liu, F., and Sirignano, W.~A.,
  \enquote{Turbulent Accelerating Combusting Flows with a Methane-Vitiated Air
  Flamelet Model,} \emph{AIAA Paper-xxxx}, January 2024 (To be available).
\newblock \doi{xx.xxxx/x.xxxxx}.

\bibitem[{Mcbride et~al.(October 1993)Mcbride, Gordon, and
  Reno}]{mcbride1993coefficients}
Mcbride, B.~J., Gordon, S., and Reno, M.~A., \enquote{Coefficients for
  Calculating Thermodynamic and Transport Properties of Individual Species,}
  NASA Technical Memorandum 4513, October 1993.

\bibitem[{White(2006)}]{white2006viscous}
White, F.~M., \emph{Viscous Fluid Flow}, 3\textsuperscript{rd} ed.,
  McGraw-Hill, New York, 2006, Chap. Appendix A.

\bibitem[{Menter et~al.(2003)Menter, Kuntz, and Langtry}]{menter2003ten}
Menter, F.~R., Kuntz, M., and Langtry, R., \enquote{Ten Years of Industrial
  Experience with the SST Turbulence Model,} \emph{Turbulence, Heat and Mass
  Transfer 4}, edited by K.~Hanjalic, Y.~Nagano, and M.~Tummers, Begell House,
  Danbury, CT, 2003, pp. 625--632.

\bibitem[{Westbrook and Dryer(1984)}]{westbrook1984chemical}
Westbrook, C.~K., and Dryer, F.~L., \enquote{Chemical Kinetic Modeling of
  Hydrocarbon Combustion,} \emph{Progress in Energy and Combustion Science},
  Vol.~10, No.~1, 1984, pp. 1--57.
\newblock \doi{10.1016/0360-1285(84)90118-7}.

\bibitem[{Liu and Jameson(1993)}]{liu1993multigrid}
Liu, F., and Jameson, A., \enquote{Multigrid Navier-Stokes Calculations for
  Three-Dimensional Cascades,} \emph{AIAA Journal}, Vol.~31, No.~10, 1993, pp.
  1785--1791.
\newblock \doi{10.2514/3.11850}.

\bibitem[{Yao et~al.(2001)Yao, Jameson, Alonso, and Liu}]{yao2001development}
Yao, J., Jameson, A., Alonso, J.~J., and Liu, F., \enquote{Development and
  Validation of a Massively Parallel Flow Solver for Turbomachinery Flows,}
  \emph{Journal of Propulsion and Power}, Vol.~17, No.~3, 2001, pp. 659--668.
\newblock \doi{10.2514/2.5793}.

\bibitem[{Sadeghi and Liu(2005)}]{sadeghi2005computation}
Sadeghi, M., and Liu, F., \enquote{Computation of Cascade Flutter by Uncoupled
  and Coupled Methods,} \emph{International Journal of Computational Fluid
  Dynamics}, Vol.~19, No.~8, 2005, pp. 559--569.
\newblock \doi{10.1080/10618560500508367}.

\bibitem[{Zhu et~al.(2018{\natexlab{a}})Zhu, Luo, and Liu}]{zhu2018flow}
Zhu, Y., Luo, J., and Liu, F., \enquote{Flow Computations of Multi-Stages by
  URANS and Flux Balanced Mixing Models,} \emph{Science China Technological
  Sciences}, Vol.~61, 2018{\natexlab{a}}, pp. 1081--1091.
\newblock \doi{10.1007/s11431-017-9262-9}.

\bibitem[{Zhu et~al.(2018{\natexlab{b}})Zhu, Luo, and Liu}]{zhu2018influence}
Zhu, Y., Luo, J., and Liu, F., \enquote{Influence of Blade Lean Together with
  Blade Clocking on the Overall Aerodynamic Performance of a Multi-Stage
  Turbine,} \emph{Aerospace Science and Technology}, Vol.~80,
  2018{\natexlab{b}}, pp. 329--336.
\newblock \doi{10.1016/j.ast.2018.07.016}.

\bibitem[{Jameson et~al.(June 1981)Jameson, Schmidt, and
  Turkel}]{jameson1981numerical}
Jameson, A., Schmidt, W., and Turkel, E., \enquote{Numerical Solution of the
  Euler Equations by Finite Volume Methods Using Runge Kutta Time Stepping
  Schemes,} \emph{AIAA Paper 1981-1259}, June 1981.
\newblock \doi{10.2514/6.1981-1259}.

\bibitem[{Yoon and Jameson(1988)}]{yoon1988lower}
Yoon, S., and Jameson, A., \enquote{Lower-Upper Symmetric-Gauss-Seidel Method
  for the Euler and Navier-Stokes Equations,} \emph{AIAA Journal}, Vol.~26,
  No.~9, 1988, pp. 1025--1026.
\newblock \doi{10.2514/3.10007}.

\bibitem[{Strang(1968)}]{strang1968construction}
Strang, G., \enquote{On the Construction and Comparison of Difference Schemes,}
  \emph{SIAM Journal on Numerical Analysis}, Vol.~5, No.~3, 1968, pp. 506--517.
\newblock \doi{10.1137/0705041}.

\bibitem[{Wu et~al.(2019)Wu, Ma, and Ihme}]{wu2019efficient}
Wu, H., Ma, P.~C., and Ihme, M., \enquote{Efficient Time-Stepping Techniques
  for Simulating Turbulent Reactive Flows with Stiff Chemistry,} \emph{Computer
  Physics Communications}, Vol. 243, 2019, pp. 81--96.
\newblock \doi{10.1016/j.cpc.2019.04.016}.

\bibitem[{Mott et~al.(2000)Mott, Oran, and van Leer}]{mott2000quasi}
Mott, D.~R., Oran, E.~S., and van Leer, B., \enquote{A Quasi-Steady-State
  Solver for the Stiff Ordinary Differential Equations of Reaction Kinetics,}
  \emph{Journal of Computational Physics}, Vol. 164, No.~2, 2000, pp. 407--428.
\newblock \doi{10.1006/jcph.2000.6605}.

\bibitem[{Wilcox(1988)}]{wilcox1988reassessment}
Wilcox, D.~C., \enquote{Reassessment of the Scale-Determining Equation for
  Advanced Turbulence Models,} \emph{AIAA Journal}, Vol.~26, No.~11, 1988, pp.
  1299--1310.
\newblock \doi{10.2514/3.10041}.

\bibitem[{Menter(1992)}]{menter1992influence}
Menter, F.~R., \enquote{Influence of Freestream Values on k-omega Turbulence
  Model Predictions,} \emph{AIAA Journal}, Vol.~30, No.~6, 1992, pp.
  1657--1659.
\newblock \doi{10.2514/3.11115}.

\bibitem[{Arts and Lambert~de Rouvroit(1992)}]{arts1992aero}
Arts, T., and Lambert~de Rouvroit, M., \enquote{Aero-Thermal Performance of a
  Two-Dimensional Highly Loaded Transonic Turbine Nozzle Guide Vane: A Test
  Case for Inviscid and Viscous Flow Computations,} \emph{Journal of
  Turbomachinery}, Vol. 114, No.~1, 1992, pp. 147--154.
\newblock \doi{10.1115/1.2927978}.

\end{thebibliography}

\end{document}